\documentclass[%
  reprint,
superscriptaddress,
nofootinbib,
 amsmath,amssymb,
 aps,
 prc,
 showkeys,
]{revtex4-1}

\usepackage{graphicx}
\usepackage{dcolumn}
\usepackage{bm}
\usepackage{subfigure}
\usepackage{float} 
\usepackage{hyperref}
\usepackage[mathlines]{lineno}
\usepackage{tabularx}
\usepackage{enumitem}
\usepackage{array}
\usepackage{longtable}
\usepackage{blindtext}

\newcommand{\be}{\begin{equation}}
\newcommand{\ee}{\end{equation}}
\newcommand{\bea}{\begin{eqnarray}}
\newcommand{\eea}{\end{eqnarray}}

\begin{document}

\preprint{BeAGLE-BNL}

\title{BeAGLE: Benchmark $e$A Generator for LEptoproduction \\ 
in high energy lepton-nucleus collisions}
\author{Wan~Chang}
\email{changwan@mails.ccnu.edu.cn}
\affiliation{%
 Key Laboratory of Quark and Lepton Physics (MOE) and Institute of Particle Physics, Central China Normal University, Wuhan 430079,China
}%
\affiliation{%
 Department of Physics, Brookhaven National Laboratory, Upton, New York 11973, USA
}%
\author{Elke-Caroline~Aschenauer }
\email{elke@bnl.gov}
\affiliation{%
 Department of Physics, Brookhaven National Laboratory, Upton, New York 11973, USA
}%
\author{Mark~D.~Baker}
 \email{mdbaker@mdbpads.com}
\affiliation{%
 Mark D. Baker Physics and Detector Simulations LLC, Miller Place, New York 11764, USA
}%
\author{Alexander~Jentsch}
\affiliation{%
 Department of Physics, Brookhaven National Laboratory, Upton, New York 11973, USA
}%
\author{Jeong-Hun~Lee}
\affiliation{%
 Department of Physics, Brookhaven National Laboratory, Upton, New York 11973, USA
}%
\author{Zhoudunming~Tu}
\email{zhoudunming@bnl.gov}
\affiliation{%
 Department of Physics, Brookhaven National Laboratory, Upton, New York 11973, USA
}%
 \author{Zhongbao~Yin}
 \affiliation{%
  Key Laboratory of Quark and Lepton Physics (MOE) and Institute of Particle Physics, Central China Normal University, Wuhan 430079,China
 }%
\author{Liang~Zheng}
\email{zhengliang@cug.edu.cn}
\affiliation{
 School of Mathematics and Physics, China University of Geosciences (Wuhan), Wuhan 430074, China
}%

\date{\today}

\begin{abstract}
The upcoming Electron-Ion Collider (EIC) will address several outstanding puzzles in modern nuclear physics. Topics such as the partonic structure of nucleons and nuclei, the origin of their mass and spin, among others, can be understood via the study of high energy electron-proton ($ep$) and electron-nucleus ($e$A) collisions. Achieving the scientific goals of the EIC will require a novel electron-hadron collider and detectors capable to perform high-precision measurements, but also dedicated tools to model and interpret the data. To aid in the latter, we present a general-purpose $e$A Monte Carlo (MC) generator - BeAGLE. In this paper, we provide a general description of the models integrated into BeAGLE, applications of BeAGLE in $e$A physics, implications for detector requirements at the EIC, and the tuning of the parameters in BeAGLE based on available experimental data. Specifically, we focus on a selection of model and data comparisons in particle production in both $ep$ and $e$A collisions, where baseline particle distributions provide essential information to characterize the event. In addition, we investigate the collision geometry determination in $e$A collisions, which could be used as an experimental tool for varying the nuclear density. 
\end{abstract}

\maketitle

\section{\label{sec:intro}Introduction}
One of the pillars of the Standard Model~\cite{Gaillard:1998ui, Oerter:2006iy} is the theory of Quantum Chromodynamics (QCD), which describes the mechanism for the interactions between quarks and gluons~\cite{Politzer:1974fr}. It is a self-contained fundamental theory of quark and gluon fields that is rich in symmetries~\cite{PhysRevLett.30.1343,PhysRevD.8.3633}. However, despite the successes of QCD, many fundamental questions remain open to-date, some of which will have to be addressed by a highly anticipated new machine - the Electron-Ion Collider (EIC)~\cite{Accardi:2012qut,AbdulKhalek:2021gbh}. 

The upcoming U.S.-based EIC is being designed to achieve a wide range of center-of-mass energies from 20 to 140 GeV, ion beams from deuteron to heavy nuclei (e.g. lead), high luminosities of $10^{33-34} \rm{cm}^{-2} \rm{s}^{-1}$, and highly polarized (70\%) electron, proton, and light-ion beams~\cite{ref:EICCDR}. The EIC will be the world's first dedicated electron-nucleus collider and the first collider to scatter polarized electrons off polarized light ions. The EIC science covers a broad range of topics from detailed investigations of hadronic structure with unprecedented precision to exploring new regimes of strongly interacting matter~\cite{Gelis:2010nm, Jalilian-Marian:2014ica}. The EIC will allow us to investigate the full three-dimensional dynamics of the proton, going well beyond the information about the longitudinal momentum nuclear structure contained in colinear parton distributions. With the unique capability to reach a wide range of momentum transfer $Q^{2}$ and Bjorken-$x$ ($x_{\rm {Bj}}$) values, the EIC can offer the most powerful tool to precisely quantify how the spin of gluons and that of quarks of various flavors contribute to the proton spin. Another frontier of the EIC science is to understand the formation of nuclei and their partonic structure. Particularly, the nucleus itself is an unprecedented QCD laboratory, where novel nuclear phenomena can be systematically studied by colliding electrons with different nuclear species~\cite{Accardi:2012qut}. 

However, the challenge of achieving the entire EIC science program via a single machine and a general-purpose detector is also unprecedented. The design of the Interaction Region (IR) and integration of a general purpose collider detector, along with its ancillary detectors over $\pm$40 meters along the beam-lines requires careful planning. This design has to be guided and optimized via simulations of the physics processes and their kinematics to achieve the optimal placement of the detectors to maximize geometric acceptance, and to aid in identification of the best technologies. Therefore, a general-purpose $e$A Monte Carlo (MC) model suitable for both investigating the physics and the impact of the machine design is sorely needed. 


The BeAGLE (\textbf{B}enchmark \textbf{eA} \textbf{G}enerator for \textbf{LE}ptoproduction) general-purpose MC generator simulates $e$A collisions with the production of exclusive final-state particles, including the fragments from the nuclear remnant breakup process~\cite{Beagle}. Prior to the present paper, it has already been used extensively for exploring physics with final-state particles at pseudorapidities $>$ 4.5, e.g., diffractive and spectator-tagging physics, and the associated detector/IR integration requirements for the ``far-forward" region (ion-going direction) at the EIC~\cite{AbdulKhalek:2021gbh}. Key physics topics at the EIC, which are very demanding on far-forward detection, include tagging and vetoing of incoherent Vector Meson (VM) production in $e$Pb collisions to enable studies of gluon imaging in nuclei~\cite{Chang:2021jnu} and tagging of the spectator nucleon in $e$D scattering to allow for the extraction of free nucleon structure~\cite{Jentsch:2021qdp}, as well as to study Short-Range Correlations~\cite{Hen:2016kwk, Atti2015InmediumSD} in the deuteron~\cite{Tu:2020ymk}. 

The design of the far-forward detectors and subsequent IR integration issues are urgent at this time because the EIC accelerator design will soon be settled, and the detector technology choices are happening in parallel, with both efforts requiring input from the other. Therefore, in order to maximize the EIC physics output and design the interaction region that is optimized for the aforementioned scientific goals, a reliable MC generator that can describe a wide range of final-states with different kinematic regions is needed.  In this paper, we will significantly extend our focus from studies on exclusive observables~\cite{Chang:2021jnu} in the far-forward region to inclusive particle production for both forward and central regions based on BeAGLE simulations. Moreover, we will compare BeAGLE simulations with available fixed-target $e$A data to further improve the model, and systematically study the model parameter dependence on various observables. 

The outline of this paper is as follows. A detailed introduction of BeAGLE is given in Sec.~\ref{sec:generator}. In Sec.~\ref{sec:compare}, we discuss the validation process on the PYTHIA-6 MC model~\cite{Sjostrand:2006za} using the HERA leading proton data~\cite{basile1981leading}. Based on the established PYTHIA parameters, we compare the BeAGLE simulations with fixed target $\mu$A data from the E665 experiment~\cite{E665:1993trt}. In Sec.~\ref{sec:centrality}, we present results from a systematic investigation of the collision geometry, determined via the detection of neutrons from the nuclear breakup. In Sec.~\ref{sec:discussions}, we describe the future opportunities and challenges of the BeAGLE model. Finally, a summary is provided in Sec.~\ref{sec:summary}.

\section{\label{sec:generator} BeAGLE}

\begin{figure}[tbh]
\centering  
\includegraphics[width=0.49\textwidth]{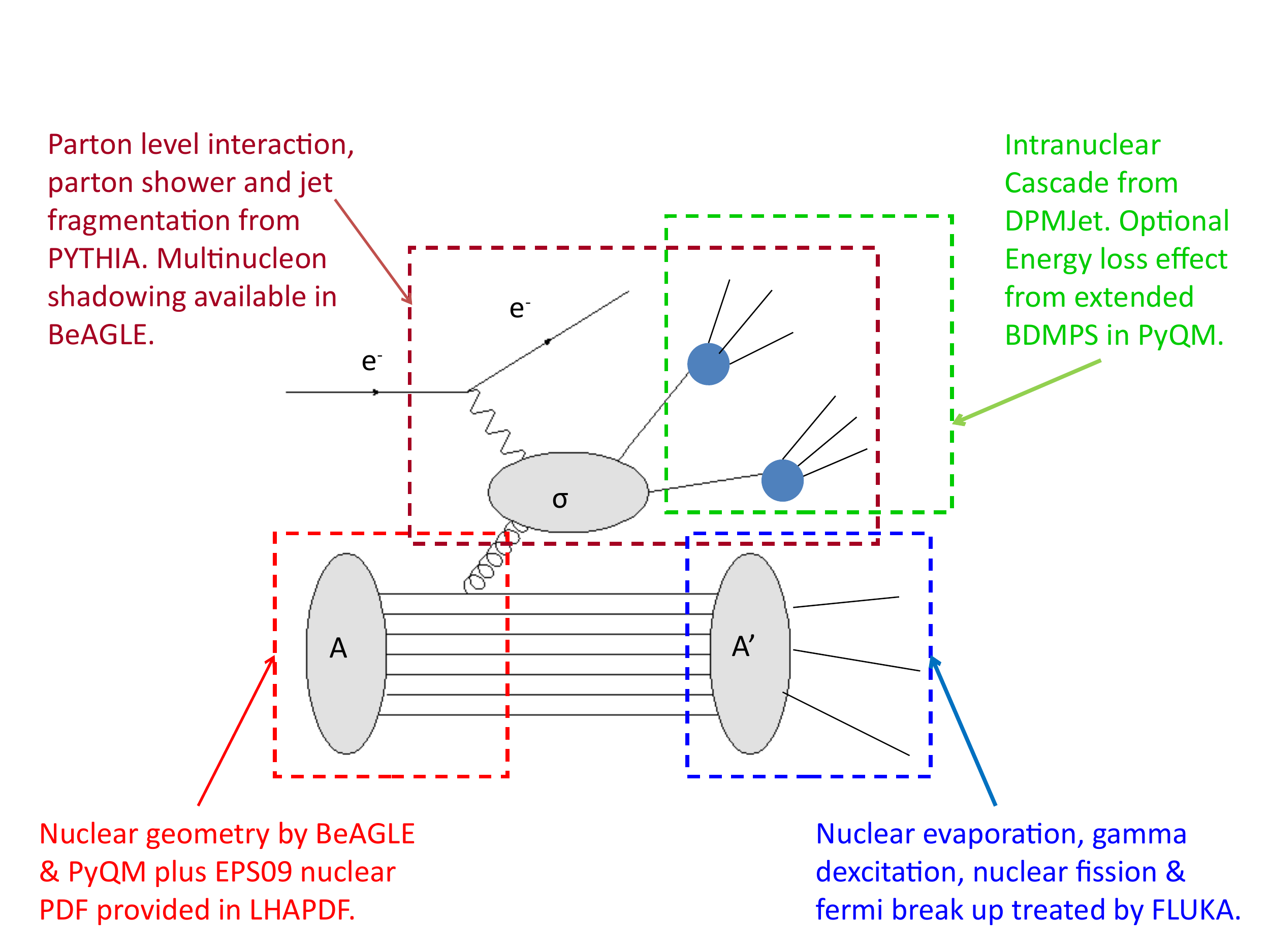}
\caption{The BeAGLE MC event generator with its main components, e.g., PYTHIA-6, DPMJet, PyQM, and FLUKA.} 
\label{fig:BEAGLE}
\end{figure}

BeAGLE is a hybrid model that uses modules from DPMJet~\cite{Roesler:2000he}, 
PYTHIA-6~\cite{Sjostrand:2006za}, PyQM~\cite{Dupre:2011afa}, FLUKA~\cite{Bohlen:2014buj,Ferrari:2005zk} and LHAPDF5~\cite{Whalley:2005nh} to describe high-energy lepto-nuclear scattering. Overall steering and optional multi-nucleon scattering (shadowing) is provided in BeAGLE, as well as an improved description of Fermi momentum distributions of nucleons in the nuclei (compared to DPMJet). DPMJet is not designed for light nuclei, so substantial changes had to be made for the case when the nucleus is a deuteron; details are described below. The geometric density distribution of nucleons in the nucleus is provided primarily by PyQM while the parton distributions within that geometry are taken from the EPS09 nuclear parton distribution functions (nPDFs)~\cite{Eskola:2009uj}. BeAGLE also allows the user to provide ``Woods-Saxon'' parameters, including non-spherical terms, to override the default geometric density description. The partonic interactions and subsequent fragmentation process is carried out by PYTHIA-6. The optional PyQM module implements the Salgado-Wiedemann quenching weights to describe partonic energy loss~\cite{SW:2003}.  Hadron formation and interactions with the nucleus through an intra-nuclear cascade are described by DPMJet. The decay of the excited nuclear remnant is described by FLUKA, including nucleon and light ion evaporation, nuclear fission, Fermi breakup of the decay fragments, and finally de-excitation by photon emission. See Fig.~\ref{fig:BEAGLE} for an illustration and the User's Guide here: \url{https://eic.github.io/software/beagle.html}.

Due to the structure of the BeAGLE generator coherent diffraction is currently not included. Since the primary interaction is modeled by PYTHIA-6 at the nucleon level, for any nuclear beam, the target nucleus will break up or at least be excited in the final state. Furthermore, for diffractive interactions, the lepton-nucleus cross section is assumed to be $A$ times the lepton-nucleon cross-section, rather than calculated from first principles. As observed in the data in $ep$ collisions at HERA, coherent diffraction in DIS was found to be 15\% of the total inclusive DIS cross section~\cite{Abramowicz:1998ii}, while in the nucleus, it has been predicted that coherent diffractive processes can be enhanced due to possible gluon saturation effects at high energy~\cite{Toll:2012mb}. Measurements of coherent diffraction in nuclei are expected to be one of the golden channels to study non-linear QCD effects\cite{Iancu:2002xk} at the EIC. 


In this framework, the lepton-nucleus collision can be illustrated in several steps as follows:
\begin{enumerate}[label=\Alph*.]
    \item The collision is simulated by selecting a struck nucleon in the nucleus according to a Glauber-type model, where the nucleon level cross section is weighted by the EPS09 nPDFs leading to an event at the partonic level; optional gluon radiation by PyQM~\cite{RoblesGajardo:2022efe}, accounting for nuclear medium effects, is available; finally, the fragmentation/hadronization is performed with the Lund string model provided by PYTHIA-6;
    \item Hadrons produced during the previous stage participate in  a ``formation-zone'' Intra-Nuclear Cascade (INC)~\cite{PhysRev.131.1801}, which produces secondary particles; 
    \item The breakup of the excited nuclear remnants will be treated by the FLUKA model. 
\end{enumerate}

\subsection{\label{subsec:hard_interaction}Hard interactions and Fermi momentum}
Initial nucleons are placed in coordinate space according to the Woods-Saxon distribution~\cite{Miller:2007ri} with intrinsic Fermi momentum, some of which will be struck off the nucleus by the exchanged virtual photon emitted by the lepton. The corresponding nucleon level cross section, $\sigma_{\gamma*N}(x,Q^2)$, is obtained from PYTHIA-6, where the magnitude of $\sigma_{\gamma*N}(x,Q^2)$ is parametrized such that the $\sigma_{\gamma*A}/(A\sigma_{\gamma*N})$ follows the EPS09 nuclear modification factor $R(x,Q^2)=f^A(x,Q^2)/f^p(x,Q^2)$~\cite{Eskola:2009uj}. This scaling feature based on the nPDFs at the cross section level enables general studies of nuclear effects.

For the hard scattering between a virtual photon and the struck nucleon discussed above, three different options of an initial collision geometry, including multiple nucleon scattering and shadowing effects~\cite{Frankfurt:1988nt, Frankfurt:2003jf, Frankfurt:2006am}, are available. The BeAGLE framework also allows a user-defined parameter, $genShd$, to switch between the different modes: i) $genShd=1$, only one nucleon is probed by the virtual photon and participates in the primary scattering simulated by PYTHIA-6; ii) $genShd=2$, if the impact parameter between the virtual photon and any nucleon is less than a distance, $D=\sqrt{\sigma_{\gamma*N}/\pi}$, one random selected nucleon will be simulated by PYTHIA-6 for an inelastic interaction, while the other nucleons will undergo elastic interactions; iii) $genShd=3$ is the same as $genShd=2$ except the order is fixed such that the first struck nucleon always undergoes inelastic scattering, while the rest scatter elastically. 

Due to nuclear binding, nucleons inside of a nucleus have internal momentum, commonly known as Fermi motion~\cite{Bodek:1980ar}. In BeAGLE, we adopt a non-relativistic model of the nucleon spectral function, provided by Ref.~\cite{CiofidegliAtti:1995qe}. This parametrization applies to all nuclei, ranging from deuterons to heavy nuclei, e.g., lead (Pb). For the case of the deuteron, the parametrization has been extended with the Light-Front formalism by Strikman \& Weiss~\cite{Strikman:2017koc}. Details from recent BeAGLE deuteron studies can be found in Refs.~\cite{Tu:2020ymk, Jentsch:2021qdp}.  

The nucleon momentum distribution in the ion rest frame is parametrized as follows, 
\be
n(k) = n_{0}(k) + n_{1}(k).
\ee
\noindent For nucleus $A=2,3,4$, 
\be
n_{0}(k)= \sum^{m_{0}}_{i=1}{A^{(0)}_{i}}\frac{e^{-B^{(0)}_{i}k^{2}}}{(1+C^{(0)}_{i}k^{2})^{2}}.
\ee
\noindent For nucleus $A>4$,

\begin{align}
& n_{0}(k) = A^{(0)}e^{-B^{(0)}k^{2}} [1+C^{(0)}k^{2}+D^{(0)}k^{4}+E^{(0)}k^{6}+\\
\nonumber
& F^{(0)}k^{8}].
\end{align}
\noindent Here $k$ is the internal nucleon momentum and $A^{(0)}$ to $F^{(0)}$ are parameters given in Table A1-A2 in Ref.~\cite{CiofidegliAtti:1995qe}. Note that $n_{0}(k)$ describes the low momentum part of the wave function, or the Mean-Field region, while $n_{1}(k)$ describes the high momentum tail, known as the Short-Range Correlation region. Currently, only $n_{0}(k)$ has been implemented in BeAGLE. However, for the deuteron case, $n(k) = n_{0}(k)$, where Short-Range Correlations in the high momentum tail have been studied in Ref.~\cite{Tu:2020ymk}. In addition, for $A>4$, the parametrizations are based on a few typical nuclei $A_{\rm{typical}}$, e.g., Carbon-12, Oxygen-16, Calcium-40, Iron-56, Lead-208, and above. Any nucleus between them, $A_{\rm{select}}$, will use one of the nearest typical nuclei for the mass number, such that $A_{\rm{select}}<A_{\rm{typical}}$. Differences in $n_{0}(k)$ for various mass numbers $A$ are generally small for $A >$ 12. 

BeAGLE currently does not account for the Fermi motion in the DIS cross section calculations, where kinematic distributions, e.g., $x$ and $Q^{2}$, are unmodified from the PYTHIA-6 generator. Accounting for the Fermi momentum would violate energy-momentum conservation because the primary interaction simulated by PYTHIA-6 assumes an on-shell nucleon mass. The higher the off-shell mass as determined from the nuclear wave function, the more violation in energy and momentum it will cause. In order to correct for this artifact, the excess energy and momentum are absorbed by the remnant nucleus. 

However, in the case of deuteron (or light nuclei in general) this correction will not be reasonable because there is only one spectator nucleon (or a few spectator nucleons) in the system. The correction would artificially distort the spectator momentum distribution. Therefore, for the deuteron, we leave the spectator unmodified, where the energy and momentum are corrected by the outgoing particles from the current fragmentation\footnote{This correction can be switched on and off in the BeAGLE control card.}. By using this approach, the spectator tagging and related physics topics can be studied with the genuine information from the wave function. No Final-State Interactions (FSI) are present in the BeAGLE generator for lepton-deuteron collisions.

\begin{figure*}[tbh]
\centering
\includegraphics[width=0.4\textwidth]{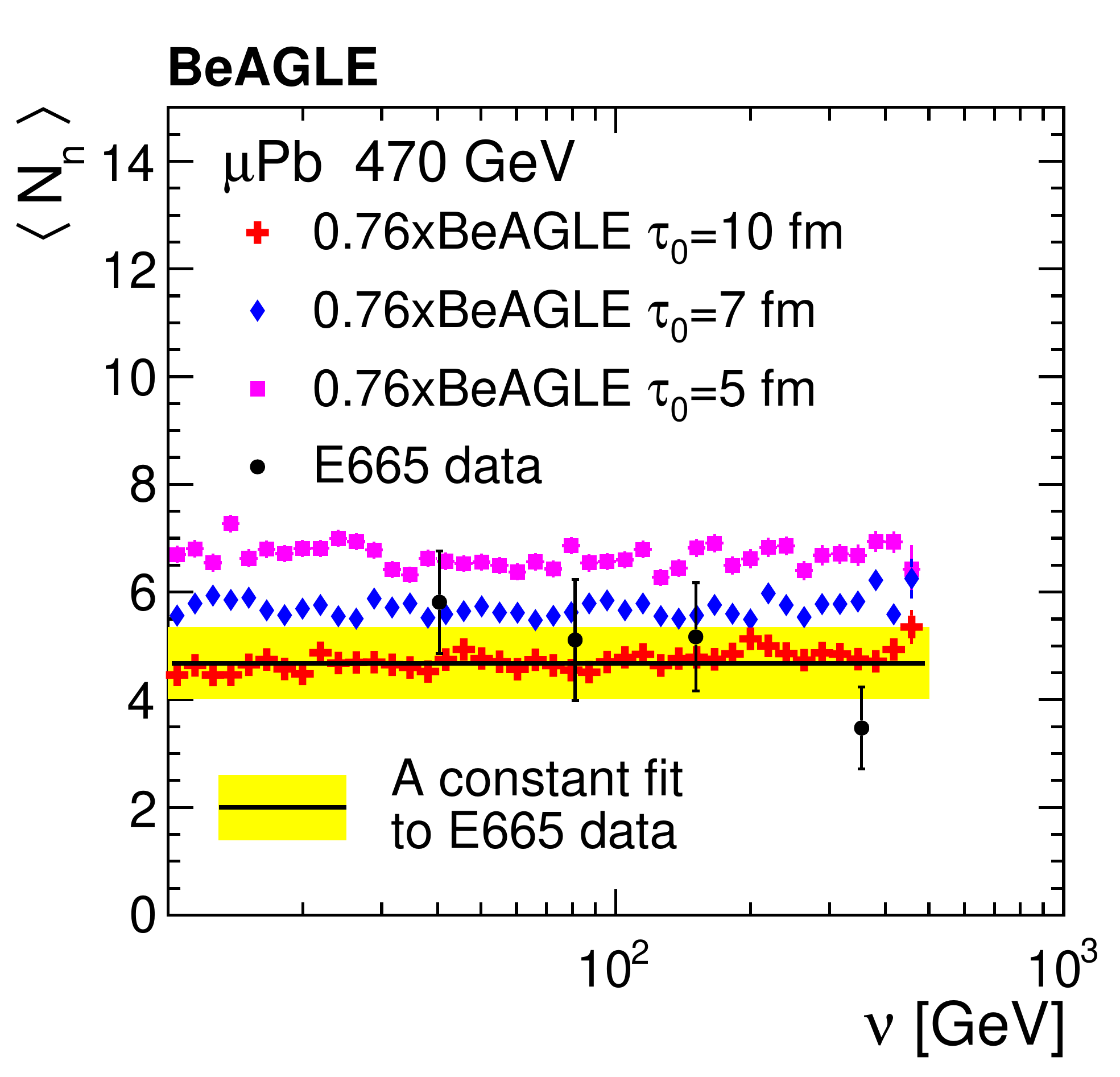}
\includegraphics[width=0.4\textwidth]{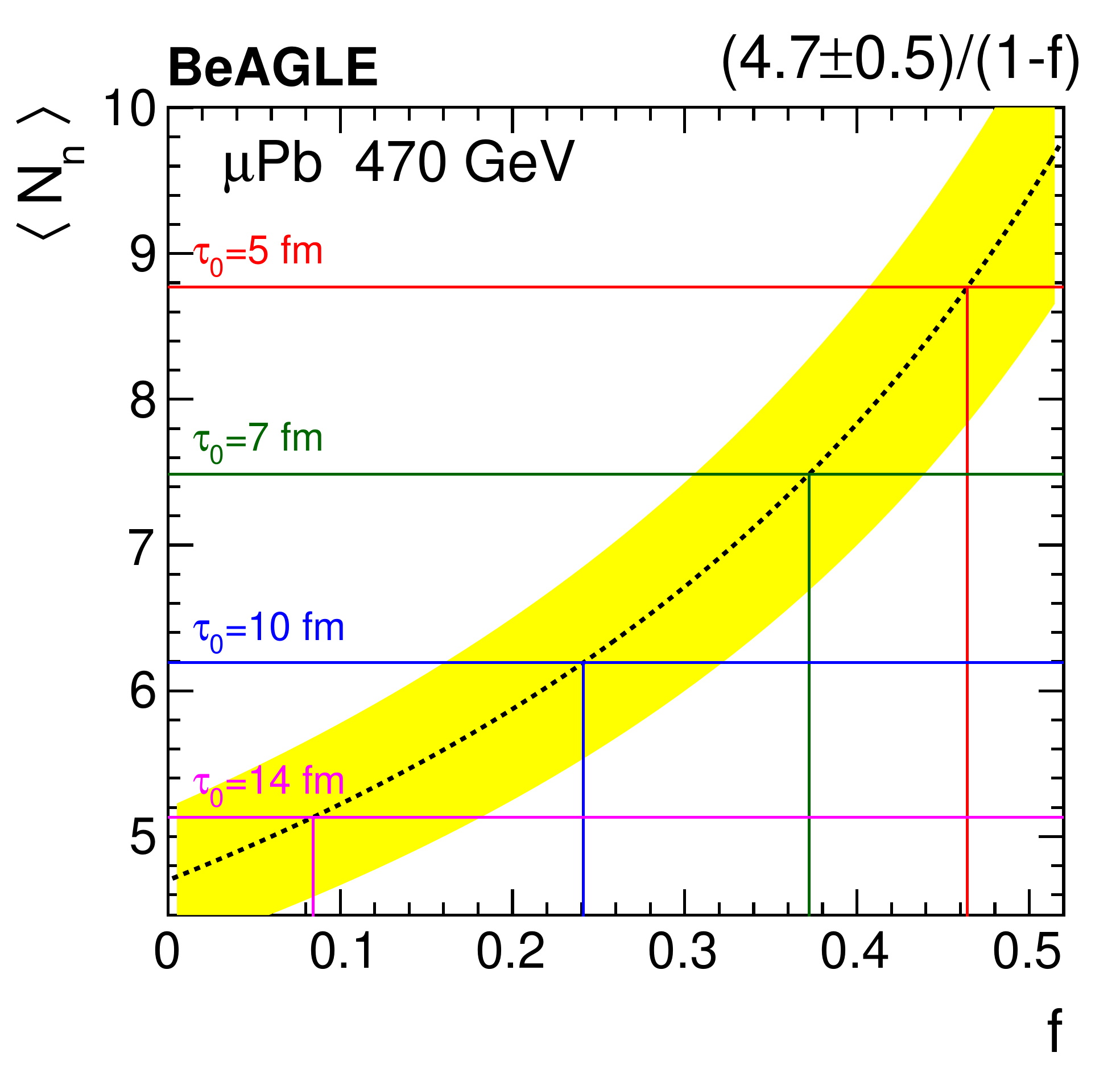}
\caption{(Left) Neutron multiplicity as a function of $\nu$ for $\mu$Pb collisions with 470 GeV muons. The results with different values of $\tau_{0}$ from the BeAGLE generator with a weight ($f=N_{\rm{coherent}}/N_{\rm{total}}$, see text for details) of 0.76 represented by different color markers. The neutron data from Ref.~\cite{E665:1995utr} are represented by the black points, the black solid line is the result of a fit to a constant function for the data result, the band shows the statistical uncertainty from the fit. (Right) The average neutron multiplicity $\left \langle N_{n} \right \rangle$ vs.\ $f$ for a variety of different values of $\tau_0$ in the BeAGLE model. }
\label{fig:tau0}
\end{figure*}

\subsection{\label{subsec:INC}Intra-nuclear Cascade} 

The generated particles from the primary scattering will be placed at the struck nucleon position and transported through the INC following the formation zone formalism implemented in DPMJet~\cite{Roesler:2000he}. Each primary particle is assigned a formation time sampled from an exponential distribution with the characteristic time scale $\tau$~\cite{Ferrari:1995cq, Zheng:2014cha} defined in the lab frame as follows:
\begin{equation}  
\tau =\tau_{0}\frac{E}{m}\frac{m^{2}}{m^{2}+p_{\perp}^{2}},
\label{equation:formationtime} 
\end{equation}
\noindent where $E$, $m$ and $p_{\perp }$ are the energy, mass, and transverse momentum of the produced particle, respectively. The parameter $\tau_{0}$ is treated as a free parameter to be determined/tuned by the experimental data. 

These produced particles can induce secondary interactions (a cascade process) if they are formed inside the nucleus. Particles with higher energy and smaller transverse mass are more likely to be formed outside and leave the nucleus without a secondary interaction. The value of the $\tau_{0}$ parameter has been systematically extracted from the experimental data in our previous publication~\cite{Chang:2021jnu}, and in the present work.   

Since forward neutron production from the evaporation process is sensitive to the INC, we use the multiplicity data of neutron emission in $\mu$Pb collisions from the E665 experiment at Fermilab~\cite{E665:1995utr} to tune the $\tau_{0}$ parameter. BeAGLE does not simulate coherent diffractive events, which do not produce neutrons in the final state. However, the E665 data do include  contributions from the coherent diffractive process. In order to properly use the neutron multiplicity to determine the $\tau_{0}$ parameter, a weight ($f=N_{\rm{coherent}}/N_{\rm{total}}$) is needed for the BeAGLE model to account for the coherent contribution in the cross section data. 

\be
N_{n}({\rm E665}) = 0* f + N_{n}({\rm BeAGLE})*(1-f).
\label{equation:E665}
\ee 
\noindent Note that the coherent event fraction $f$ was not explicitly determined in the E665 experiment~\cite{E665:1995utr}, thus a few assumptions on $f$ were made in order to determine the value of $\tau_{0}$.

Figure~\ref{fig:tau0} (left) shows the average neutron multiplicity $\left \langle N_{n} \right \rangle$ as a function of photon energy $\nu$ as measured by the E665 experiment and simulated by BeAGLE with $f=24\%$, where different values of $\tau_0$ are presented. A constant fit is performed to the E665 data, where the yellow band shows a statistical uncertainty corresponding to one standard deviation. With the assumption of $f=24\%$, the best value of $\tau_{0}$ is found to be 10 fm. The $\left \langle N_{n} \right \rangle$ from E665 is found to be $4.7\pm0.5$. The choice of $f$ is inspired by HERA measurements~\cite{Abramowicz:1998ii}, where one finds a large fraction of diffractive events, contributing about 15\% to the total deep inelastic cross-section for $ep$ collisions~\cite{Wolf:2009jm, Armesto:2019gxy}. However, theoretical studies, e.g., in Ref~\cite{Toll:2012mb}, indicate that the ratio of diffractive events to the total cross section in $e$A could be larger than what is observed in $ep$ collisions, due to non-linear QCD effects. 

In Fig.~\ref{fig:tau0} (right), the average neutron multiplicity as a function of $f$ is presented, where the dotted line with the yellow band represents a match to the E665 data with different assumptions for $f$, given by $(4.7\pm0.5) /(1-f)$. The straight colored lines show a few selected $\tau_{0}$ values with their corresponding $\left \langle N_{n} \right \rangle$ in BeAGLE. As shown in Fig.~\ref{fig:tau0} (right), for $\tau_{0} = 14$ fm, the corresponding $f$ value is less than that in $ep$ collisions at HERA, while for $\tau_{0}=5$ fm or 7 fm, the corresponding $f$ needs to be larger than 0.3, which exceeds current theoretical predictions~\cite{Toll:2012mb, Toll:2013gda}. Therefore, we use $\tau_{0}=10$~fm as the default setting in the BeAGLE model, while in the following analysis we perform systematic studies using other $\tau_{0}$ values.

\subsection{\label{subsec:evapo}Nuclear remnant breakup} 
After all possible secondary interactions are exhausted, excitation energies of the nuclear remnant can be calculated by summing up the recoil momenta transferred to the remnant by the particles leaving the nuclear potential. The breakup of the nuclear remnant is modeled using fission, the evaporation of nucleons and light nuclei, and photon emissions within the FLUKA machinery~\cite{Bohlen:2014buj,Ferrari:2005zk} for a given excitation energy. Since FLUKA is not an open-source program, the BeAGLE event generator has no handle on changing the evaporation process and can only adjust the INC in the previous step. 

\section{\label{sec:compare} Data and MC comparison}

In this section, comparisons of experimental data and the BeAGLE MC will be presented. We start with the case of $ep$ by using the PYTHIA-6 model, which is independent of the $e$A modeling in BeAGLE. The target fragmentation of the leading proton distribution has been investigated, and a good set of baseline parameters regarding the nucleon target fragmentation are established. The leading proton data are based on the ZEUS experiment at HERA~\cite{ZEUS:2008ipu}. These improvements on the PYTHIA-6 parameters will be used later in the BeAGLE event generator. After that, we will show a comparison of BeAGLE and E665 data~\cite{E665:1993trt} for inclusive charged particle rapidity distributions in both the forward and backward regions for $\mu$Xe collisions. 

\subsection{\label{pythiaandzeus}  Comparison between PYTHIA-6 and $ep$ data at ZEUS}

Since PYTHIA-6 is used to model the primary interaction in BeAGLE, it is crucial to optimize the parameters used in this stage of the framework. Leading proton data collected by the ZEUS experiment at HERA~\cite{ZEUS:2008ipu} were examined in order to optimize the PYTHIA parameters for the fragmentation $p_{\rm{T}}$ and intrinsic $k_{\rm{T}}$. Three parameters were investigated in PYTHIA and are detailed in Table~\ref{tablepythiapara}. MSTP(94) controls the energy partitioning in the beam remnant cluster decay. The default value of 3 uses the regular fragmentation function, while MSTP(94)$=$2 uses the function $P(\chi) = (k+1)(1-\chi )^{k}$, where $\chi$ is the light cone energy fraction taken by the hadron or diquark. The fragmentation function corresponding to MSTP(94)$=$2 and PARP(97)$=$6 is, $P(\chi) = 7(1-\chi )^{6}$. PARJ(170) is a parameter which we added to PYTHIA to allow separate control of the Gaussian rms $p_\mathrm{T}$ for hadrons in the recoil, which in standard PYTHIA is set to be the same as that for the string fragmentation: PARJ(21).

\begin{figure}[tbh]
\centering  
\includegraphics[width=0.4\textwidth]{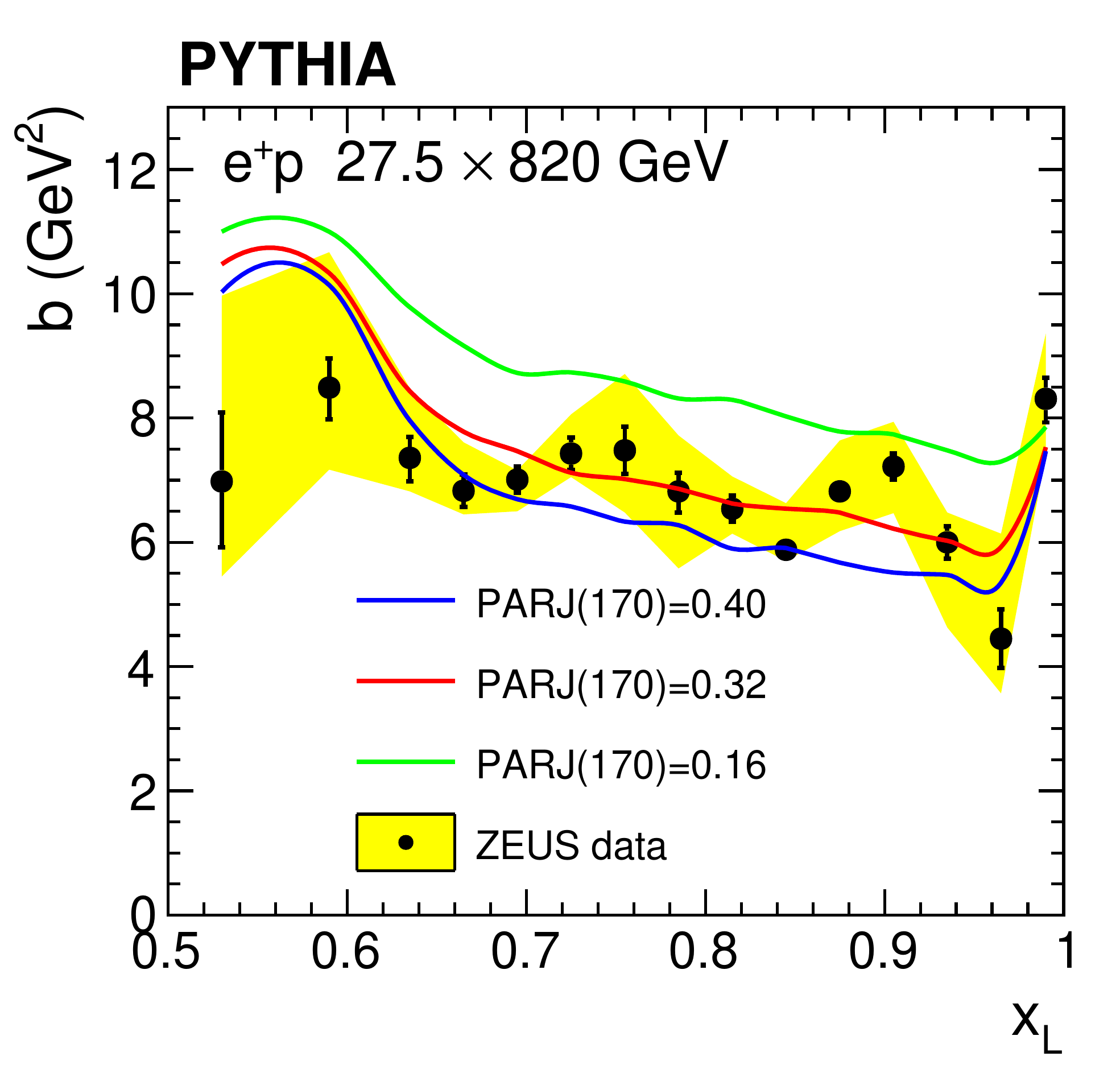}
\includegraphics[width=0.4\textwidth]{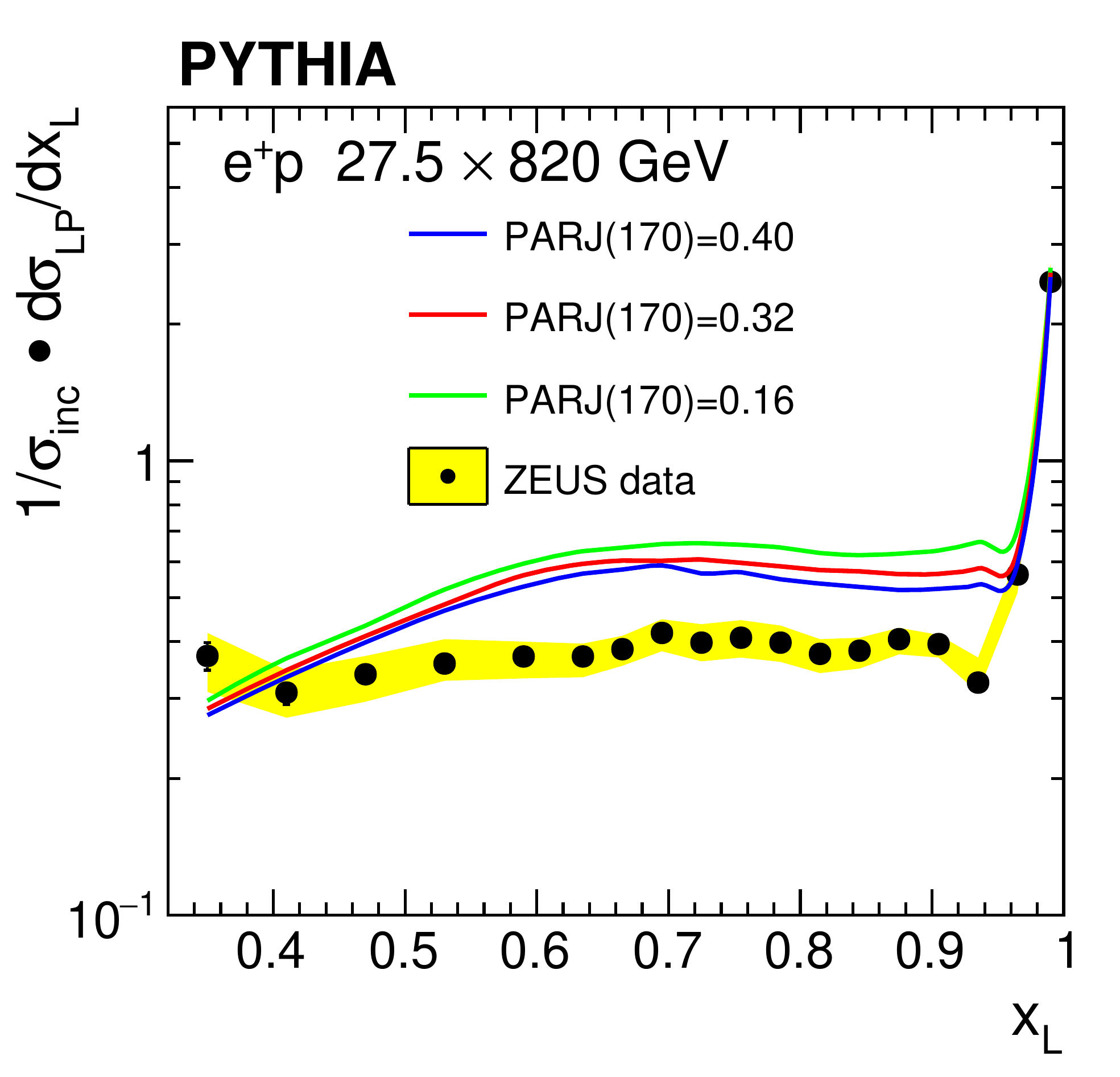}
\caption{(Top) The $p_{\rm_{T}}^{2}$-slope, b, of the cross-section $d^{2} \sigma_{LP} / dx_{L}dp_{\rm{T}}^{2}$ of leading proton,  as defined by the parameterisation $A\cdot e^{-b\cdot p_{\rm_{T}}^{2}}$ and obtained from a fit to the data in bins of $x_{L}$.
(Bottom) Single differential cross section of the leading proton normalized to the total DIS cross section $1/\sigma_{\rm{inc}}\cdot d\sigma_{\rm{LP}}/dx_{L}$. The MC results represented by lines are compared to data from Ref.~\cite{ZEUS:2008ipu}.} 
\label{FigcompZEUSxLvsb}
\end{figure}

\begin{table}[t]
\fontsize{11}{13}\selectfont
\centering
     \caption{The parameters of PYTHIA tuned by ZEUS leading proton data at HERA.}
    \begin{tabular}{c c c}
        \hline
        \hline
         Parameter & Default & Tuned  \\ \hline
         MSTP(94) & 3 & 2 \\
         PARP(97) & - & 6 \\
         PARJ(170) & - & 0.32 \\
         \hline
        \hline
    \end{tabular}    
    \label{tablepythiapara}
\end{table}

Figure~\ref{FigcompZEUSxLvsb} shows comparisons of different calculated distributions with the measurements of leading protons from the ZEUS experiment~\cite{ZEUS:2008ipu} for positron-proton scattering with beam energies of $E_{e}=27.5~\mathrm {GeV}$ and $E_{p}=820 ~ \mathrm {GeV}$. In Ref.~\cite{ZEUS:2008ipu}, the semi-inclusive reaction $e^{+}p \rightarrow e^{+}Xp$ was studied with the ZEUS detector with an integrated luminosity of 12.8 $\rm{pb}^{-1}$. The final-state proton, carrying a large fraction of the incoming proton energy, but a small transverse momentum, was detected by the ZEUS leading proton spectrometer (LPS). The selection of the LPS proton sample requires a dedicated LPS trigger and acceptance cuts to omit tracks very close to beam line or the edge of LPS detector. 

\begin{figure*}[tbh]
\centering  
\includegraphics[width=0.36\textwidth]{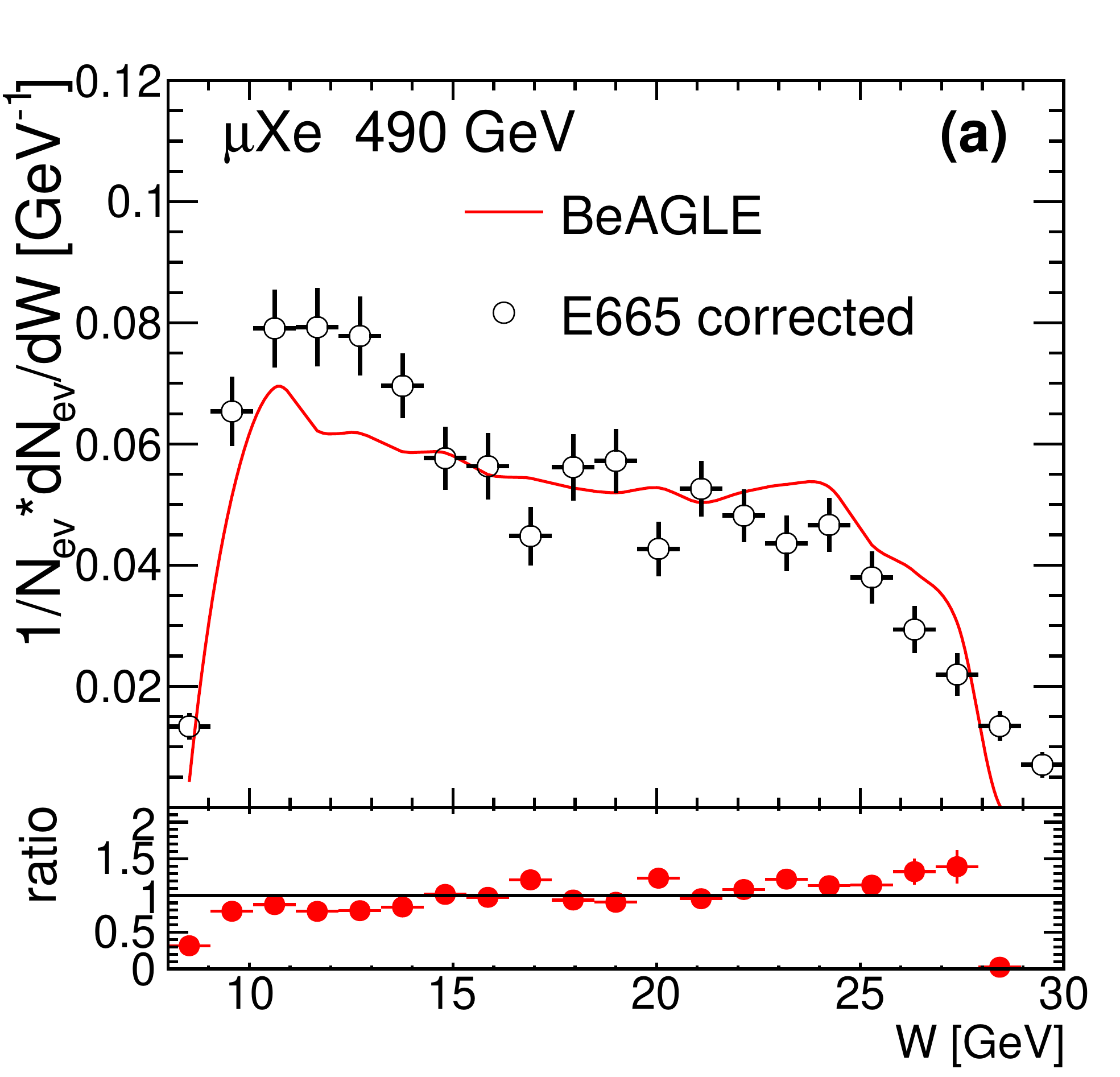}
\includegraphics[width=0.36\textwidth]{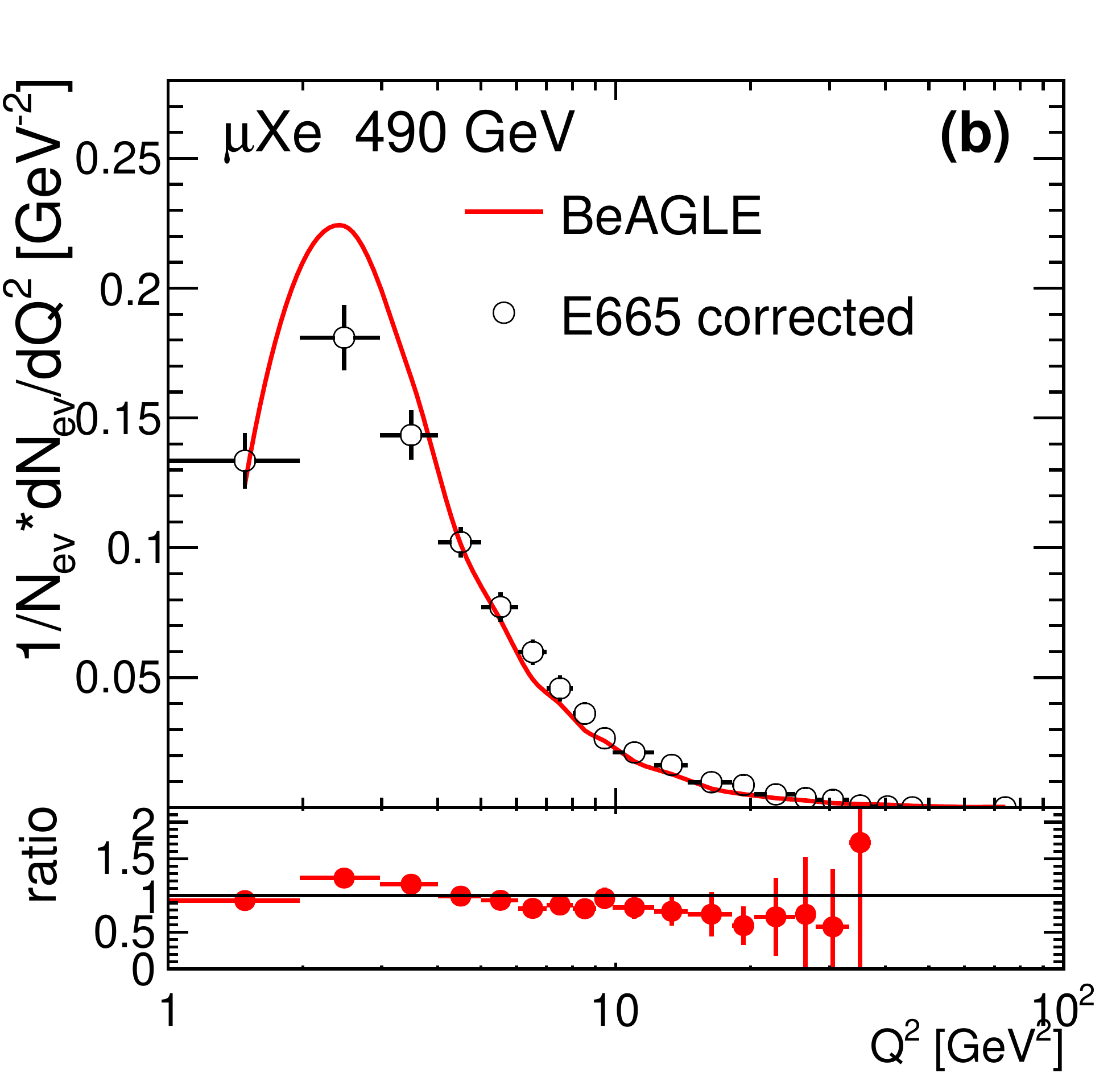}
\includegraphics[width=0.36\textwidth]{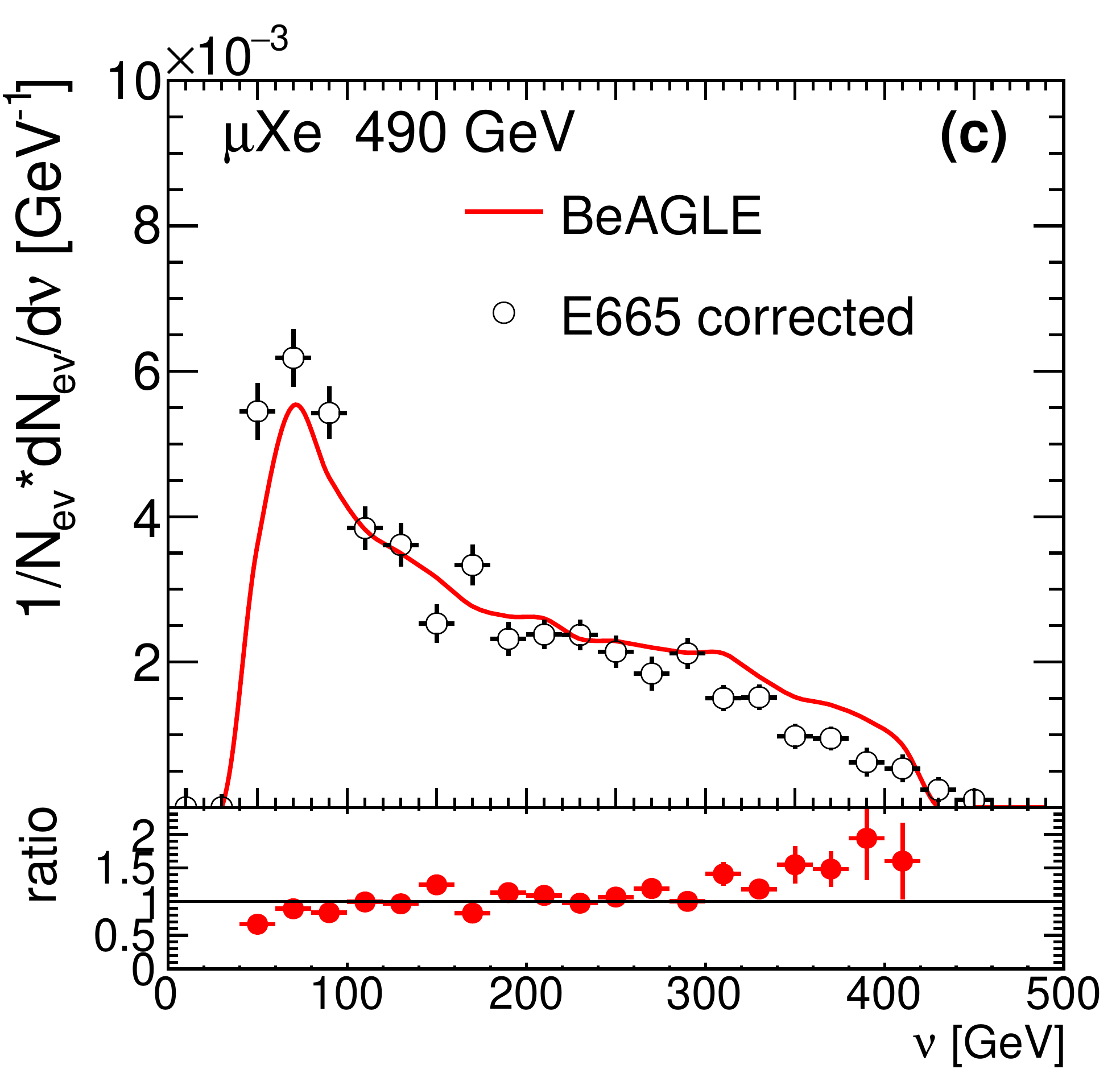}
\includegraphics[width=0.36\textwidth]{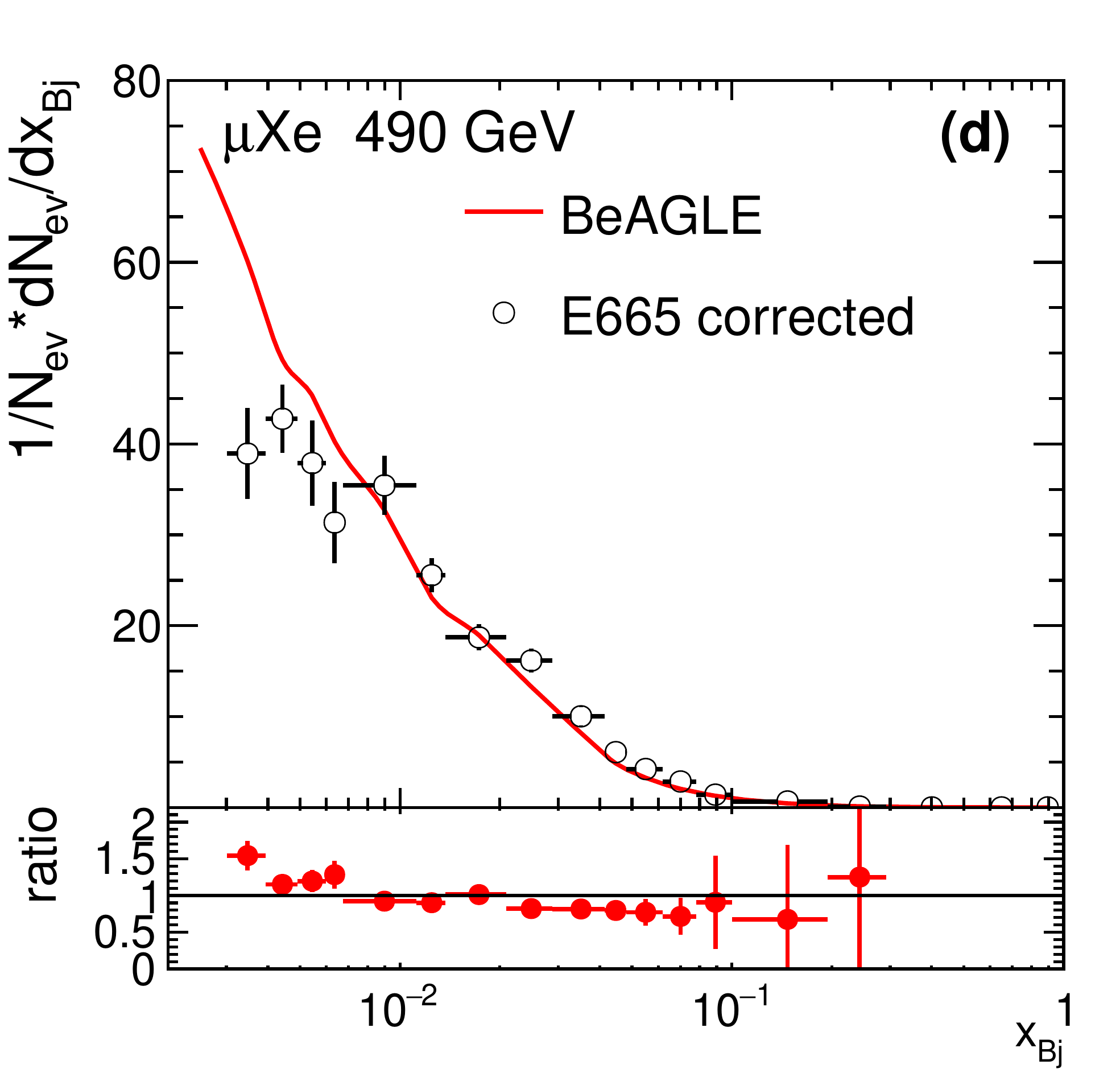}
\caption{Distributions of $W, Q^{2}, \nu, x_{\rm{bj}}$ for BeAGLE compared to the E665 data~\cite{E665:1993trt} for $\mu$Xe events. } 
\label{Figcomp}
\end{figure*}

These measurements were carried out in the kinematic range $Q^{2}>3 ~ \mathrm {GeV}^{2}, 45<W<225 ~ \mathrm {GeV}, y>0.03$ and the leading proton is measured with $p_{\rm{T}}^{2}<0.5 ~ \mathrm {GeV^{2}}, x_{L} > 0.32 $, where $x_{L}$ is the longitudinal momentum fraction of the measured proton and the incoming proton beam momentum.  Figure~\ref{FigcompZEUSxLvsb} (top) shows the $p_{\rm{T}}^{2}$-slope, $b$, of the cross-section $d^{2} \sigma_{LP} / dx_{L}dp_{\rm{T}}^{2}$ for leading protons, as defined by the parametrization $A\cdot e^{-b\cdot p_{\rm{T}}^{2}}$ and obtained from a fit to the data in bins of $x_{L}$. The black points represent the data, while the yellow band is the experimental systematic uncertainty~\cite{ZEUS:2008ipu}. The other colored lines represent the distributions for different values of PARJ(170) in PYTHIA, corrected for the leading proton spectrometer (LPS) acceptance effects. The distribution of the single differential cross section normalized to the total DIS cross section $1/\sigma_{\rm{inc}}\cdot d\sigma_{\rm{LP}}/dx_{L}$ is shown in  Fig.~\ref{FigcompZEUSxLvsb} (bottom), and the $\sigma_{\rm{inc}} = 223.9 ~ \rm{nb}$. It is found that the optimal parameter in this comparison is PARJ(170) = 0.32. 

There are two additional parameters that are sensitive to the leading proton distribution: i) PARJ(21), the width of the transverse momentum distribution in the fragmentation, and ii) PARP(91), the Gaussian width of the intrinsic $k_{\rm {T}}$ distribution. We find that the result with PARJ(21) = PARP(91) = 0.32 agrees best with the ZEUS leading proton data. However, a value of 0.4, tuned to data collected by the HERMES experiment at HERA~\cite{HERMES:2010nas, Liebing:2004us}, does a better job in the current fragmentation region. It is noted that to describe the European Muon Collaboration (EMC) data~\cite{Sloan:2690197}, higher values of the fragmentation $p_{\rm{T}}$ and intrinsic $k_{\rm{T}}$ are preferred. We use PARJ(21) = PARP(91)= 0.4 as the default settings for results presented in this study. The summary of the PYTHIA parameters used in this paper which are different from the default values are listed in the Appendix in Table~\ref{tablepythia}. 

\subsection{\label{beaglee665} Comparison between BeAGLE and the E665 $\mu$Xe data}

\begin{figure*}[tbh]

\includegraphics[width=0.36\textwidth]{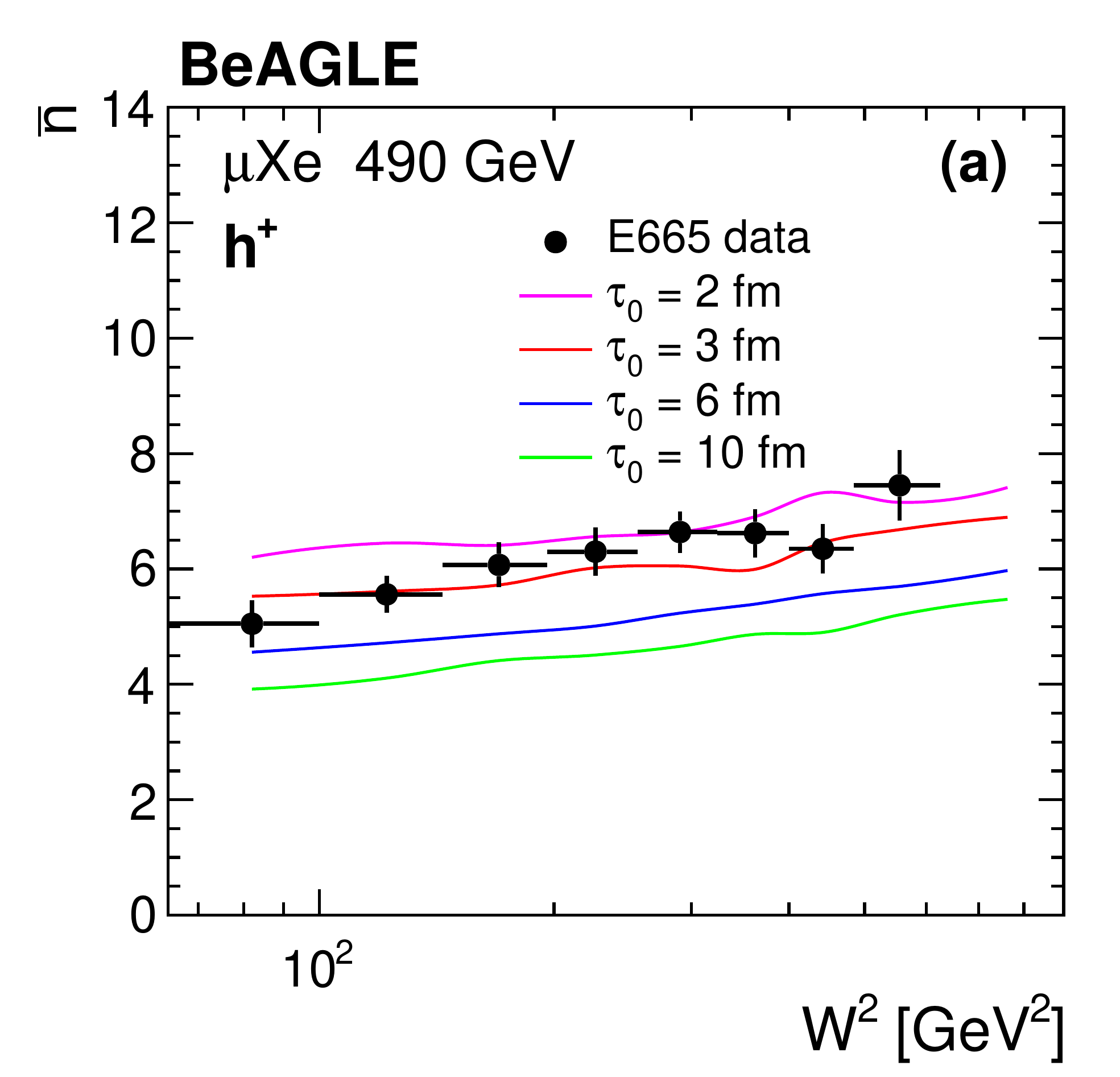}
\includegraphics[width=0.36\textwidth]{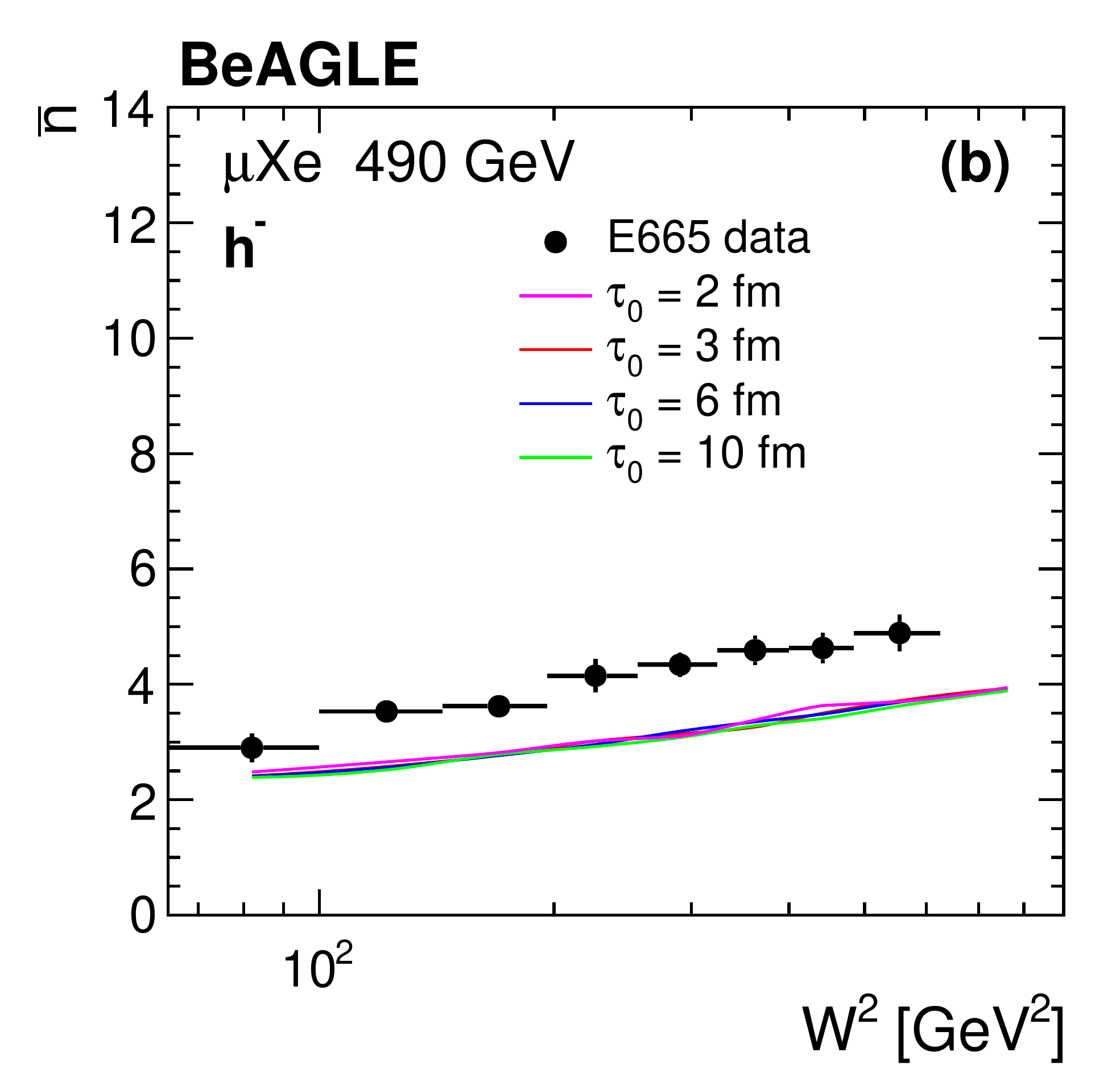}
\includegraphics[width=0.36\textwidth]{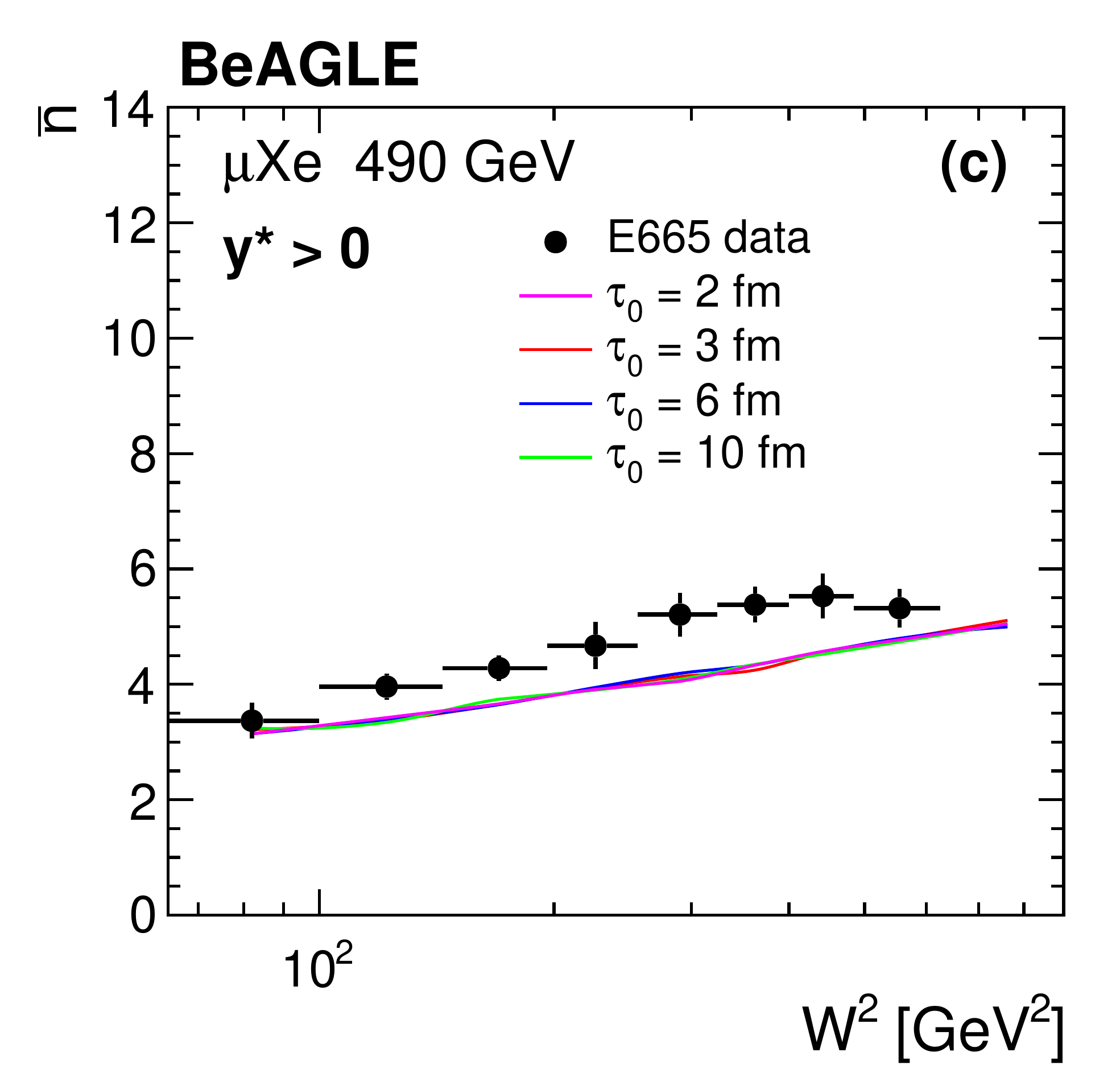}
\includegraphics[width=0.36\textwidth]{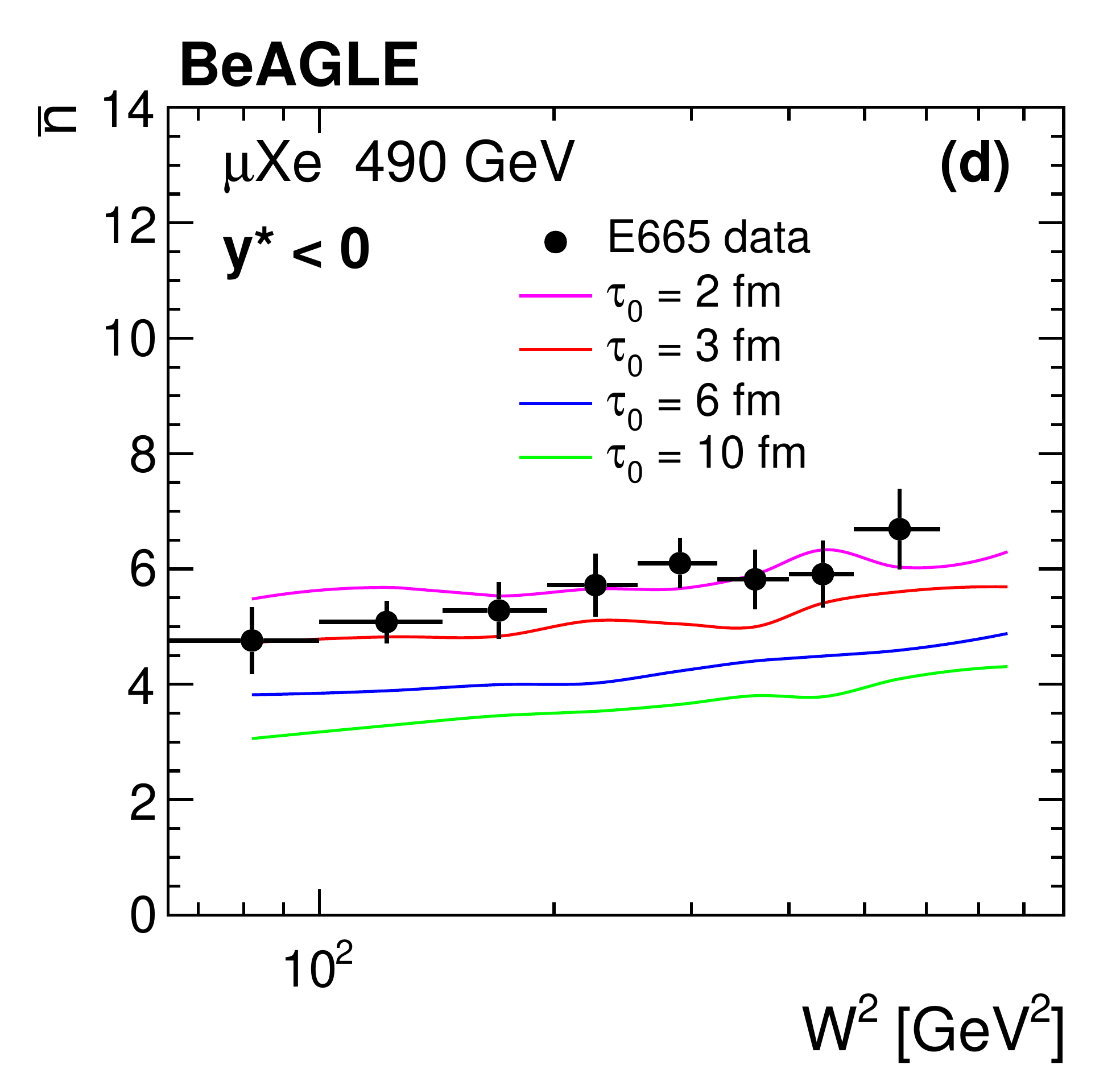}
\caption{
Multiplicity distributions as a function of $W^{2}$ for positively and negatively charged hadrons (top row) and for positive and negative $y^{*}$ (bottom row) in $\mu$Xe interactions at 490 GeV. The results are shown with different values of the parameter $\tau_{0}$ in the BeAGLE generator, represented by different colors, and are compared to data from Ref.~\cite{E665:1993trt}. }
\label{multiplicityPosiNega}
\end{figure*}

A challenge in validating the BeAGLE generator is that there are only limited $e$A collision data available to compare with. The best available data are measurements of particle production from the E665 experiment~\cite{E665:1993trt} at Fermilab. In Ref.~\cite{E665:1993trt}, the data were collected with the E665 spectrometer, which used the 490~GeV muon beam from the Tevatron at Fermilab. The experiment used a streamer chamber as a vertex detector, providing nearly $4\pi$ acceptance for charged particles. Results of charged hadron production in muon-xenon ($\mu$Xe) and muon-deuteron ($\mu$D) collisions~\cite{E665:1993trt} are used to compare with the BeAGLE model. The general picture of the interaction is the virtual photon, emitted by the incoming muon, interacts with a parton of a nucleon in the target nucleus. The hadronic center-of-mass frame (cms) is defined as the system formed by the virtual photon and the target nucleon: the struck parton is scattered into the forward direction, while the target remnant travels into the backward direction.\footnote{Note that in collider physics, the terminology of forward and backward is reversed with respect to the fixed target experiments. For comparison to the E665 data, we adopt the convention of fixed target experiments.} Given a limited particle identification capability in the E665 experiment~\cite{E665:1993trt}, all positively charged hadrons in the data with $x_{F}$($m_{\pi}$) less than $-0.2$ are assigned the proton mass, while all other positively and negatively charged hadrons are treated as pions. The variable, $x_{F}$($m_{\pi})$, is defined as $x_{F}=2p^{\ast}_{L}/W$, with $p^{\ast}_{L}$ being the longitudinal momentum of the hadron in the cms frame, assuming all particles are pions.  In order to properly compare BeAGLE simulations with the results from Ref.~\cite{E665:1993trt}, the partial identification of particles is performed in the same way on the BeAGLE simulated events. The version of the BeAGLE event generator used here is 1.01.03. The BeAGLE control card is shown in Table~\ref{tablebeagle} in the Appendix.

\begin{figure*}[tbh]
\includegraphics[width=0.36\textwidth]{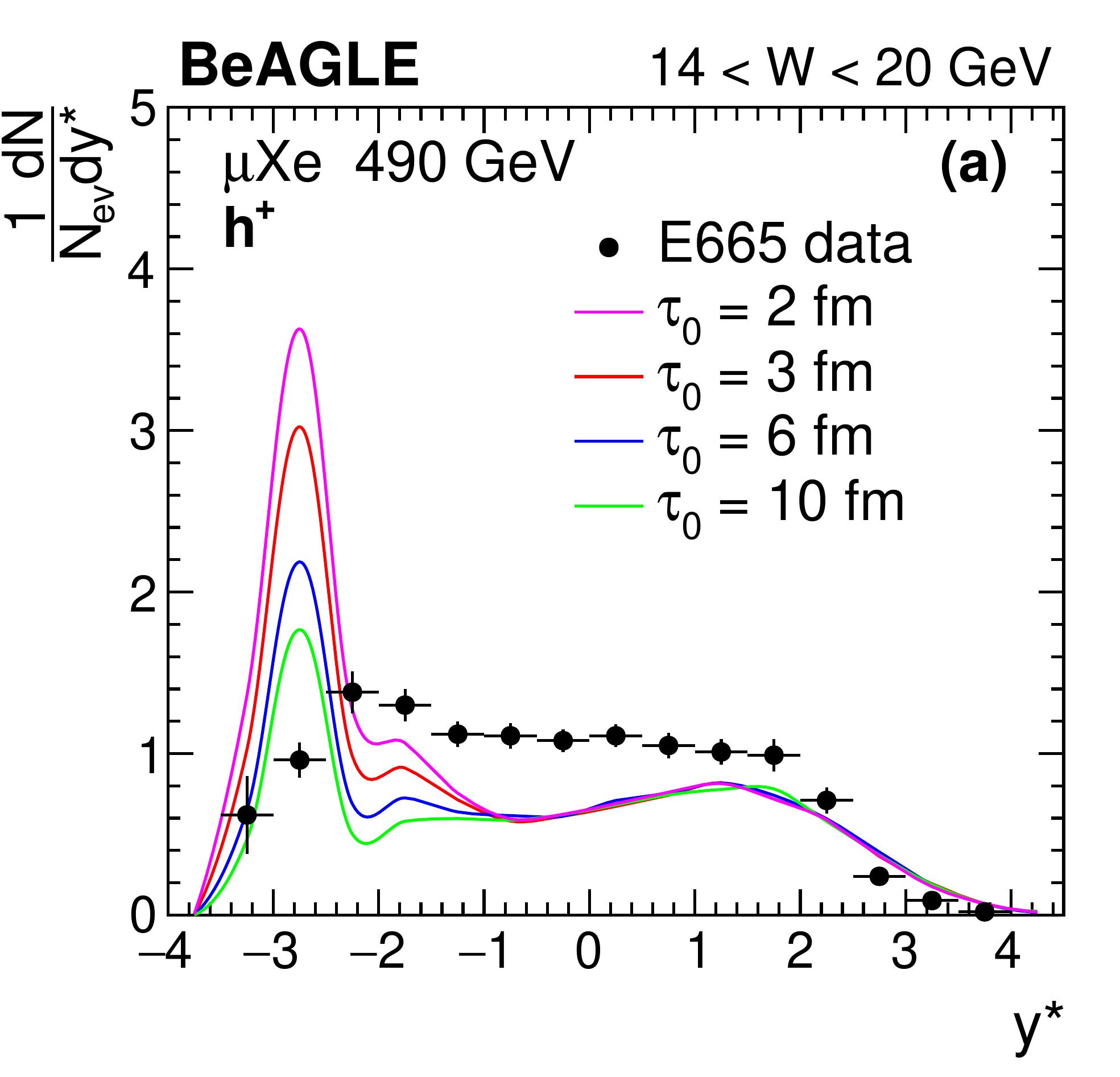}
\includegraphics[width=0.36\textwidth]{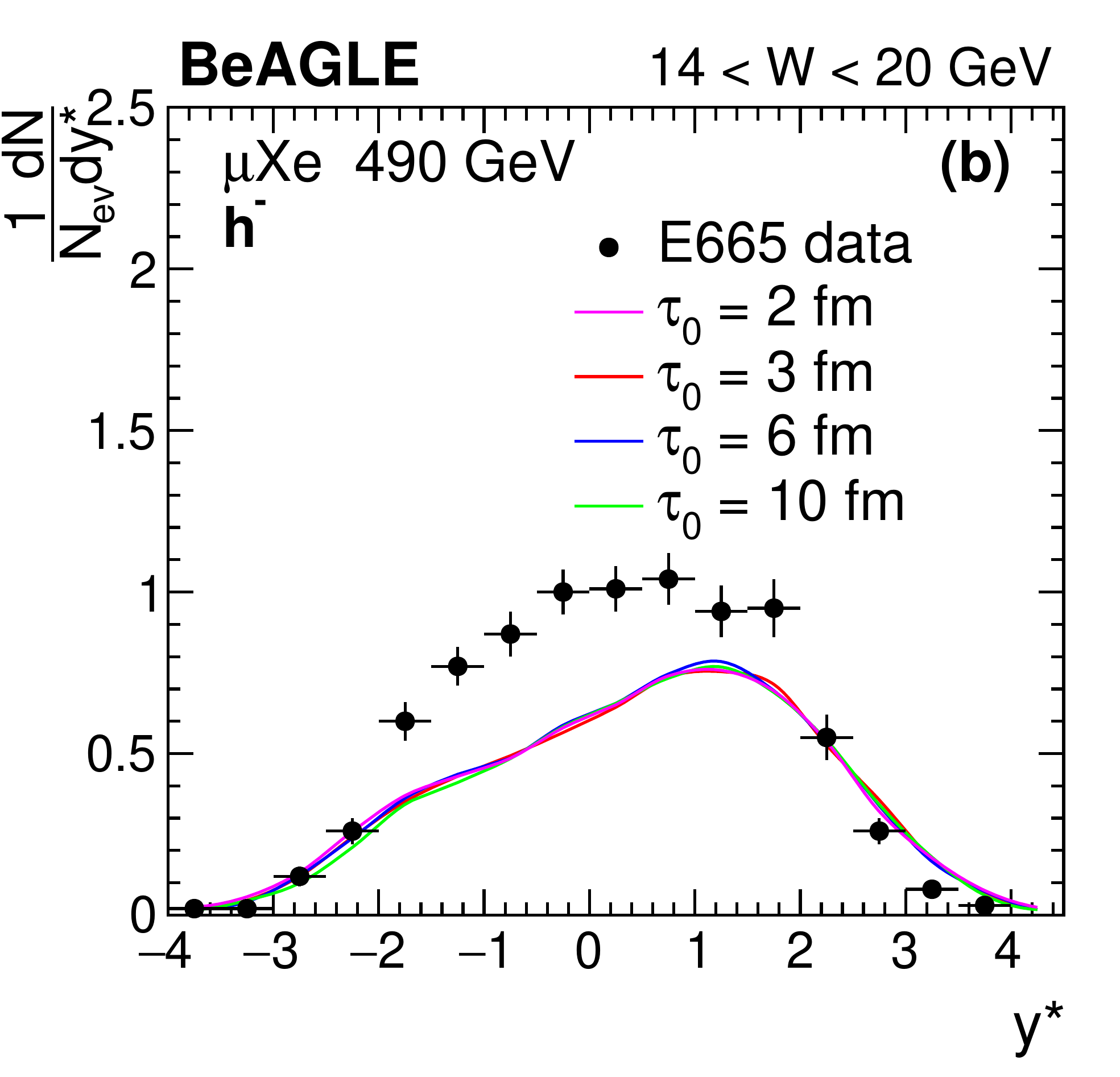}
\includegraphics[width=0.36\textwidth]{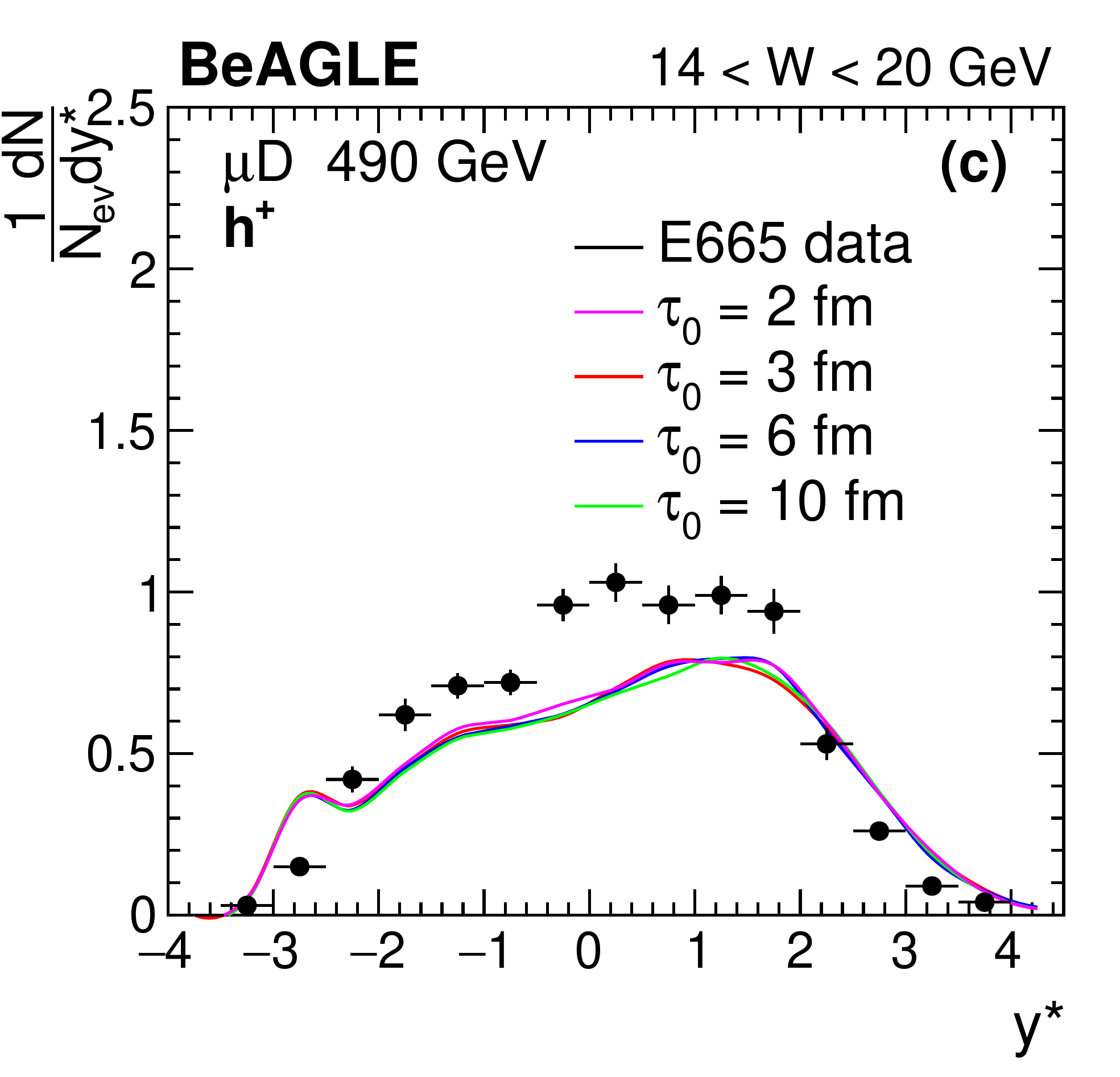}
\includegraphics[width=0.36\textwidth]{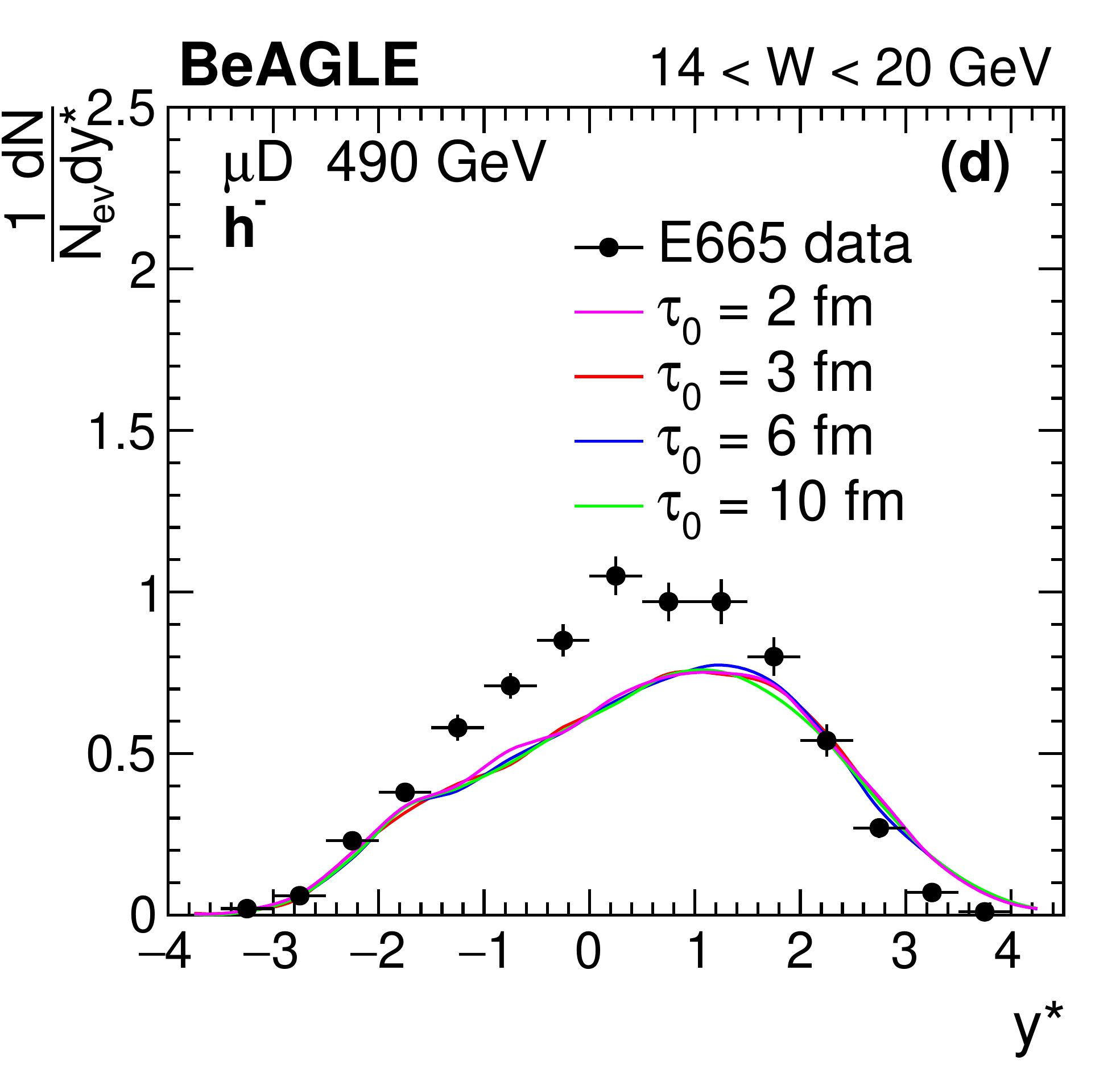}
\caption{Rapidity distribution in cms frame $y^{\ast}$ for positively (a) and negatively (b) charged hadrons in W bin: $14<W<20~\rm{GeV}$ in $\mu$Xe interactions at 490 GeV. The results with different value of $\tau_{0}$ from BeAGLE generator represented by different colors are compared to data from Ref.~\cite{E665:1993trt}. The comparison results in $8<W<14~\rm{GeV}$ and $20<W<30~\rm{GeV}$ are the same as that in $14<W<20~\rm{GeV}$.}
\label{fig:Rapidity}
\end{figure*}

Figure \ref{Figcomp} shows the distributions of $W, Q^{2}, \nu, x_{\rm{bj}}$ for BeAGLE compared with the E665 $\mu$Xe data~\cite{E665:1993trt}. The positive muon beam with an energy of 490 GeV is scattered off a Xe target. A set of kinematic cuts had to be applied to select events: $ \theta >3.5~ {\rm{mrad}}$, $Q^{2}>1~{\rm{GeV}^{2}}$, $8<W<30~{\rm{GeV}}$, and $0.1<\nu/E_{\mu}<0.85$.  The red solid lines represent the generated MC events from BeAGLE, while the E665 data~\cite{E665:1993trt} after correcting for detector acceptance effects are shown in black open circles. The ratio between the BeAGLE data and the corrected E665 results are shown in the bottom of each plot. The comparison shows that BeAGLE can do a reasonable job of describing the kinematics of the E665 data, while large deviations can be seen at the small $x_{\rm{Bj}}$ and high $Q^{2}$.


Figure~\ref{multiplicityPosiNega}(a) and (b) show the average multiplicity, $\bar{n}$, of positively and negatively charged hadrons produced in $\mu$Pb collisions with 490 GeV muons. BeAGLE simulations with different values of $\tau_{0}$ are compared with the E665 data. It is found that the distribution for negatively charged hadrons does not show any $\tau_{0}$ dependence in the BeAGLE model and the E665 data are underestimated. For positively charged hadrons, the average multiplicity increases with decreasing $\tau_{0}$. A lower $\tau_{0}$ value, e.g., 2--3 fm, reproduces the data better, indicating a contradiction with respect to our default value of $\tau_{0}$ = 10 fm, which was determined from the evaporated neutron multiplicity data (see Fig. \ref{fig:tau0} (right)). Lower $\tau_{0}$ values in Fig.~\ref{fig:tau0} (right) would suggest a very large fraction of diffractive events in $\mu$A collisions. However, the discrepancy in negatively charged particle production needs to be considered, for a clear understanding of the $\tau_{0}$ dependence. 

In Fig.~\ref{multiplicityPosiNega}(c) and (d), the average charged particle multiplicity for positive and negative $y^{\ast}$ are shown based on BeAGLE simulations. Different $\tau_{0}$ parameters in the BeAGLE model and E665 data are also presented for comparison. The distribution for charged hadrons from the BeAGLE simulations underestimates the E665 data. However, with a lower $\tau_{0}$ value, the average multiplicity distributions for charged hadrons in the target fragmentation region are improved. The quantitative dependence on $\tau_{0}$ is similar to that of the positively charged hadrons shown in Fig.~\ref{multiplicityPosiNega}(a), as positively charged hadrons dominate in the target fragmentation region. 

We find that the BeAGLE model underestimates the multiplicity everywhere, especially for negatively charged particles, and in the current fragmentation region, where neither of them has a $\tau_{0}$ dependence. Although the data may suggest a lower $\tau_{0}$ parameter in $\mu$A collisions, this comparison also implies that something other than $\tau_{0}$ plays an important role in the particle production. Recent results from the H1 experiment at HERA have reported a measurement of the charged particle multiplicity distribution~\cite{H1:2020zpd} in a wide range of DIS kinematics, where the PYTHIA-8 model~\cite{Sjostrand:2007gs, Sjostrand:2014zea} also underestimates the data almost everywhere. While a separate analysis on this subject in $ep$ DIS is highly important, e.g., a Rivet analysis~\cite{Bierlich:2019rhm}, the analysis in this paper is only focused on parameters sensitive to nuclear effects.

In order to quantitatively understand the difference between the small amount of available experimental data and the BeAGLE model, we investigate the particle production in a differential way. In Fig.~\ref{fig:Rapidity}, the normalized cms-rapidity $y^{*}$ distributions for positively and negatively charged hadrons are shown in $\mu$Xe ((a) and (b)) and $\mu$D ((c) and (d)) with 490 GeV muon beams. The selected kinematic phase space is within $14 < W < 20~\rm GeV$. In $\mu$Xe events, for positively charged hadrons, there is no $\tau_{0}$ dependence found at forward rapidities, while a strong dependence is observed in the backward region. In the E665 $y^{*}$ distribution comparisons with BeAGLE simulations shown in Fig.~\ref{fig:Rapidity}, BeAGLE underestimates the forward particle production, and predicts a different peak position of the backward production. Additionally, BeAGLE underestimates the negatively charged particles and all charged particles in $\mu$D almost everywhere in rapidity except for the very forward and backward regions. For both $\mu$Xe and $\mu$D systems, similar observations are found in other $W$ ranges, thus it is observed that the discrepancies between the data and BeAGLE are not dependent on the kinematics.

\begin{figure}[tbh]
\centering 
\includegraphics[width=0.4\textwidth]{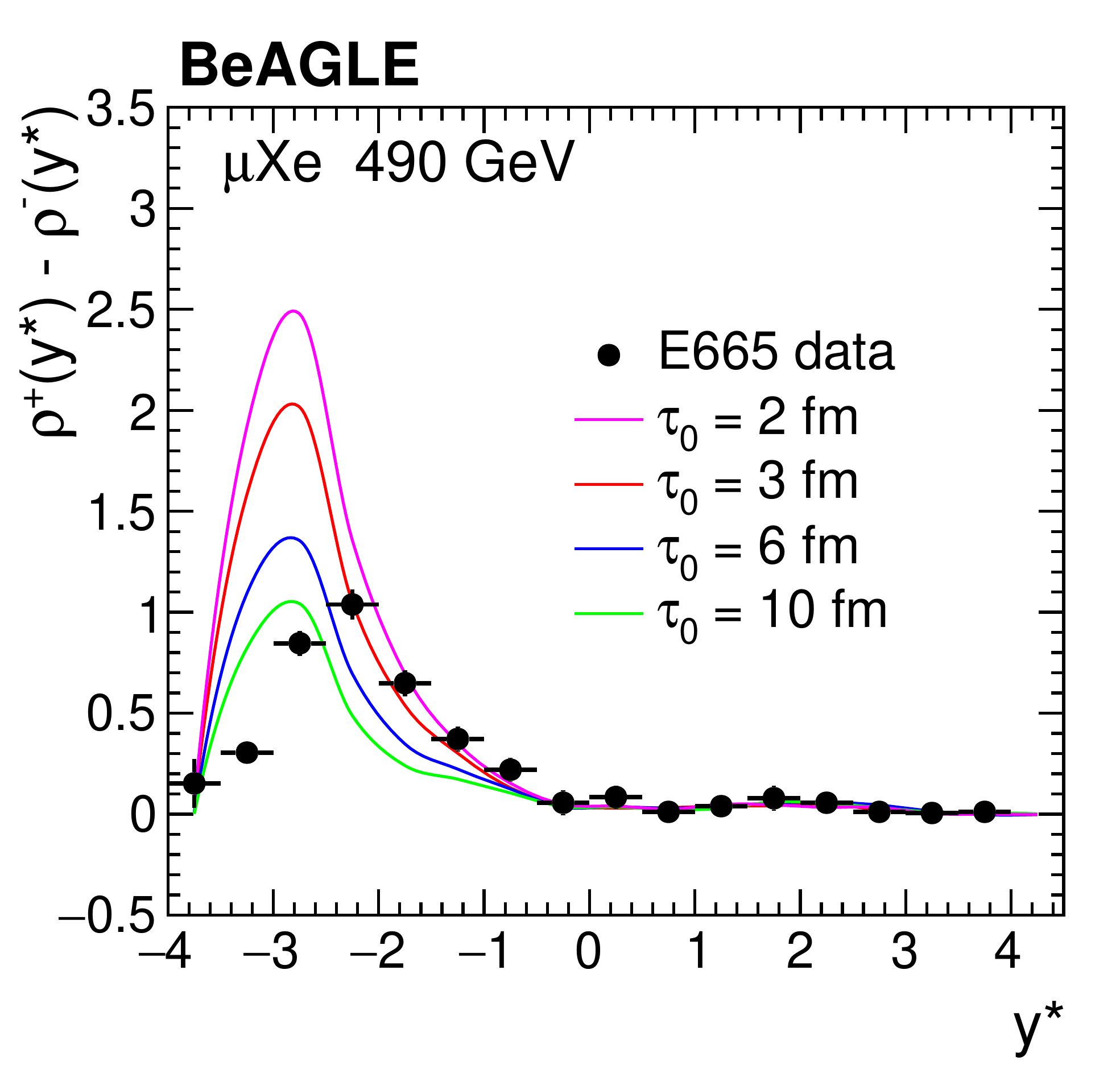}
\includegraphics[width=0.4\textwidth]{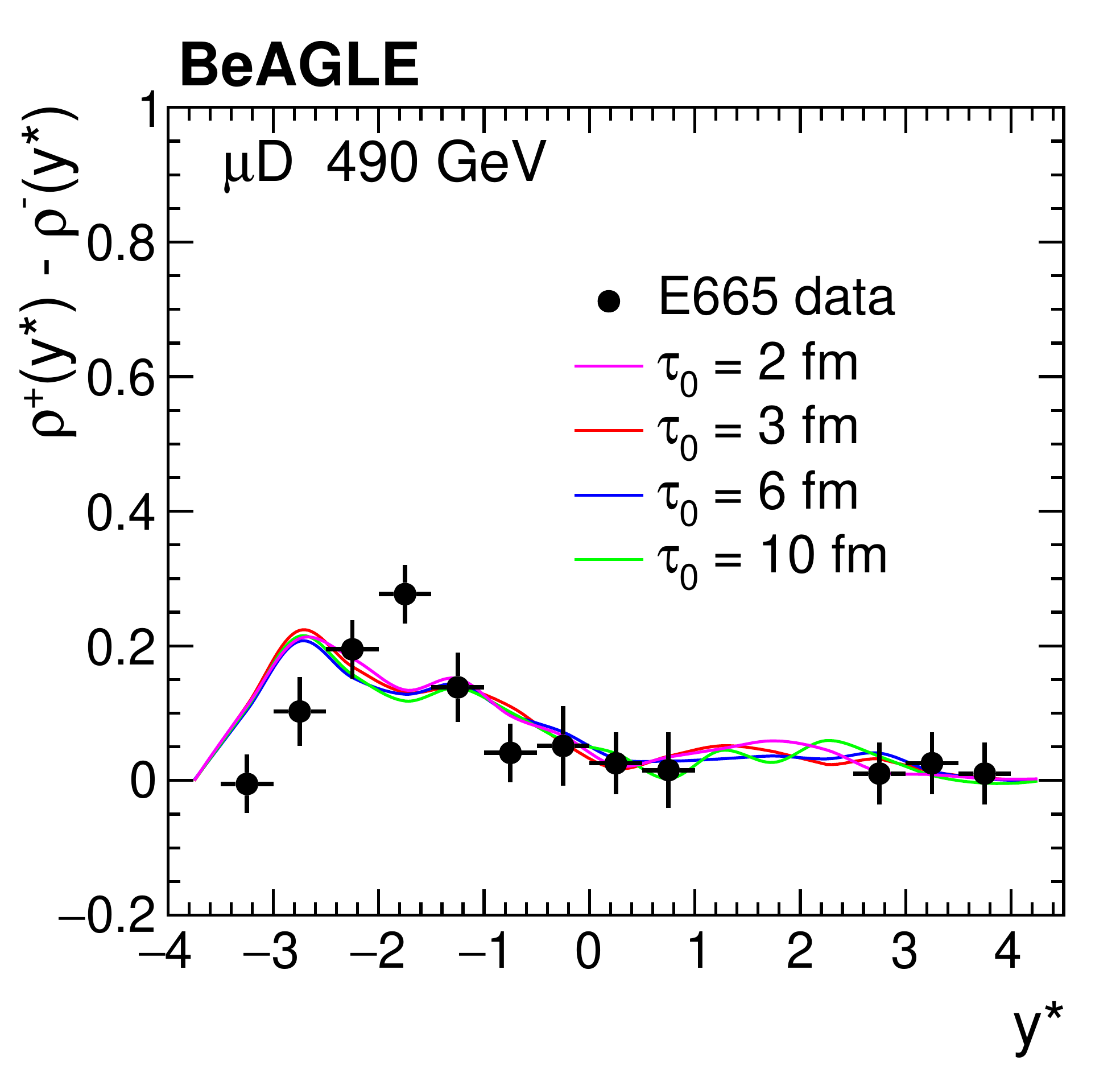}
\caption{Normalized cms-rapidity $y^{*}$ distribution of the hadronic net charge in $\mu$Xe (top) and $\mu$D (bottom) collisions using 490 GeV muon beams. The results with different values of $\tau_{0}$ from the BeAGLE generator represented by lines with different colors are compared to data from Ref.~\cite{E665:1993trt}.}  
\label{RapidityChargeNet}
\end{figure}

To further isolate the contributions from the primary interaction and the nuclear remnant fragmentation, we study the difference between positively and negatively charged particles, which is more sensitive to effects like INC. Therefore, the normalized cms-rapidity $y^{*}$ distributions of the net charge, $\rho^{+}(y^{*})-\rho^{-}(y^{*})$, are shown in Fig.~\ref{RapidityChargeNet} for both $\mu Xe$ and $\mu D$ collisions, where $\rho^{\pm}$ is defined as follows, 
\begin{equation}
    \rho ^{\pm }\left ( y^{\ast } \right )=\frac{1}{N_{ev}}\cdot \frac{dN^{\pm }}{dy^{\ast }}.
\end{equation}
\noindent Here $N_{ev}$ is the number of selected events and $N^{\pm}$ is the number of positively or negatively charged hadrons, respectively. In $\mu$Xe events, for charged hadrons, there is no $\tau_{0}$ dependence of the $\rho^{\pm}$ distribution at forward rapidity or in the current fragmentation region, similar to what has been found in Fig.~\ref{fig:Rapidity}.  However, in the backward region, despite the large $\tau_{0}$ dependence in the BeAGLE model, the peak position of the distribution is found to be stable for all $\tau_{0}$ values, and is different compared to the E665 data by about half a unit of rapidity. In $\mu$D events, $\tau_{0}$ dependence is hardly visible, while the shift in the peak position in the backward region is even larger than that in $\mu$Xe events. 



\begin{figure}[th]
\centering 
\includegraphics[width=0.4\textwidth]{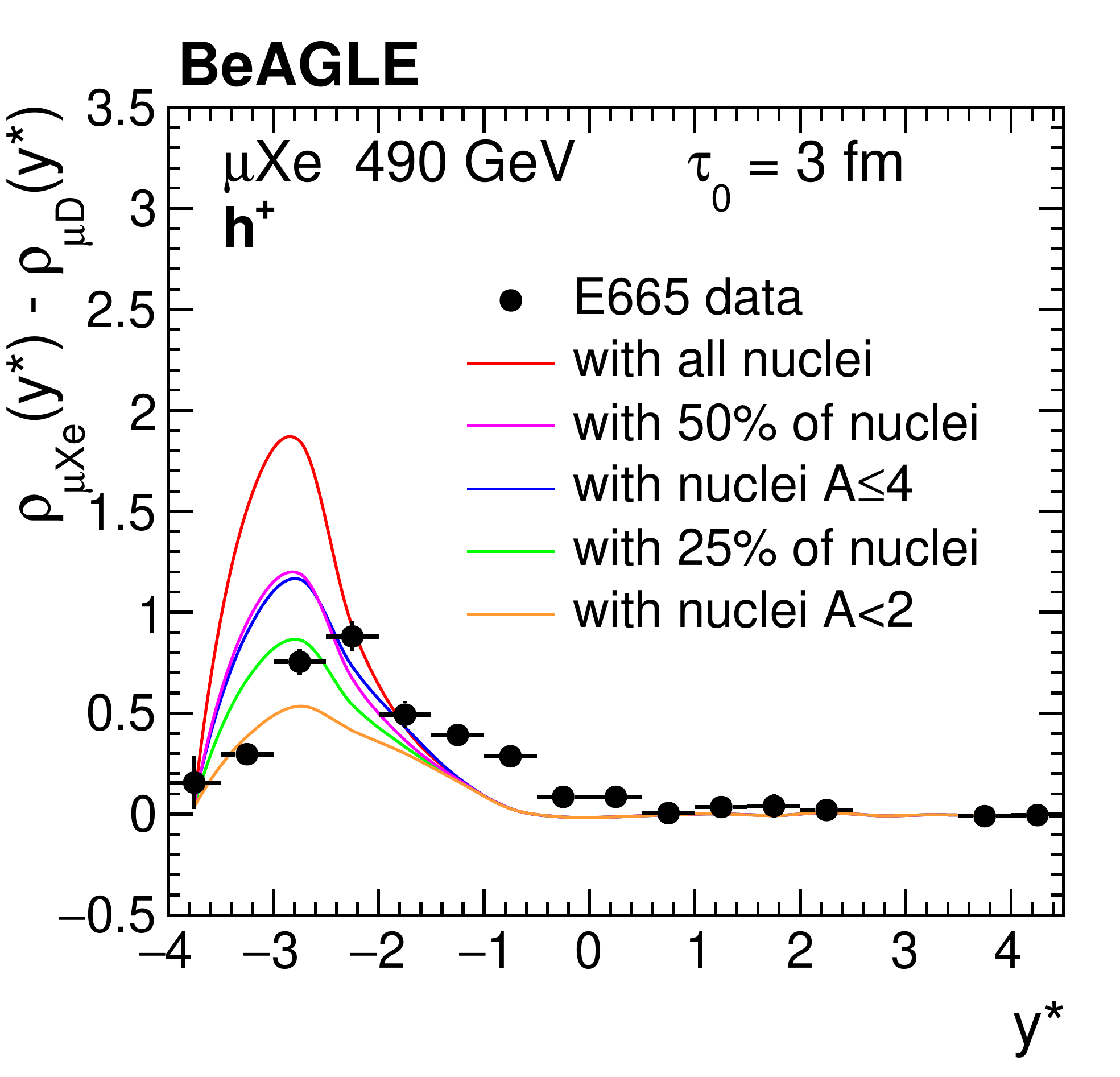}
\caption{The difference of normalized $y^{*}$ distributions for positively charged hadrons between $\mu$Xe and $\mu$D. The results from BeAGLE generator are with $\tau_{0}=3~\rm{fm}$. The lines with different color represent the results with different way treated to nucleus measurement. Black points represent the data from Ref.~\cite{E665:1993trt}.}  
\label{RapidityXeminuD}
\end{figure}

Comparisons of the normalized $y^{*}$ distributions between $\mu$Xe and $\mu$D collisions for positively charged hadrons are presented in Fig.~\ref{RapidityXeminuD} for both E665 data and BeAGLE simulations. Here, BeAGLE uses a $\tau_{0}$ of 3 fm but shows different assumptions for final-state nuclei. The discrepancy still exits in the backward region, where the peak positions from BeAGLE sit at larger negative values of $y^{*}$ compared to the E665 data. Since there is no clear description of remnant nuclei detection in Ref.~\cite{E665:1993trt}, we try a few different ways to treat the remnant nuclei in the BeAGLE simulation. The red line is the result with all remnant nuclei included. The magenta line denotes a randomly selected 50\% of all nuclei. The blue line, green line, and orange line represent only nuclei whose mass number $A$ is smaller than 4, a random selection of 25\% of all nuclei, and no nuclei, respectively. With different fractions of nuclei included, the net charge density in the region of $-4<y^{\ast }<-2$ changes, while the peak position remains the same. 

\begin{figure}[tbh]
\centering 
\includegraphics[width=0.4\textwidth]{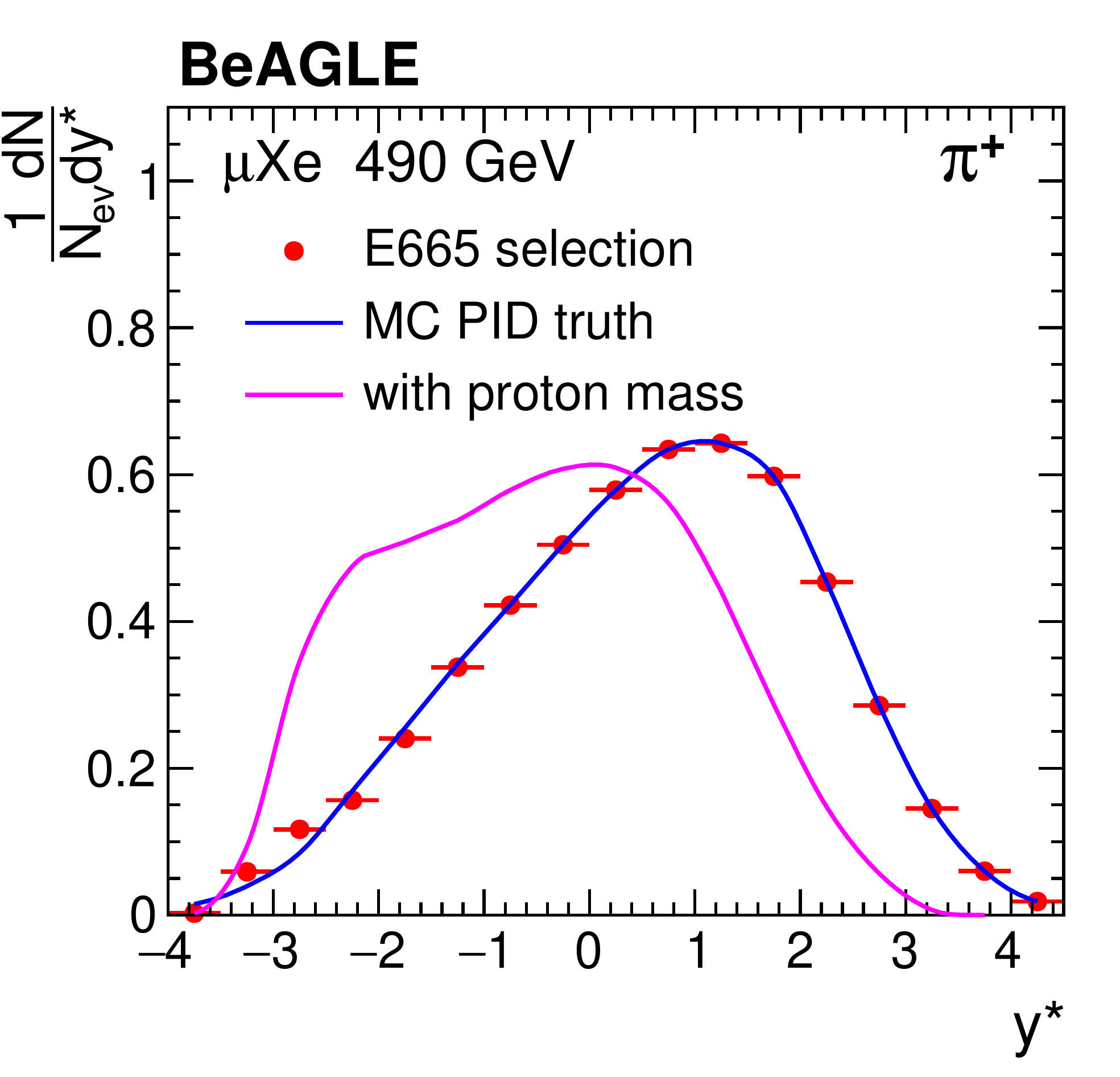}
\includegraphics[width=0.4\textwidth]{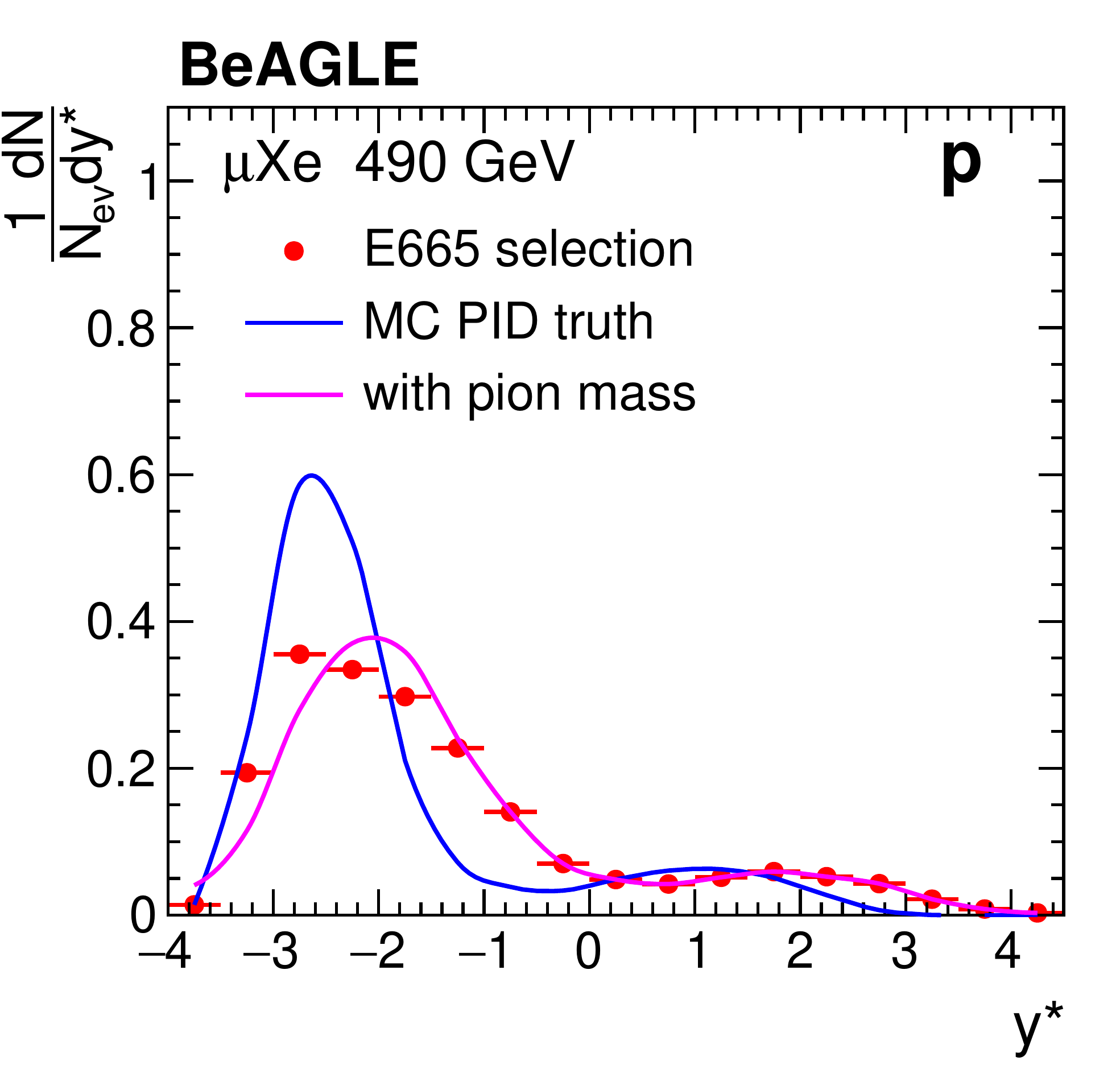}
\caption{Normalized cms-rapidity $y^{*}$ distribution of $\pi^{+}$ (top) and proton (bottom) in $\mu$Xe collisions using 490 GeV muon beams. Different colors represent the results with different methods of calculating cms-rapidity $y^{*}$ from the BeAGLE generator.}  
\label{RapidityParticle}
\end{figure}

\begin{figure}[tbh]
\centering  
\includegraphics[width=0.4\textwidth]{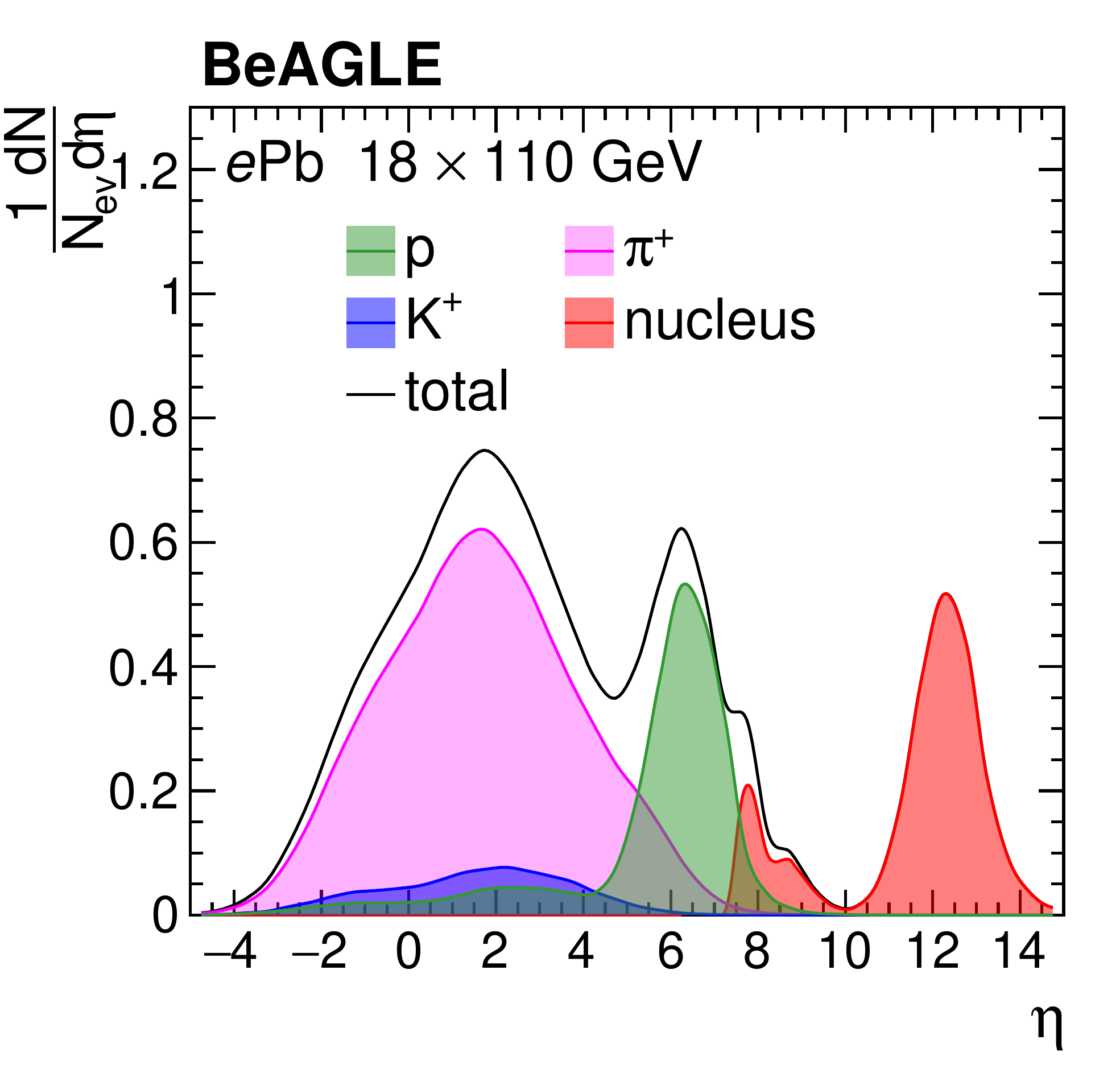}
\caption{Distribution of normalized pseudo-rapidity of positive charged particles for $e$Pb collisions with 18 GeV on 110 GeV with BeAGLE. Different colors indicate results for different particle species.} 
\label{Fig_EtaLabParticle}
\end{figure}

For the comparisons presented above, specifically from Fig.~\ref{multiplicityPosiNega} to~\ref{RapidityXeminuD}, a few things should be noted. First, the particle identification in the E665 experiment assigned either a proton or pion mass in the reconstruction, based on the $x_{F}$ value. In the absence of a more precise particle identification (PID) methodm such as those  currently in use (e.g. $dE/dx$ or time-of-flight measurements), this approach might be problematic. Figure~\ref{RapidityParticle} shows the $y^{*}$ distributions of $\pi^{+}$ mesons (upper) and protons (lower) in $\mu$Xe collisions using a 490 GeV muon beam from the BeAGLE generator. The red points, which are labeled as ``E665 selection", represent the same method as the E665 data from Ref.~\cite{E665:1993trt}, while the blue curves represent the result based on the true mass from the MC PID, and the magenta curves assume a wrong mass assignment, e.g., proton mass for pions (top), and pion mass for protons (bottom). Pions are mostly produced in the hard scattering and dominate the current fragmentation region. A large proportion of protons are generated during the INC process and its $y^{*}$ distribution is dominated in the region of $-3<y^{\ast }<-2.5$. If the protons were mis-identified as pions, the cms-rapidity $y^{*}$ would be shifted toward more-central values of rapidity. Although the data were fully corrected at the particle level, residual mis-identification in the data and a subsequent discrepancy between the data and BeAGLE in particle compositions are possible. Secondly, the E665 measurement from Ref.~\cite{E665:1993trt} did not explicitly describe the details of experimental detection of remnant nuclei. Although the peak position of $\mu \rm{Xe}-\mu \rm{D}$ shown in Fig.~\ref{RapidityXeminuD} remains the same, the details of remnant nuclei detection together with a different particle composition as described above may cause the peak position of the distribution change. Finally, the missing coherent diffractive events in the BeAGLE model could be another reason for the observed discrepancy. Naively, the diffractive DIS events would have a rapidity gap, and the $y^{*}$ distribution would be expected to be shifted more towards the forward than the backward direction. 

In addition to Ref.~\cite{E665:1993trt}, a similar result was reported by the E665 Collaboration in Ref.~\cite{adams1995nuclear}. In this study, it employed so-called ``gray tracks" to enhance proton identification. ``Gray tracks" are particles whose momenta are between 200-600 MeV$/c$, and the streamer density as observed in the streamer chamber picture is clearly higher than that of a minimum ionizing particle. Unfortunately the data reported in Ref.~\cite{adams1995nuclear} were not corrected for experimental inefficiencies and there is no reliable method to study the impact of such gray tracks in our simulations. In light of these challenges, a truly equivalent comparison between existing data and BeAGLE cannot be made. Therefore, in order to further understand particle production over a wide range of rapidity, only the EIC can provide more information about the target fragmentation in lepton-nucleus collisions. 

Figure~\ref{Fig_EtaLabParticle} shows the normalized distribution of positively charged particles as a function of pseudorapidity ($\eta$) at the top EIC energy, simulated by BeAGLE. The total distribution includes all particle species, depicted in the black curve. Other colors indicate distributions for different particle species. Almost all pions and kaons are produced during the hard collision, and their pseudorapidities range from $-4$ to 4, which falls into the acceptance of the expected general purpose detector of the EIC. Protons are distributed across a wide range of pseudorapidity, from $-4$ to 10, where three different far-forward proton detectors (B0 tracker, Off-momentum detector, and Roman Pots) can cover a large fraction of the phase space of pseudorapidity $>$ 4.5~\cite{AbdulKhalek:2021gbh}. Nuclei are produced in the last step of the BeAGLE proceeses via evaporation, but they are separated into two kinematic regions. The nuclei distributed within $7 < \eta < 10$ are light nuclei, e.g., deuterons and alpha particles. The large remnant nuclei are distributed within $10 < \eta < 15$. Detecting these nuclei is a major experimental challenge, and is one of the on-going efforts at the future EIC, hopefully achieved through optimizing the far-forward instrumentation and the 2nd IR design~\cite{AbdulKhalek:2021gbh}.

\section{\label{sec:centrality} Collision geometry determination in lepton-nucleus interactions}

In this section, we show an example of how BeAGLE can help optimize measurements with different collision geometry in lepton-nucleus interactions at the future EIC. 
Precise quantification of the nuclear effects in $e$A collisions requires knowledge of the underlying collision geometry. In fixed target DIS experiments of nuclei up to now, the collision geometry has only been qualitatively investigated by varying the target nucleus. However, at the EIC, it is possible to characterize an event-by-event collision geometry by studying the nuclear breakup, an idea initially introduced in Ref.~\cite{Zheng:2014cha}.  The collision geometry in each event can be linked to the multiplicity of evaporation neutrons at very forward rapidities ($>$ 4.5), measured by the Zero-Degree Calorimeter (ZDC) (see Ref.~\cite{AbdulKhalek:2021gbh} for details). In the following, we will introduce variables that are sensitive to the collision geometry and their correlation with experimental observables, e.g., evaporated neutrons and protons. Compared to Ref.~\cite{Zheng:2014cha}, we provide more systematic studies by varying model parameters using the BeAGLE generator to demonstrate the robustness of this measurement. This result provides an important experimental handle to all inclusive and semi-inclusive DIS measurements at the EIC. 

\begin{figure}[b]
\centering
\includegraphics[width=1.0\linewidth]{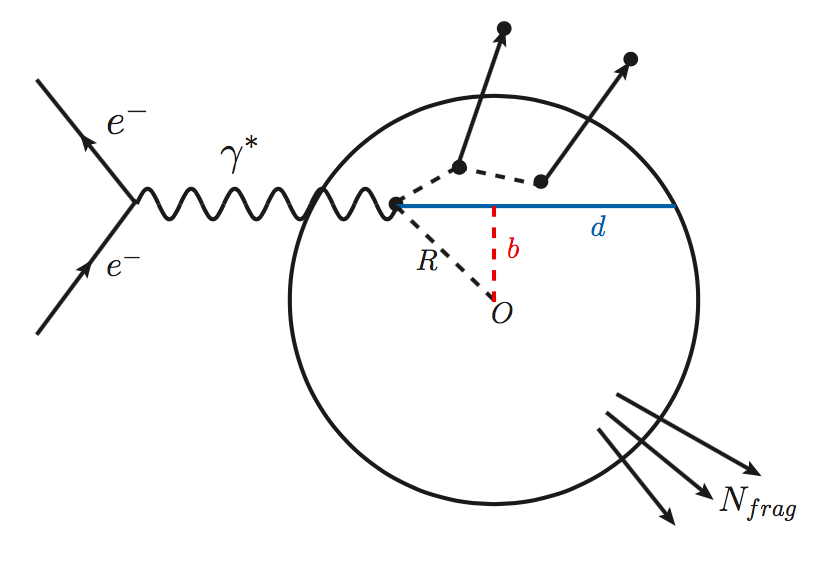}
\caption{Relevant quantities to describe the collision geometry. The effective interaction
length, $d$, is the distance between the photon-nucleon interaction point and the edge of the nucleus in the direction of the virtual photon, weighted by the nuclear density. The variable $b$ is the impact parameter between $d$ and the center of the nucleus.}
\label{Fig1_0}
\end{figure}

\begin{figure*}[tbh]
\centering  
\includegraphics[width=0.36\textwidth]{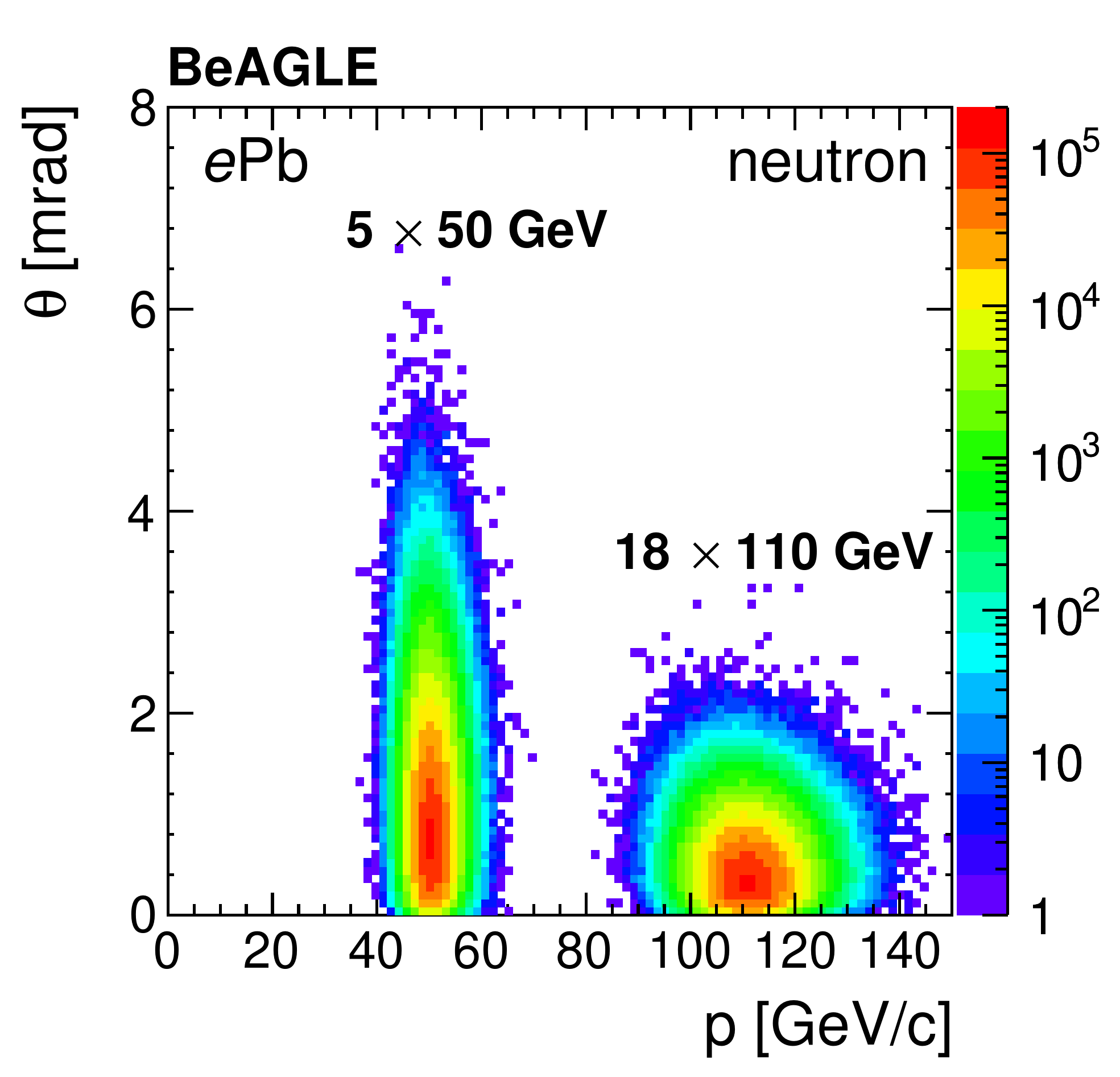}
\includegraphics[width=0.36\textwidth]{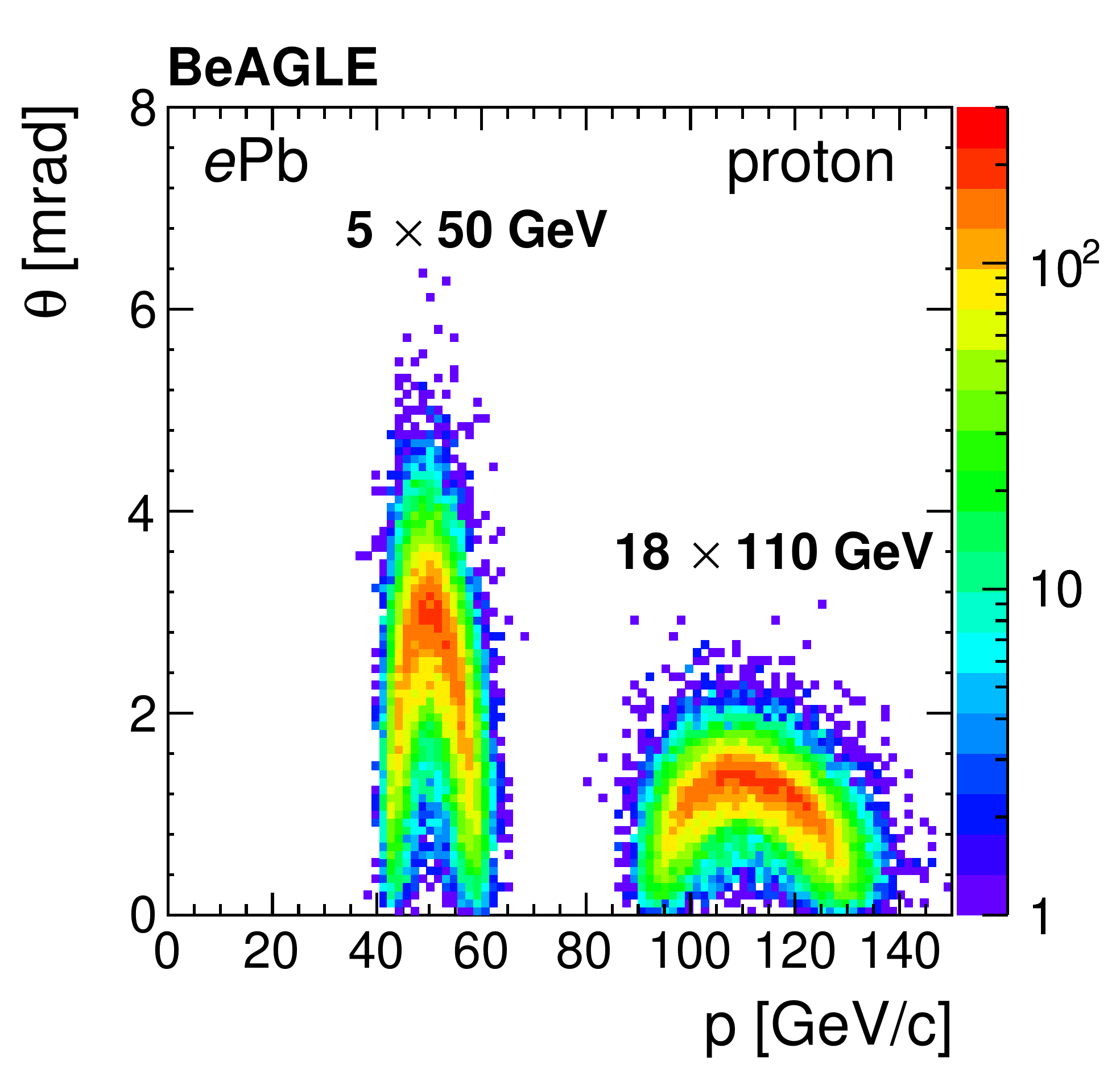}
\caption{The correlation of momentum and scattering angle for two beam energies for neutrons and protons.} 
\label{Fig1_1}
\end{figure*}

\subsection{\label{sec: centrality_1 }  Definition of $e$A collision geometry}

The collision geometry in DIS reveals the underlying spatial information of the nuclear matter probed by the exchanged virtual photon with respect to the rest of the nuclear target. Depending on the physics process under study, different collision geometry quantities can be defined. In Ref.~\cite{Zheng:2014cha}, the fiducial traveling length and the impact parameter were used as important controls to quantify the effect of parton energy loss and gluon saturation. In this paper, we define the effective interaction length, $d$, as the distance between the photon-nucleon interaction point and the edge of the nucleus in the direction of the virtual photon, weighted by the nuclear density $\rho_{0}$,
\begin{equation}  
d (b, z_0) = \int_{z_0 }^{+\infty }\mathrm{d} z \rho (b,z)/\rho _{0}.
\label{equation:d} 
\end{equation}
Here, $z_0$ is the position of the nucleon involved in the scattering along the photon moving direction, and $b$ is the impact parameter. If multiple nucleons are participating in the interaction, we use the effective interaction length averaged over all the involved nucleons. This definition avoids the possible negative region\footnote{If one scattered nucleon is outside of the geometric nuclear radius due to fluctuation, the fiducial $d$ becomes negative.} of fiducial $d$ used in Ref.~\cite{Zheng:2014cha} and is more directly connected to the amount of nuclear material. These geometric variables are depicted in Fig.~\ref{Fig1_0}. In addition, we use the scaled thickness function $T(b)$ as an alternative to characterize the collision geometry as follows,
\begin{equation}  
T(b)/\rho _{0} = \int_{-\infty }^{+\infty }\mathrm{d} z \rho (b,z)/\rho _{0},
\label{equation:Tb} 
\end{equation}
\noindent in units of fm. This quantity can be explicitly studied together with the gluon saturation physics in $e$A collisions. 

\subsection{\label{sec: centrality_2 }  Measuring forward nuclear fragments}


The event sample used for the collision geometry study is generated from the BeAGLE model for $e$Pb collisions at 18 $\times$ 110 GeV with $\tau_{0}=10~$fm, shadowing model $genShd=3$, $1~\rm{GeV^{2}}<Q^{2}<100~\rm{GeV^{2}}$, and $0.01<y<0.95$.

The most abundant final-state products produced during the nuclear breakup are evaporated protons and neutrons. The left and right panel in Fig.~\ref{Fig1_1} show the distribution of the neutrons and protons, respectively, as a function of momentum and scattering angle at two collision energies (18 $\times$ 110 GeV and 5 $\times$ 50 GeV). The evaporation momenta are close to the beam momentum and their scattering angles are small ($\sim$ few milliradians). At a beam energy of 50 GeV, the largest scattering angle is about 6 mrad, while at 110 GeV, the maximum scattering angle is about 3 times smaller. In contrast to neutrons, there are only a few protons emitted at very small angles because the protons need to overcome the Coulomb barrier to leave the nucleus. As the number of emitted protons during the nuclear evaporation is significantly lower than that of neutrons, it is best to study the properties of the nuclear evaporation process by measuring neutrons.

\begin{figure*}[tbh]
\centering
\includegraphics[width=0.36\textwidth]{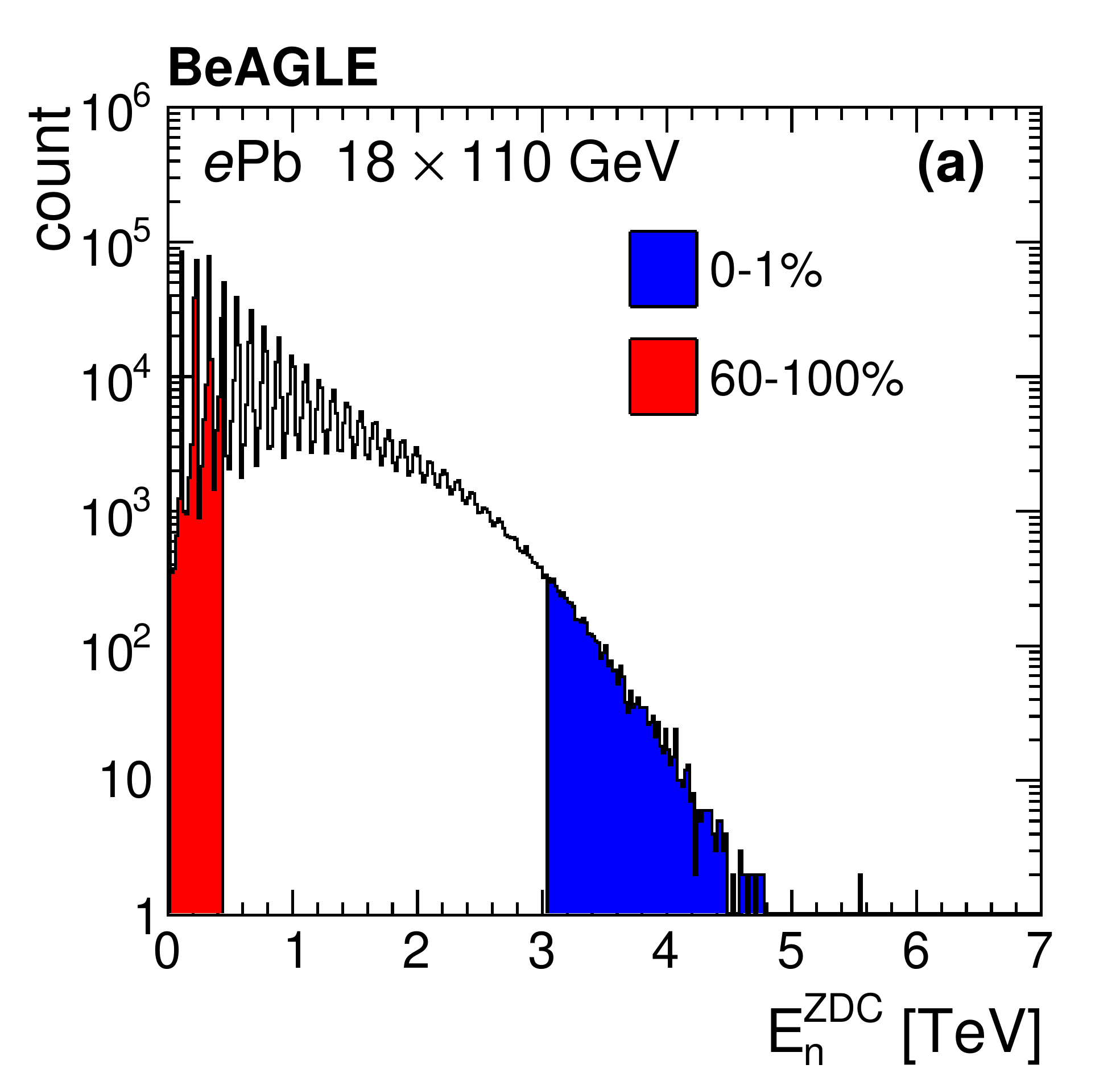}
\includegraphics[width=0.36\textwidth]{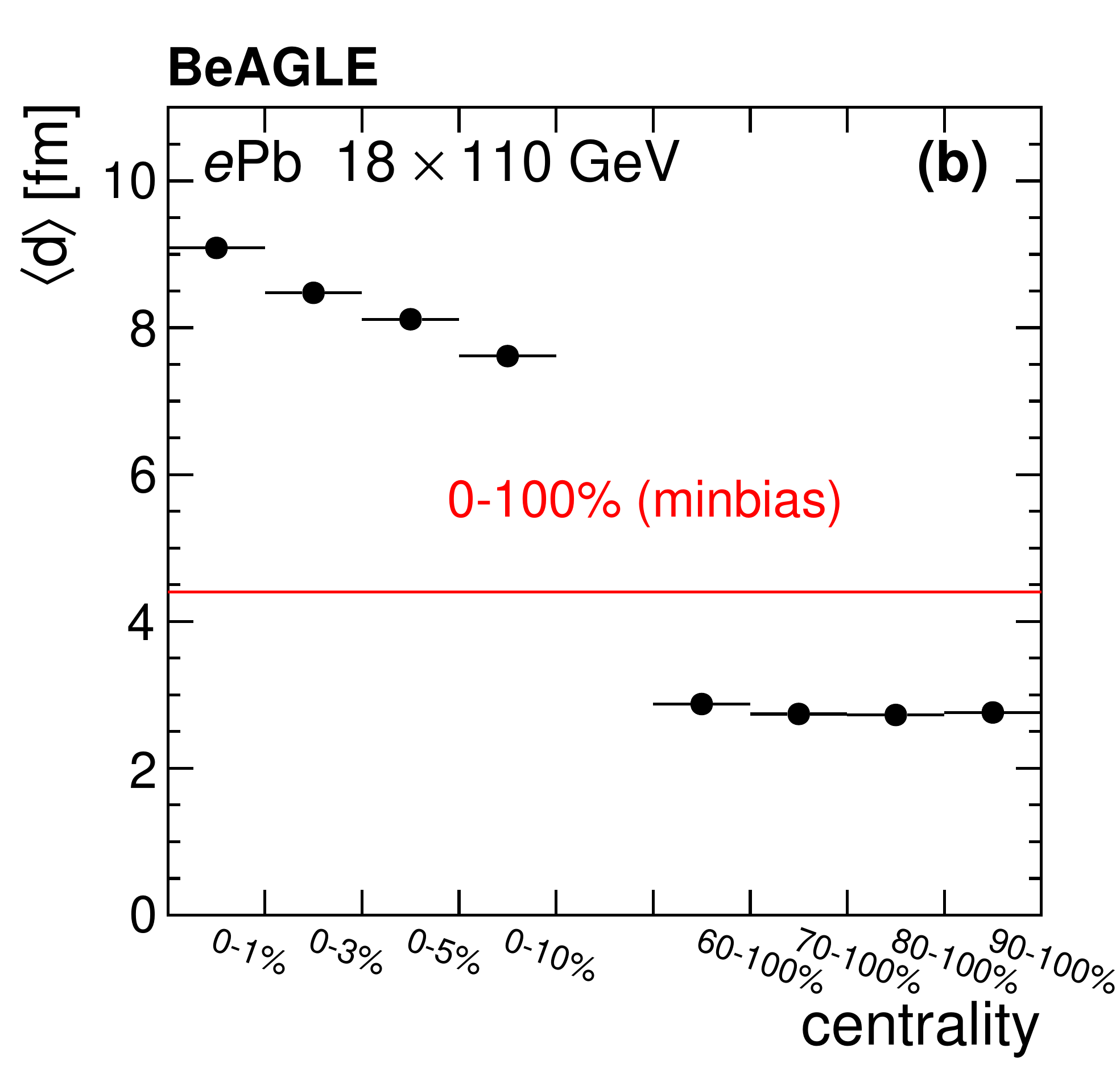}
\caption{(a) the centrality selection by the energy deposition in the ZDC. (b) the distribution of $\left \langle d \right \rangle$ for different choices of central: 0-1\%, 0-3\%, 0-5\%, 0-10\%, as well as for the peripheral centrality: 60-100\%, 70-100\%, 80-100\%, 90-100\%. Note that both of these two distributions are in MC generate level, without detector smearing.}
\label{Fig1_17}
\end{figure*}

The neutron multiplicity has been demonstrated to be a tool to access the collision geometry variables~\cite{Zheng:2014cha}, inspired by how centrality is determined in heavy-ion collisions~\cite{Alver:2008aq}. As it is difficult to directly measure a large number of neutrons, we use the energy deposition in the ZDC, similar to Ref.~\cite{Zheng:2014cha}. Higher energy deposited in the ZDC (large multiplicity of evaporation neutrons) is expected to correspond to more-central events. The distribution of energy deposition in the ZDC $E_{\rm{n}}^{\rm{ZDC}}$ (in generated level) is shown in Fig.~\ref{Fig1_17}(a), where the blue area corresponds to the events with a centrality of 0-1$\%$, representing the top 1$\%$ events with the highest energy deposition being greater than 2.82 TeV. The red area corresponds to the events with a centrality of 60-100 $\%$ with the energy being less than 0.44 TeV. In Fig.~\ref{Fig1_17}(b), the average traveling distance $\left \langle d \right \rangle$ is shown as a function of the ZDC energy percentage class. The value of $\left \langle d \right \rangle$ for minimum-bias (0--100\%) events is 4.402. $\left \langle d \right \rangle$ decreases clearly going from a centrality of 0-1\% to 0-10\%, but one loses a factor of 10 in statistics, this decreasing trend is not obvious in peripheral collisions. For the following analysis, we choose 0-1\% as a central collision, and 60-100\% as the most-peripheral collision.

\begin{figure*}[th]
\centering 
\includegraphics[width=0.36\textwidth]{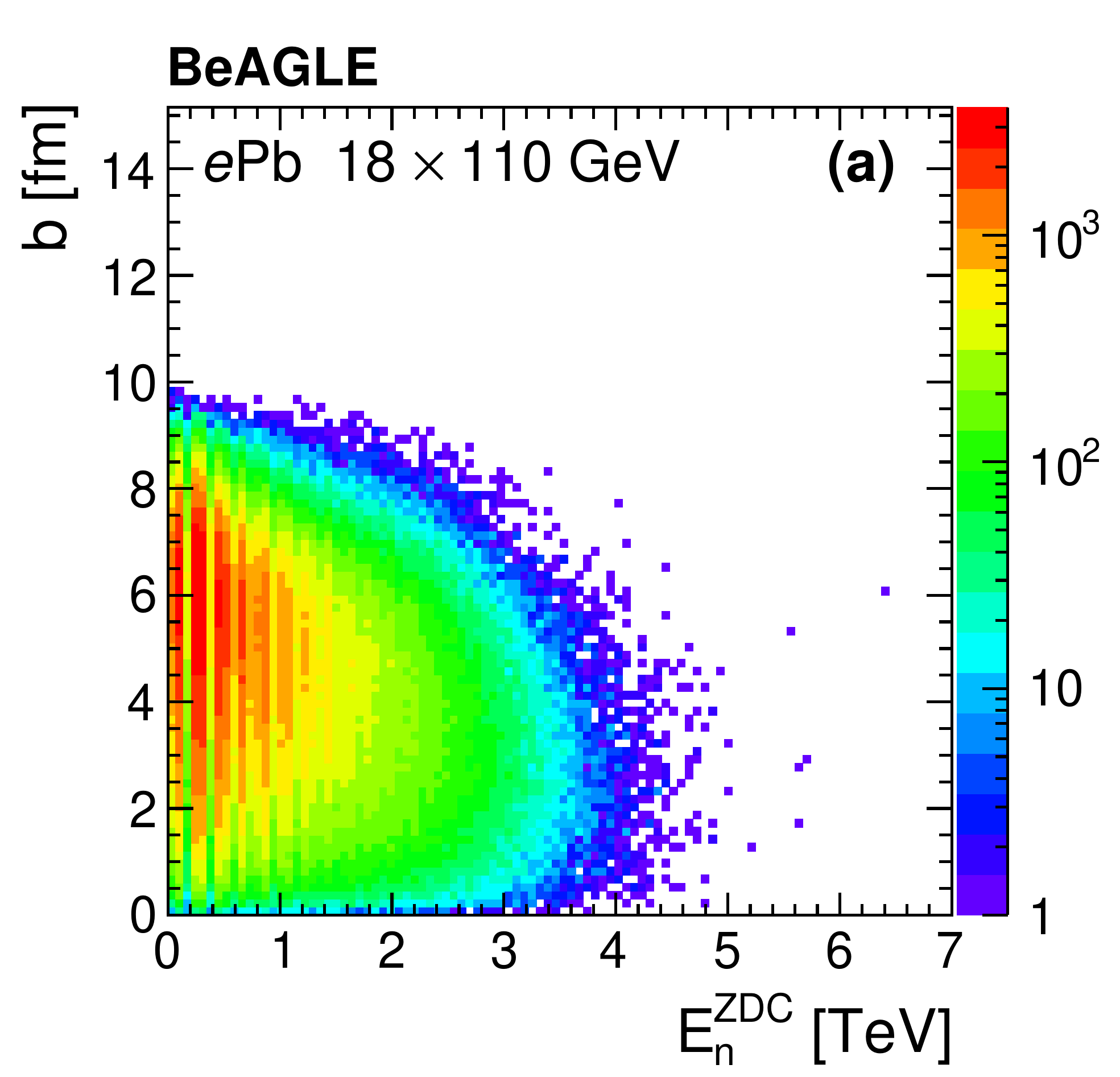}
\includegraphics[width=0.36\textwidth]{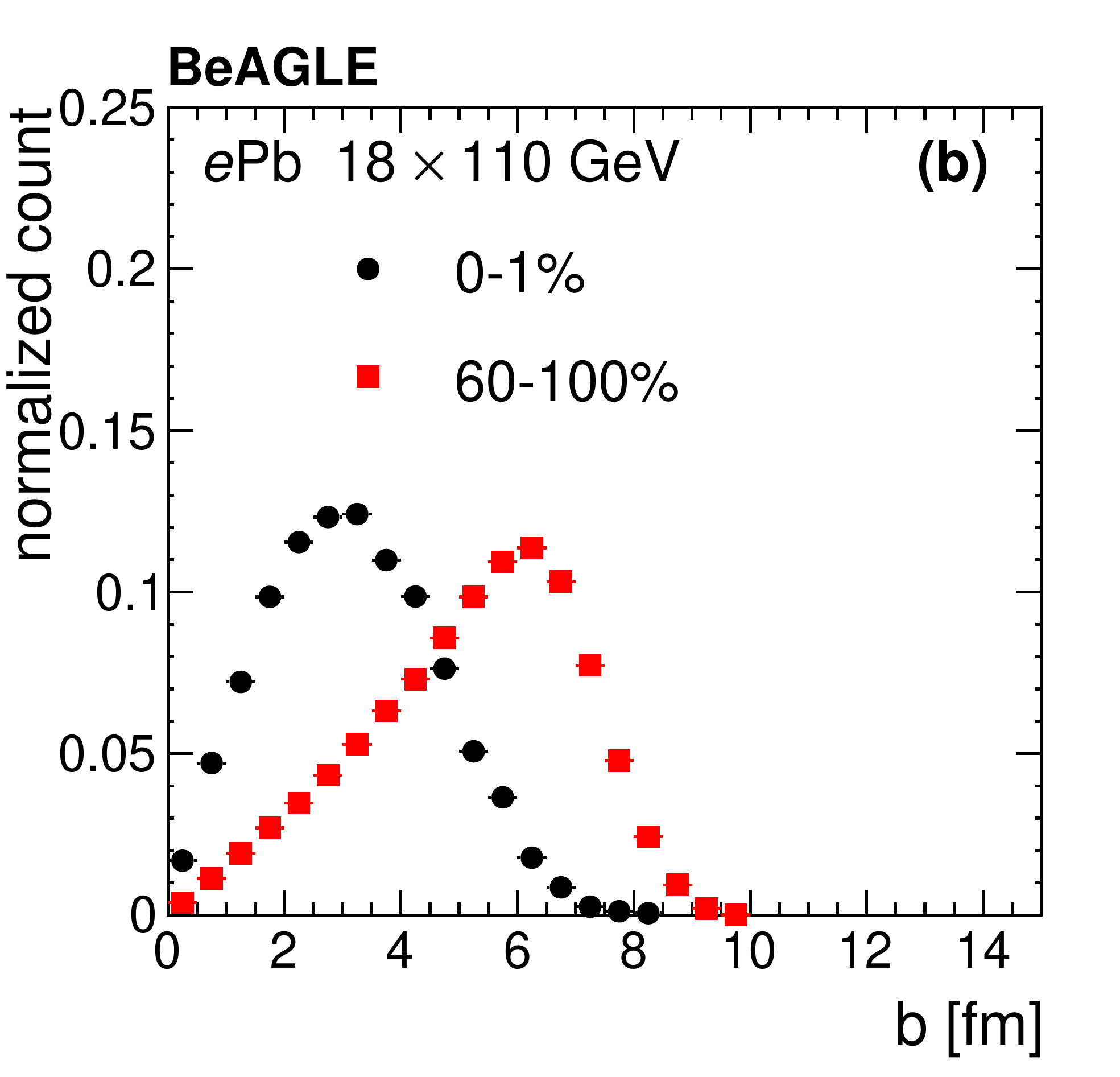}
\includegraphics[width=0.36\textwidth]{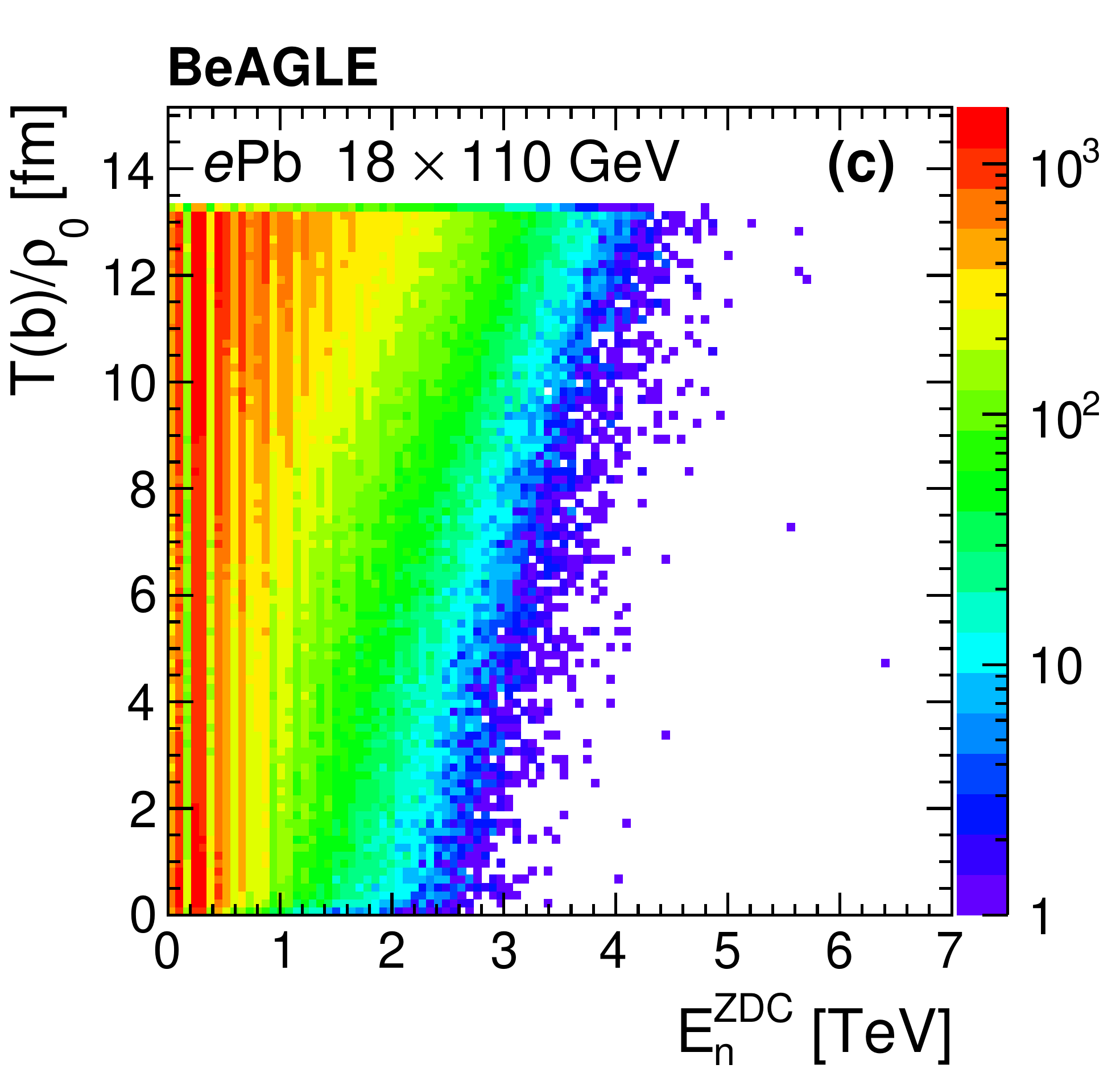}
\includegraphics[width=0.36\textwidth]{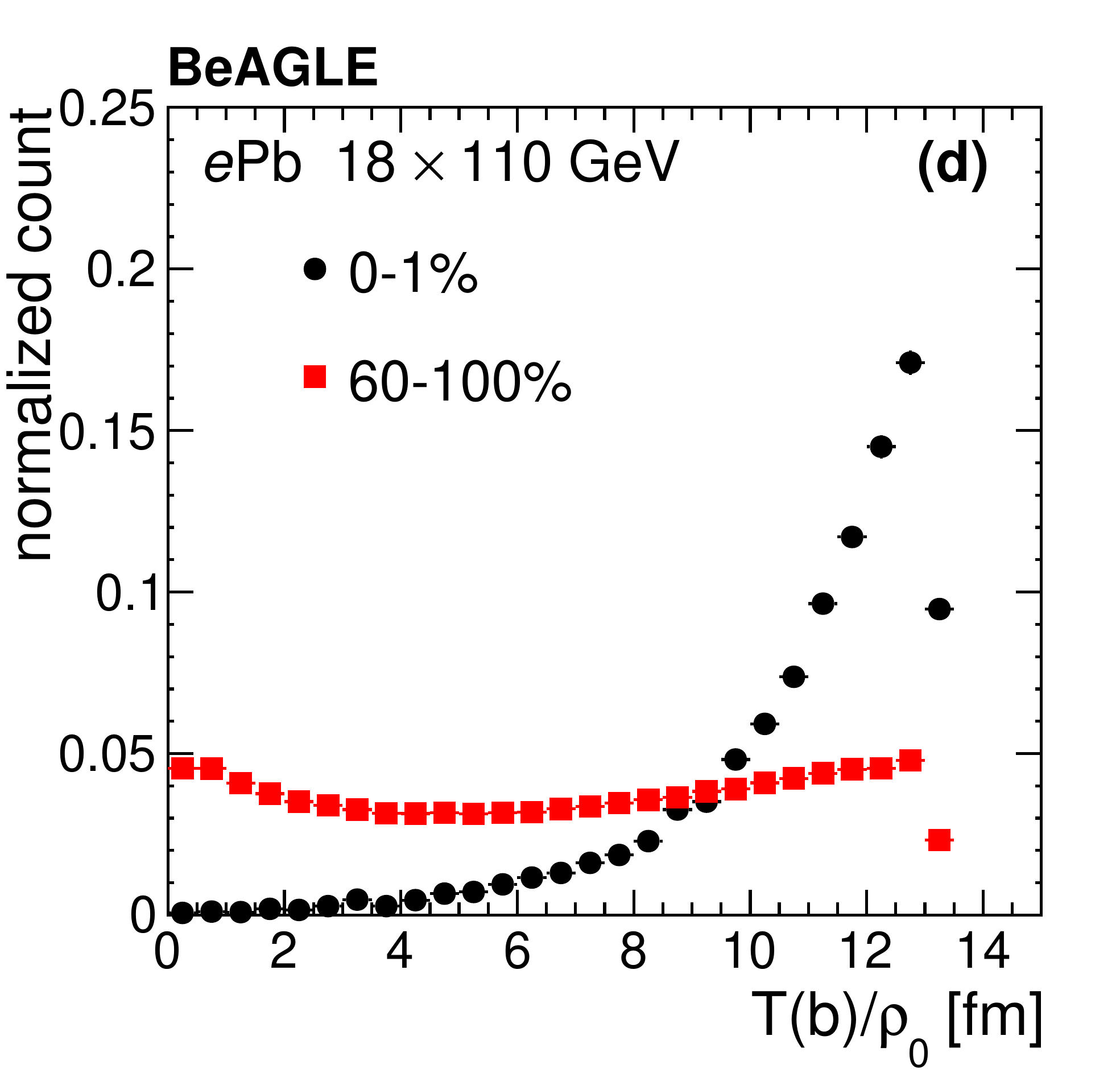}
\caption{(a) the correlation between the deposited energy in the ZDC and impact parameter $b$. (b) $b$ distributions for both central and peripheral collisions. (c) the correlation between the deposited energy in the ZDC and the nuclear thickness $T(b)/\rho _{0} $. (d) $T(b)/\rho _{0}$ distributions for both central and peripheral collisions. Note that all of these distributions are in MC generated level, without detector smearing.} 
\label{Fig1_6}
\end{figure*}

 The correlation between the deposited energy in the ZDC and impact parameter $b$, the nuclear thickness $T(b)/\rho _{0} $ are shown in Fig.~\ref{Fig1_6}(a) and (c), separately. With increasing energy, $b$ decreases while $T(b)/\rho _{0} $ becomes larger. Figure.~\ref{Fig1_6}(b) and (d) show the $b$ and $T(b)/\rho _{0}$ distributions in central (0-1\%) and peripheral (60-100\%) collisions, separately. They are normalized by the number of total events. A clear difference between central and peripheral collisions in both $b$ and $T(b)/\rho _{0}$ can be seen, and by selecting different centrality classes, we obtain an experimental handle on the collision geometry.

\subsection{\label{sec: centrality_5}  Systematic study of collision geometry}

\begin{figure*}[th]
\centering  
\includegraphics[width=0.36\textwidth]{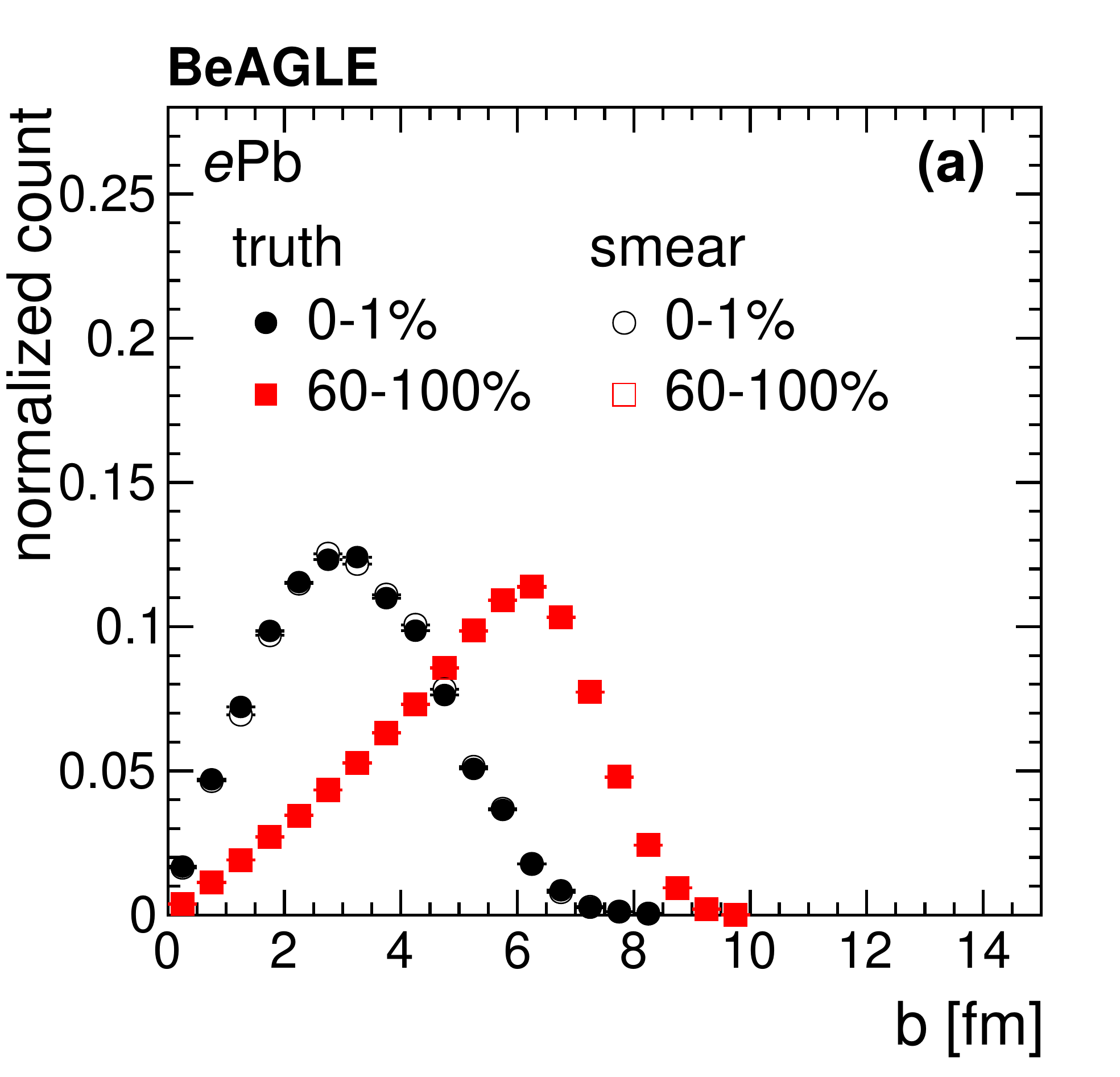}
\includegraphics[width=0.36\textwidth]{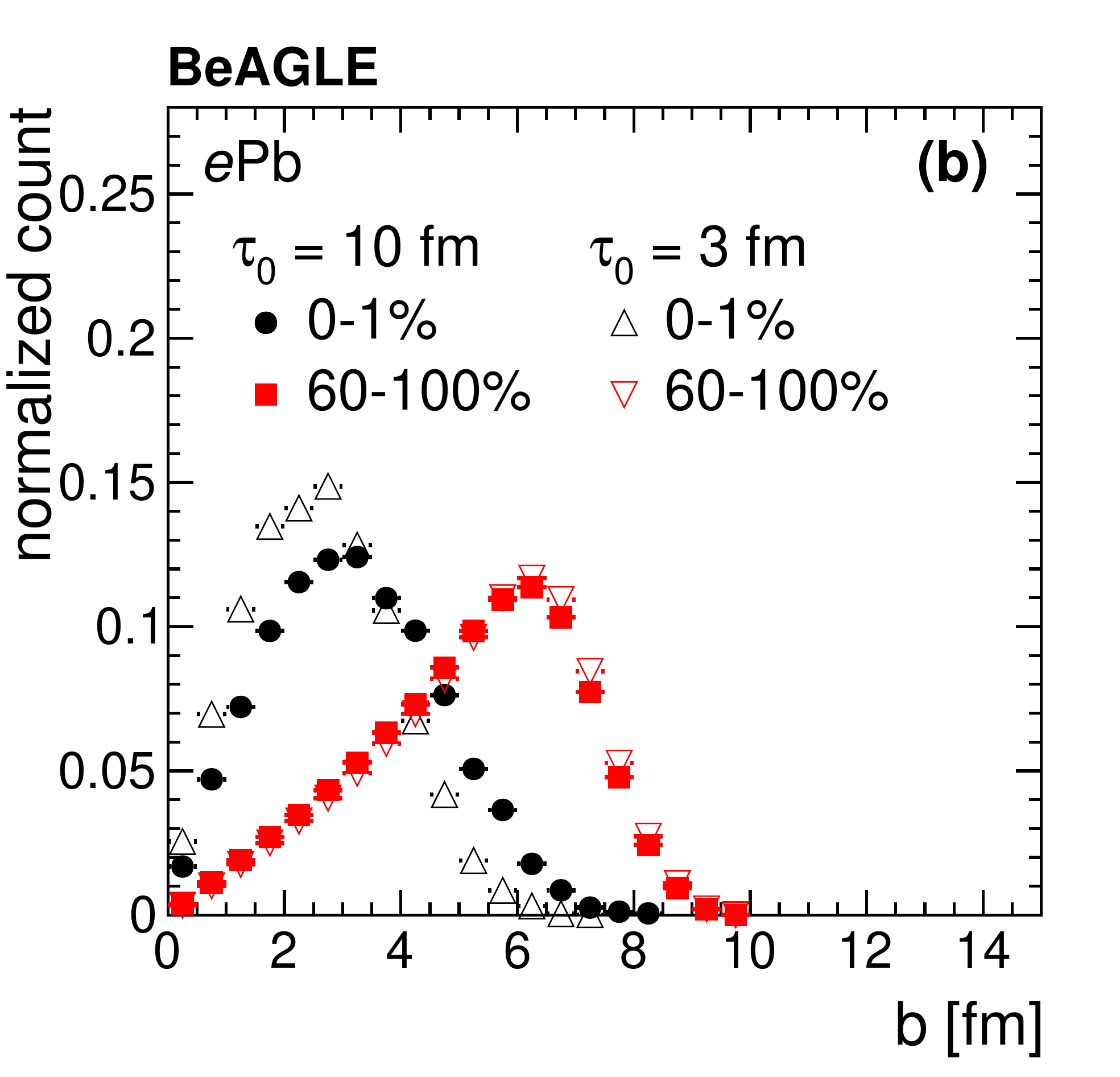}
\includegraphics[width=0.36\textwidth]{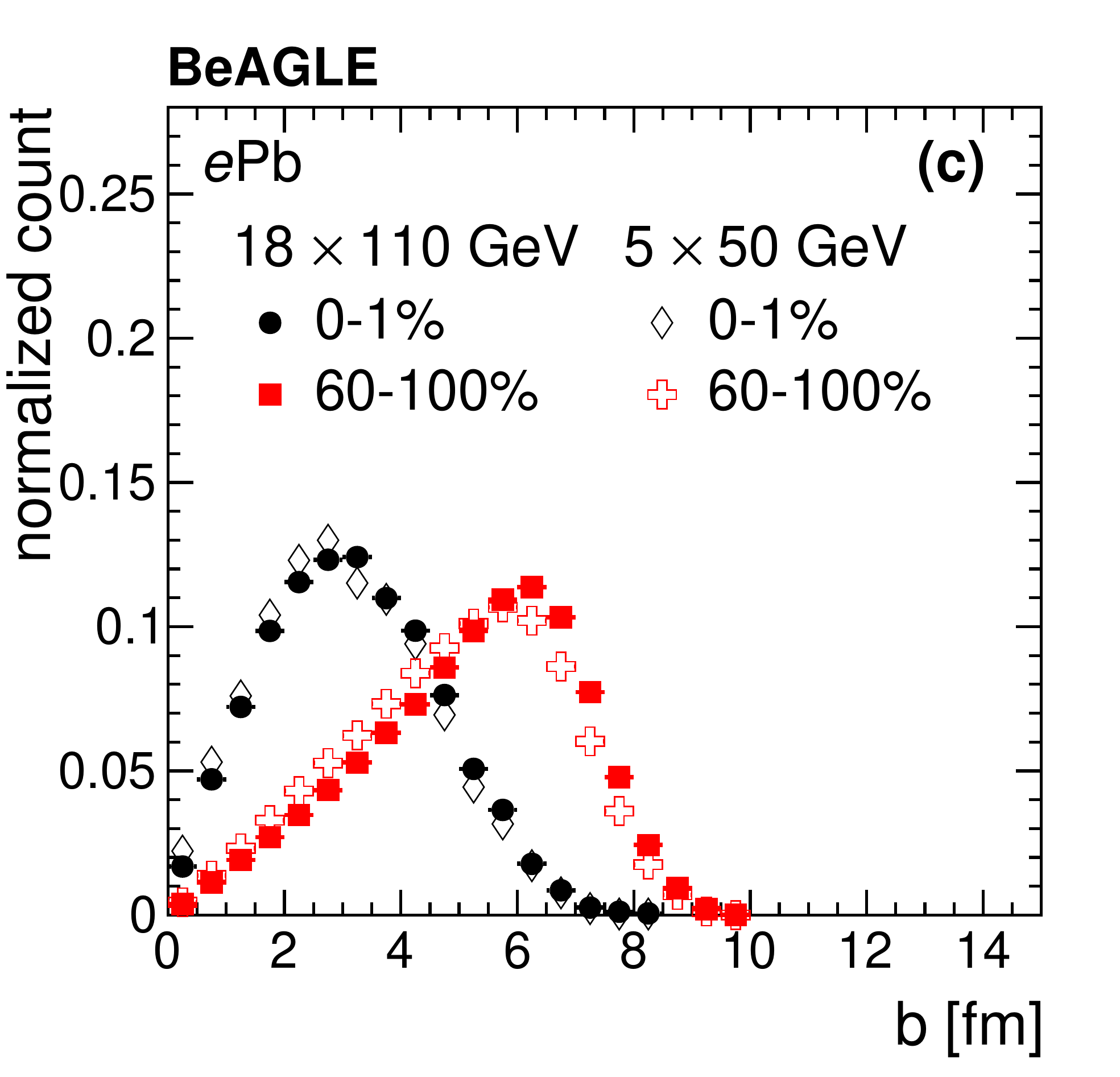}
\includegraphics[width=0.36\textwidth]{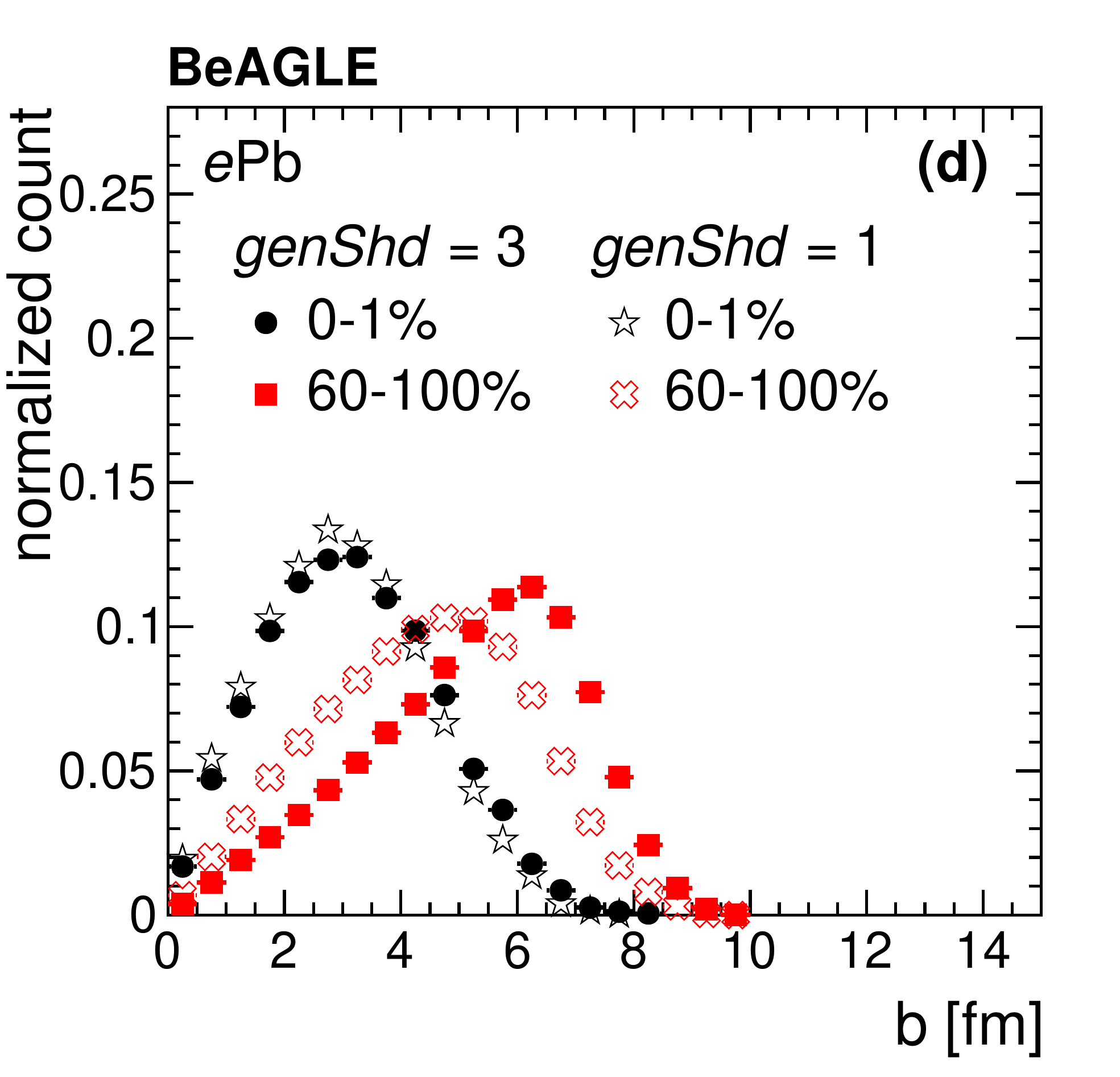}
\caption{The comparison of $b$ distributions in central and peripheral collisions for: (a) before and after accounting for detector smearing, the assumed resolution is $\frac{\sigma}{E} =  \frac{100\%}{\sqrt{E}} + 10 \%$, (b) $\tau_{0}$ = 10 fm and $\tau_{0}$ = 3 fm, (c) collision beam energy is 18 $\times$ 110 GeV and 5 $\times$ 50 GeV, (d) shadowing model 1 and 3. In all the plots, the black solid points and the red solid squares represent the results of central and peripheral collisions with the default event sample, separately. The default event sample is at 18 $\times$ 110 GeV for $e$Pb collisions with $\tau_{0}=10~$fm, shadowing model $genShd=3$, $1~\rm{GeV^{2}}<Q^{2}<100~\rm{GeV^{2}}$, and $0.01<y<0.95$. The open markers show the results with the only change as labeled in the legend.} 
\label{Fig1_9}
\end{figure*}

\begin{figure*}[th]
\centering  
\includegraphics[width=0.36\textwidth]{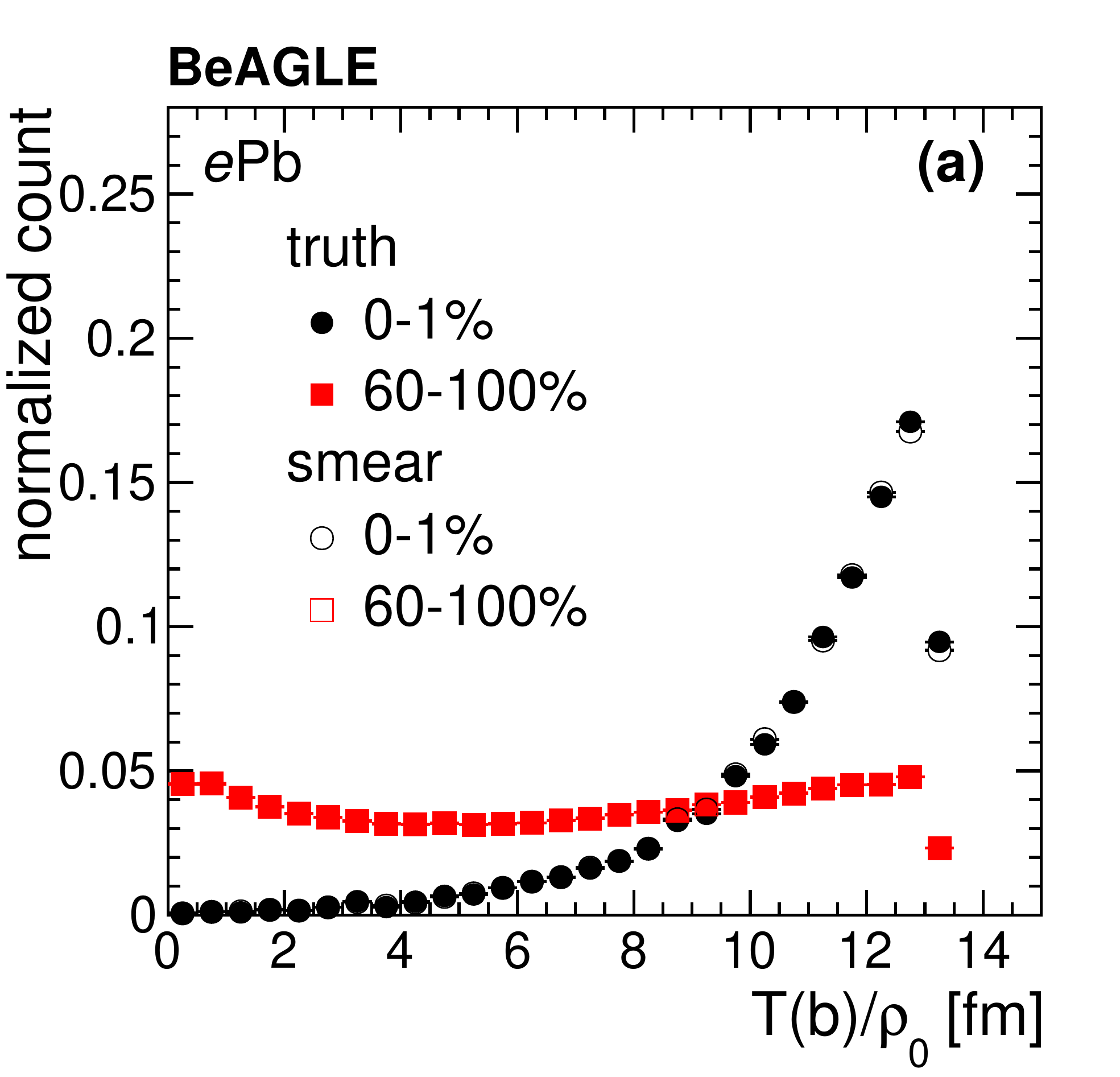}
\includegraphics[width=0.36\textwidth]{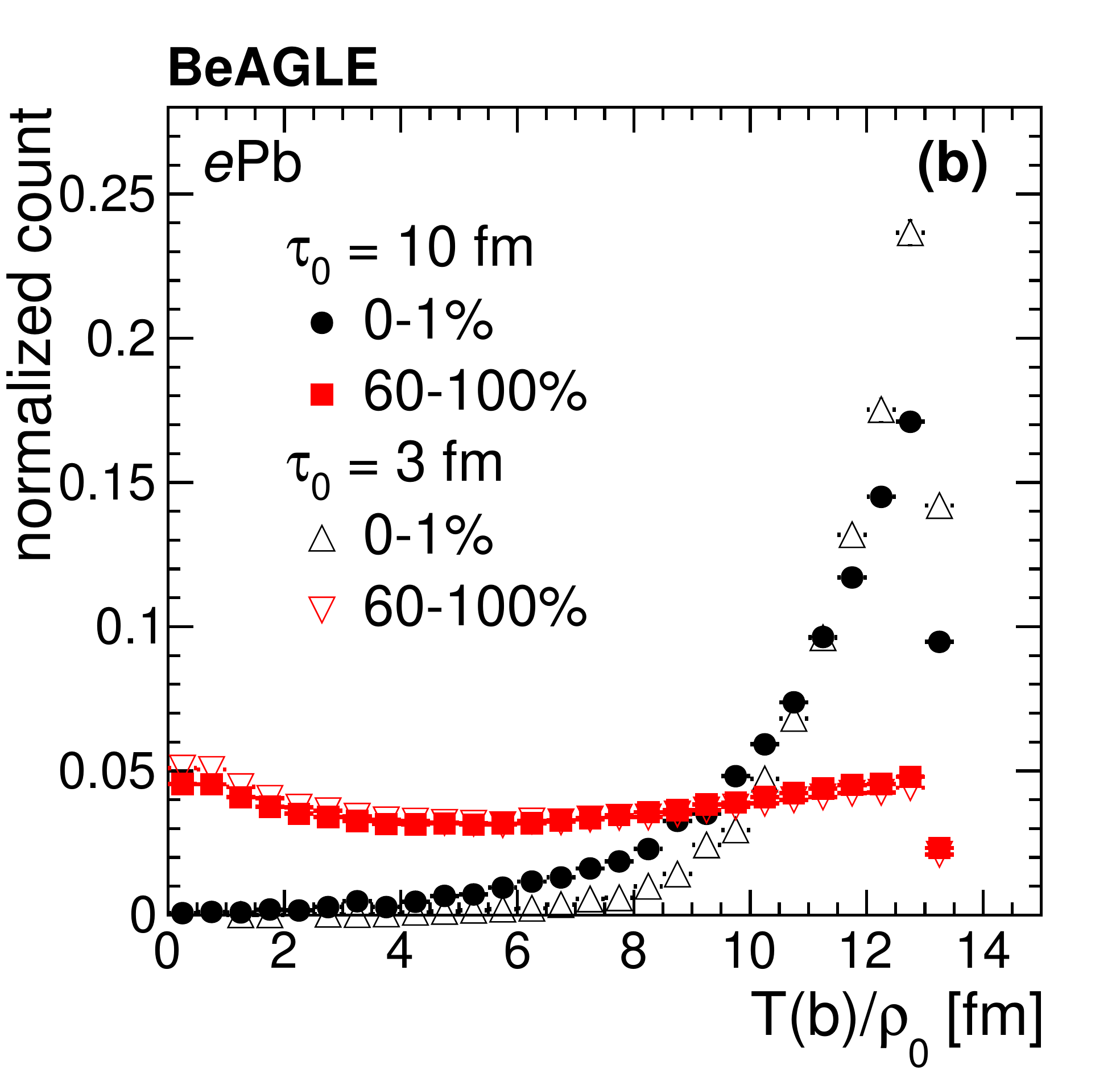}
\includegraphics[width=0.36\textwidth]{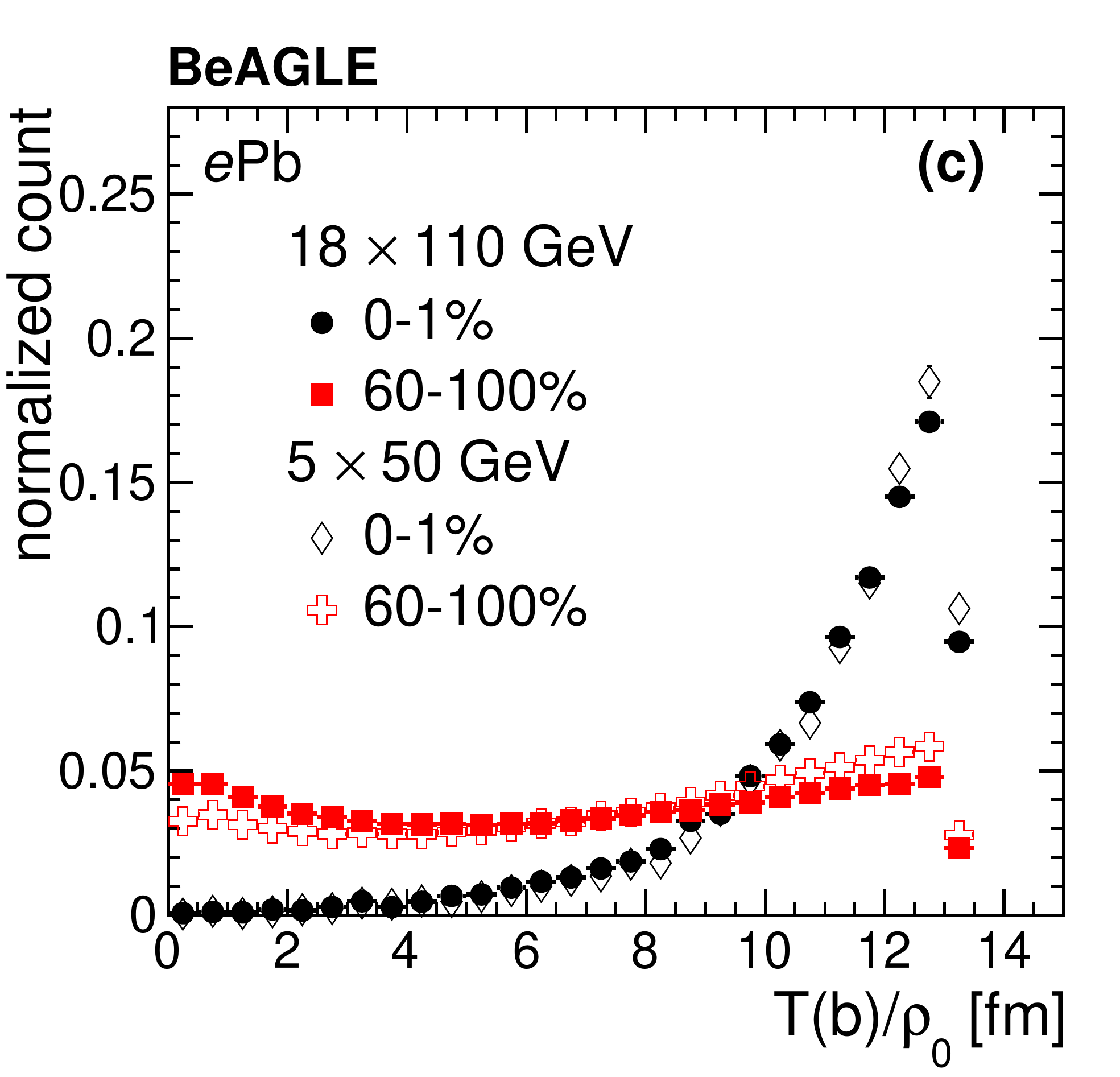}
\includegraphics[width=0.36\textwidth]{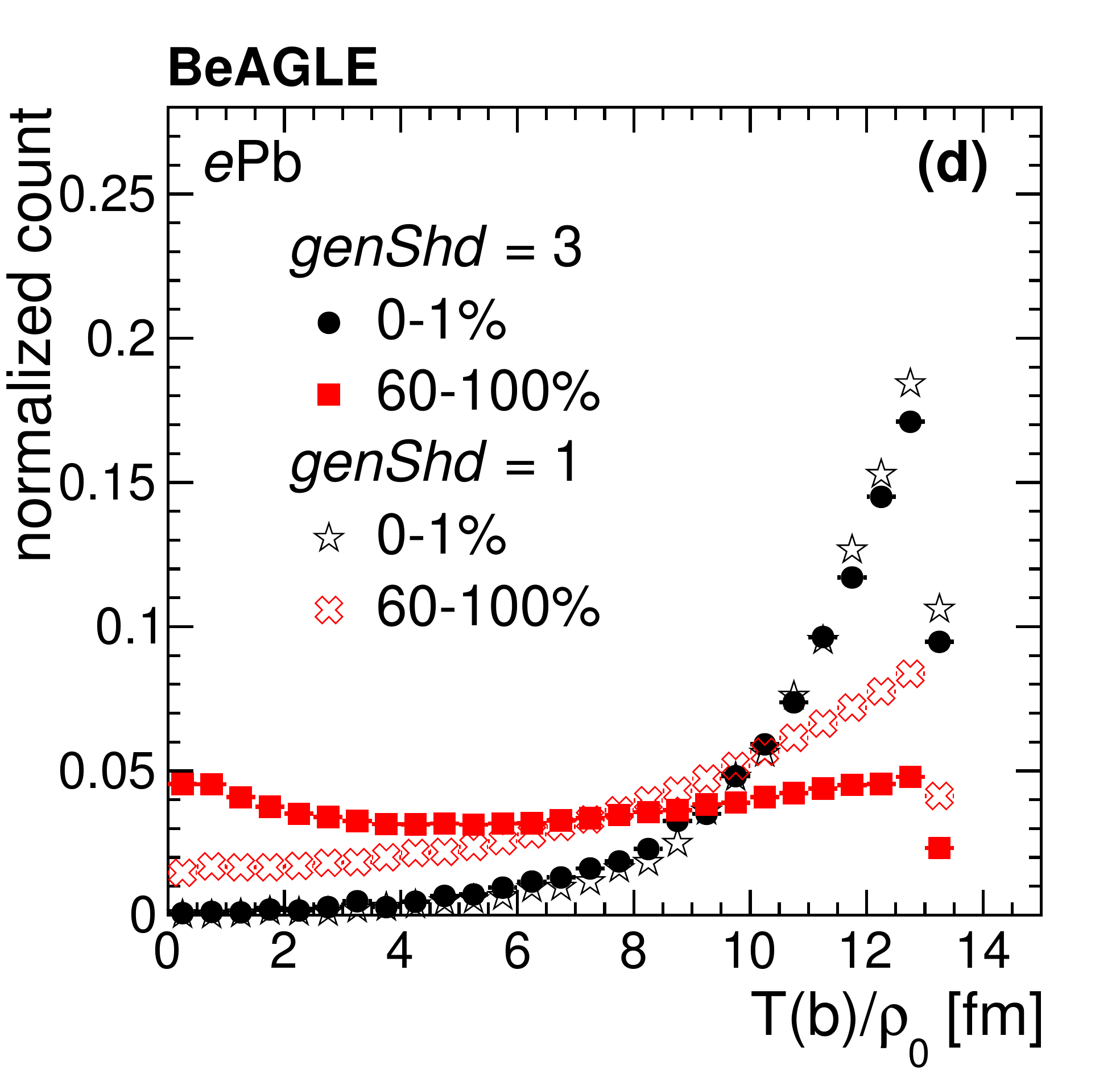}
\caption{The comparison of $T(b)/\rho _{0}$ distributions in central and peripheral collisions for: (a) before and after accounting for detector smearing, the assumed resolution is $\frac{\sigma}{E} =  \frac{100\%}{\sqrt{E}} + 10 \%$, (b) $\tau_{0}$ = 10 fm and $\tau_{0}$ = 3 fm, (c) collision beam energy is 18 $\times$ 110 GeV and 5 $\times$ 50 GeV, (d) shadowing model 1 and 3. In all the plots, the black solid points and the red solid squares represent the results of central and peripheral collisions with the default event sample, separately. The default event sample is at 18 $\times$ 110 GeV for $e$Pb collisions with $\tau_{0}=10~$fm, shadowing model $genShd=3$, $1~\rm{GeV^{2}}<Q^{2}<100~\rm{GeV^{2}}$, and $0.01<y<0.95$. The open markers show the results with the only change as labeled in the legend.} 
\label{Fig1_8}
\end{figure*}

In this subsection, we perform four different systematic tests, shown in Fig.~\ref{Fig1_9} and Fig.~\ref{Fig1_8}, respectively: (a) detector effect; (b) parameter $\tau_{0}$ dependence; (c) energy dependence; (d) shadowing effect. In these figures, black solid circles and red solid squares represent the results of central and peripheral collisions with the default event sample, respectively. The open markers show the results with the change as labeled in the legend.

First, as the centrality is selected via the energy deposition in the ZDC, we take the ZDC energy resolution and the angular acceptance ($\theta<5.5~\rm{mrad}$) into account. We assume a ZDC energy resolution of $\frac{\sigma}{E} =  \frac{100\%}{\sqrt{E}} + 10 \% $ to smear the energy of each individual neutron with a Gaussian distribution. Figure~\ref{Fig1_9}(a) and Fig.~\ref{Fig1_8}(a) illustrate the change in the $b$ and $T(b)/\rho _{0}$ distributions after detector smearing, respectively. The black points represent central collisions, while the red points depict peripheral collisions. The solid markers show the generated distribution without smearing, while the open markers include detector smearing. One can conclude that the results at generator level and after detector smearing are almost identical. The small impact of the ZDC energy resolution on centrality does not put stringent requirements on the ZDC performance.

Secondly, in this analysis, the default option is $\tau_{0} = 10$ fm and $genShd=3$. In order to study the impact of $\tau_{0}$ on centrality, it was lowered to 3 fm. A smaller $\tau_{0}$ means more particles can be formed in the nucleus, which results in more emitted neutrons from the nuclear break up, and consequently a larger energy deposition in the ZDC. Figure~\ref{Fig1_9}(b) and Fig.~\ref{Fig1_8}(b) shows the $b$ and $T(b)/\rho_{0}$ comparison for $\tau_{0}$ = 10 and 3 fm in both central and peripheral collisions for the $genShd=3$ case, respectively. There is no significant difference between the distributions of $\tau_{0}$ = 10 and 3 fm observed for peripheral events, while some differences for central events. However, the difference between peripheral and central events is small, showing a weak dependence on $\tau_{0}$. 

Thirdly, the energy of the emitted particles scales with the beam energy. However, for the $b$ distribution, there is no significant difference between central and peripheral collisions for the various beam energies, as shown in Fig.~\ref{Fig1_9}(c). The same behavior is observed for $T(b)/\rho_{0}$, and summarized in Fig.~\ref{Fig1_8}(c). This indicates that there is no beam energy dependence for the centrality definition. Therefore, although some model parameters are not precisely determined in BeAGLE, we find the correlation between ZDC energy and collision geometry is very stable. 

To model nuclear shadowing effects, BeAGLE has 3 different models implemented, as described in Sec.~\ref{subsec:hard_interaction}. Studies indicate a very small effect of shadowing on the energy deposition in the ZDC in the BeAGLE framework. Predictions for $b$ and $T(b)/\rho _{0}$ with the different shadowing models are also studied. Fig.~\ref{Fig1_9}(d) and Fig.~\ref{Fig1_8}(d) show the comparison of $b$ and $T(b)/\rho _{0}$ between $genShd=3$ and $genShd=1$, respectively. In both distributions, no difference is observed between the two shadowing options in central collisions, but some differences are seen in peripheral collisions. The observed differences arise from the low $Q^{2}$ region. This can be understood from the fact that $Q^{2} \propto \frac{1}{\lambda }$, where $\lambda$ is the wavelength of the photon. At low $Q^{2}$, the photon has a large wavelength, and can interact with many nucleons at once. However, for high $Q^{2}$, the wavelength of the photon is small, and therefore less nucleons particpate in the interaction. No difference as a function of $x_{\rm{bj}}$ for these two shadowing models is found.  

\section{\label{sec:discussions} Discussion}

In the previous sections, we compared $ep$ DIS events from the PYTHIA event generator to data from the ZEUS experiment at HERA, as well as $\mu$Xe and $\mu$D collision results from the BeAGLE event generator and E665 data at FermiLab. The results show that we can tune the PYTHIA model to describe target fragmentation in $ep$ collisions, while BeAGLE can not fully describe the target fragmentation region in $e$A at E665. Model uncertainties, e.g., $\tau_{0}$, and insufficient knowledge of the experimental selection in E665 might be responsible for the observed discrepancy. In order to further improve our understanding on the way toward the EIC, currently available Ultra-Peripheral Collisions (UPC) data at the Relativistic Heavy-Ion Collider, e.g., the recent data of $J/\psi$ photoproduction in the deuteron-gold UPC~\cite{STAR:2021wwq},  and UPC data from the Large Hadron Collider, will be extremely valuable, along with tagged target fragmentation studies at the Continuous Electron Beam Accelerator Facility at Jefferson Lab. These data provide a new pathway for study and validatation and improvement of the BeAGLE generator. 

In addition, BeAGLE currently cannot simulate coherent diffraction in $e$A due to the construction of the model. This is closely related to the determination of the formation time parameter, e.g., $\tau_{0}$. Another future plan for the BeAGLE development will be in this area, where coherent diffraction will provide important insights into the underlying gluon dynamics in the nucleus. 

In parallel to this work, there are recent efforts in improving the parton energy loss model PyQM in a different study~\cite{RoblesGajardo:2022efe}, modification of the DIS kinematics in light nuclei to account for Fermi momentum, implementation of the EMC effect\cite{Hen:2013oha, weinstein11, Seely:2009gt, Norton:2003cb, Malace:2014uea}, and short-range correlations using a generalized-contact formalism~\cite{PhysRevC.92.054311, Cruz-Torres:2019fum,Hauenstein:2021zql} for lower energy scattering. All past studies, the current work, and future studies have positioned BeAGLE as the prime MC tool for studying lepton-nucleus collisions at high energy, particularly towards the upcoming EIC.  

\section{\label{sec:summary} Summary}

In this work, we provide a comprehensive description of a high energy lepton-nucleus collision MC event generator - BeAGLE. We validate the model by comparing simulated observables from BeAGLE to available experimental data. The comparison of the PYTHIA-6 model calculations with the ZEUS experimental data in $ep$ collisions shows that we have a good PYTHIA model with refined tunes for target fragmentation in lepton-proton collisions. The BeAGLE event generator describes the E665 lepton-nucleus data for various kinematic variables. However, it only gives a fair description of the charged particle production spectra as measured by the E665 experiment. In order to obtain a full understanding of particle production in the current and target fragmentation region,  a future facility of high-energy lepton-nucleus collisions, e.g., the EIC, is required.

Based on the BeAGLE event generator, a systematic investigation of collision geometry determination using the detection of neutrons from the nuclear breakup is presented. We find the forward neutron production can provide a good experimental handle on the effective interaction length and nuclear thickness. These parameters will be important for the quantitative study of partonic energy loss in a nuclear medium, and for studies of non-linear gluon dynamics. Detector requirements for a ZDC are discussed, where the energy resolution has a small impact on the centrality determination and thus does not put stringent requirements on the detector, in contrast to studies of spectator tagging~\cite{Tu:2020ymk, Jentsch:2021qdp}. In addition, we present the dependence of the collision geometry on shadowing effects, the formation time parameter $\tau_{0}$, and the beam energy. All systematic variations are found to have small impact on the determination of the collision geometry, showing that this robust experimental measurement has minimal model dependence. The study reported in this paper provides an important baseline for developing a general-purpose MC event generator for high energy lepton-nucleus collisions. 

\begin{acknowledgments}
We thank Vasiliy Morozov, Pawel Nadel-Turonski, Charles Hyde, and Douglas Higinbotham for fruitful discussions on EIC sciences and experiments. We thank Alberto Accardi, Rapha\"el Dupr\'e, and Nestor Armesto for their insights on the nuclear geometry and DPMJET used in BeAGLE. The work of W. Chang is supported by the U.S. Department of Energy under Contract No. de-sc0012704 and the National Natural Science Foundation of China with Grant No. 11875143. The work of E.C. Aschenauer, A. Jentsch, JH. Lee, and Z. Tu is supported by the U.S. Department of Energy under Contract No. de-sc0012704, and A. Jentsch is also supported by the Program Development program at Brookhaven National Laboratory. The work of M.~D.~Baker is supported by DOE contracts de-sc0012704, DE-AC05-06OR23177, and by Jefferson Lab LDRD project LDRD1706. The work of L. Zheng is supported by National Natural Science Foundation of China under Grant No. 11905188. The work of Z. Yin is supported by the National Natural Science Foundation of China with Grant No. 11875143. 
\end{acknowledgments}

\appendix

\section{\label{sec:appendix} PYTHIA parameters and BeAGLE control card}

Table~\ref{tablepythia} summarizes the PYTHIA parameters used in this paper which are not the same as the default value. The meaning of each parameter and the default value can be found in Ref.~\cite{Sjostrand:2006za}. Except the parameters that introduced in Sec.~\ref{pythiaandzeus}, others were tuned by HERMES data~\cite{HERMES:2010nas, Liebing:2004us}. Note that the parameters of MSTP(17)$=$6 and PARP(166)$=$0.67597 are not PYTHIA-6 standard parameters. They are defined as a different parameterization of $R_{\rm{VMD}}$ with respect to the default, where $R_{\rm{VMD}}$ is the ratio of the hadronic cross sections of longitudinally to transversely polarized vector mesons and defined as,
\begin{equation}
R_{\rm{VMD}}=C(\frac{Q^{2}}{M_{\rho }^{2}})^B,
\end{equation}
\noindent with $C=$PARP(165) and $B=$PARP(166). See Refs.~\cite{Sjostrand:2006za, Liebing:2004us} for details. 

Table~\ref{tablebeagle} shows the BeAGLE input control card, including the meaning of each parameter and different value.

\begin{table*}[h]
\caption{Summary of PYTHIA parameters used in this paper that are different to the default tune.}
    \label{tablepythia}
    \begin{tabularx}{0.48\textwidth} { 
    >{\centering\arraybackslash}X 
  | >{\centering\arraybackslash}X  }
         \hline
         \hline
         parameter & value \\ \hline
         MSEL & 2 \\
         MSTP(14) & 30 \\
         MSTP(17) & 6 \\
         MSTP(19) & 1 \\
         MSTP(20) & 4 \\
         MSTP(38) & 4 \\
         MSTP(51) & 10042 \\
         MSTP(52) & 2 \\
         MSTP(81) & 0 \\
         MSTP(82) & 1 \\
         MSTP(94) & 2 \\
         MSTP(101) & 1 \\
         PARP(18) & 0.17 \\
         PARP(91) & 0.40 \\
         PARP(97) & 6.0 \\
         PARP(99) & 0.40 \\
         PARP(161) & 3.00 \\
         PARP(162) & 24.6 \\
         PARP(163) & 18.8 \\
         PARP(165) & 0.47679 \\
         PARP(166) & 0.67597 \\
         PARJ(21) & 0.40 \\
         PARJ(170) & 0.32 \\
         MSTJ(12) & 1 \\
         MSTU(113) & 5 \\
         CKIN(1) & 1.0 \\
         CKIN(65) & 1.e-09\\
         \hline
         \hline 
    \end{tabularx}
\end{table*}

\begin{table*}[h!]
\fontsize{11}{13}\selectfont
\centering
\caption{BeAGLE input control card}
    \label{tablebeagle}
    \begin{tabular}{lp{15cm}}
         \hline
         \hline
        Parameter & Descriptions   \\ \hline
        PROJPAR & Lepton beam can be ``ELECTRON" (or ``MUON$+$"). \\\hline
        TARPAR & The first number is nucleus $A$ number and the second number is charge $Z$ for the target.  \newline
                 The third number is the $n/p$ handling mode for the Pythia subevent: 
                 \begin{itemize}
                     \item 0 = sequential $n,p$. The first events are $en$, the remaining $ep$ (binomial prob.) Useful for getting $en$ and $ep$ cross-sections from Pythia.
                     \item 1 = $en$ only test mode (not really for physics).
                     \item 2 = $ep$ only test mode (not really for physics).
                     \item 3 = random mix. The events are randomly $en$ or $ep$. Useful because you can analyze just a subset of the data without bias.
                 \end{itemize}
                \\ \hline    
        TAUFOR & The first number is the formation time parameter ($\tau_0$) in $\rm {fm}/\it {c}$ for the intra-nuclear cascade, where the second number is the number of generations followed. Default=25, 0 means no cascade.\\ \hline
        MOMENTUM & The first number is for the lepton beam ($\rm {GeV}/c$), the second for the ion (or $p$) beam. Both numbers should be entered as positive, but the lepton beam will be multiplied by $-1$. \\ \hline  
        L-TAG   & These numbers are cuts: $y$Min, $y$Max, $Q^2$Min, $Q^2$Max, thetaMin, thetaMax, where $y$ and $Q^{2}$ ($\rm GeV^2$) are the leptoproduction kinematics and theta (radians) refers to the lepton scattering angle in the laboratory frame. \\ \hline
        PY-INPUT & Specifies the file (with an eight-character maximum name!) used as a pythia input file. See instructions at \url{https://eic.github.io/software/beagle.html}.  \\ \hline
        FERMI   & First number: 
        \begin{itemize}
            \item -1 = no Fermi motion at all.
            \item 1 = DPMJET Fermi motion, but Pythia subevent neglects it (DPMJetHybrid mode).
            \item 2 = Fermi motion added to Pythia subevent after the fact.
            \item 3 = Pythia subevent used correct Fermi motion (not yet implemented).
        \end{itemize}
              Second number: ``Scale factor" for Fermi momentum distribution in GeV (0.62 is default, and experts-only parameter).   \newline
              Third number: Fermi momentum distribution 0 (D) = most recommended distribution. \newline
              Fourth number: Post-processing flag  (not yet implemented except deuteron).\newline
                             $~~~~~~~~$0 (D) = no post-processing. \newline
                             $~~~~~~~~$1 = fix energy non-conservation in ion frame (for small nuclei). \\ \hline
    FSEED & Leave it as it is and only change the FSEED in the PY-INPUT for production or debugging. \\\hline
    OUTLEVEL & First 4 numbers are verbosity flags: $-1$=quiet, $\geqslant$1 increasing verbosity. \newline
                Fifth number is the number of events to print out and in some cases to be verbose about. \\ \hline
    PYVECTORS & Allowed Pythia vector mesons for diffraction: 0(D)=all, 1=$\rho$, 2=$\omega$, 3=$\phi$, 4=$J/\psi$. \\ \hline
    USERSET   & First number specifies the meaning of the variables User1,User2,User3.\newline
                Second number specifies the maximum excitation energy in GeV handed to FLUKA (D=9.0). \newline
                 Note: This should not come into play, but it protects against infinite loops. \\ \hline
    PHOINPUT  & Any options explained in the PHOJET-manual can be used in between the ``PHOINPUT" and ``ENDINPUT" cards.\\
   
    PROCESS   &       1 1 1 1 1 1 1 1\\
    
    ENDINPUT  & \\\hline
    START     & First number specifies the number of events to run. \newline
                Second number should be 0 (or missing). \\ \hline
    STOP & \\
    \hline
     \hline
    \end{tabular}
\end{table*}

\bibliography{beagle}

\begin{thebibliography}{63}%
\makeatletter
\providecommand \@ifxundefined [1]{%
 \@ifx{#1\undefined}
}%
\providecommand \@ifnum [1]{%
 \ifnum #1\expandafter \@firstoftwo
 \else \expandafter \@secondoftwo
 \fi
}%
\providecommand \@ifx [1]{%
 \ifx #1\expandafter \@firstoftwo
 \else \expandafter \@secondoftwo
 \fi
}%
\providecommand \natexlab [1]{#1}%
\providecommand \enquote  [1]{``#1''}%
\providecommand \bibnamefont  [1]{#1}%
\providecommand \bibfnamefont [1]{#1}%
\providecommand \citenamefont [1]{#1}%
\providecommand \href@noop [0]{\@secondoftwo}%
\providecommand \href [0]{\begingroup \@sanitize@url \@href}%
\providecommand \@href[1]{\@@startlink{#1}\@@href}%
\providecommand \@@href[1]{\endgroup#1\@@endlink}%
\providecommand \@sanitize@url [0]{\catcode `\\12\catcode `\$12\catcode
  `\&12\catcode `\#12\catcode `\^12\catcode `\_12\catcode `\%12\relax}%
\providecommand \@@startlink[1]{}%
\providecommand \@@endlink[0]{}%
\providecommand \url  [0]{\begingroup\@sanitize@url \@url }%
\providecommand \@url [1]{\endgroup\@href {#1}{\urlprefix }}%
\providecommand \urlprefix  [0]{URL }%
\providecommand \Eprint [0]{\href }%
\providecommand \doibase [0]{http://dx.doi.org/}%
\providecommand \selectlanguage [0]{\@gobble}%
\providecommand \bibinfo  [0]{\@secondoftwo}%
\providecommand \bibfield  [0]{\@secondoftwo}%
\providecommand \translation [1]{[#1]}%
\providecommand \BibitemOpen [0]{}%
\providecommand \bibitemStop [0]{}%
\providecommand \bibitemNoStop [0]{.\EOS\space}%
\providecommand \EOS [0]{\spacefactor3000\relax}%
\providecommand \BibitemShut  [1]{\csname bibitem#1\endcsname}%
\let\auto@bib@innerbib\@empty
\bibitem [{\citenamefont {Gaillard}\ \emph {et~al.}(1999)\citenamefont
  {Gaillard}, \citenamefont {Grannis},\ and\ \citenamefont
  {Sciulli}}]{Gaillard:1998ui}%
  \BibitemOpen
  \bibfield  {author} {\bibinfo {author} {\bibfnamefont {M.~K.}\ \bibnamefont
  {Gaillard}}, \bibinfo {author} {\bibfnamefont {P.~D.}\ \bibnamefont
  {Grannis}}, \ and\ \bibinfo {author} {\bibfnamefont {F.~J.}\ \bibnamefont
  {Sciulli}},\ }\href {\doibase 10.1103/RevModPhys.71.S96} {\bibfield
  {journal} {\bibinfo  {journal} {Rev. Mod. Phys.}\ }\textbf {\bibinfo {volume}
  {71}},\ \bibinfo {pages} {S96} (\bibinfo {year} {1999})},\ \Eprint
  {http://arxiv.org/abs/hep-ph/9812285} {arXiv:hep-ph/9812285} \BibitemShut
  {NoStop}%
\bibitem [{\citenamefont {Oerter}(2006)}]{Oerter:2006iy}%
  \BibitemOpen
  \bibfield  {author} {\bibinfo {author} {\bibfnamefont {R.}~\bibnamefont
  {Oerter}},\ }\href@noop {} {\emph {\bibinfo {title} {{The theory of almost
  everything: The standard model, the unsung triumph of modern physics}}}}\
  (\bibinfo {year} {2006})\BibitemShut {NoStop}%
\bibitem [{\citenamefont {Politzer}(1974)}]{Politzer:1974fr}%
  \BibitemOpen
  \bibfield  {author} {\bibinfo {author} {\bibfnamefont {H.~D.}\ \bibnamefont
  {Politzer}},\ }\href {\doibase 10.1016/0370-1573(74)90014-3} {\bibfield
  {journal} {\bibinfo  {journal} {Phys. Rept.}\ }\textbf {\bibinfo {volume}
  {14}},\ \bibinfo {pages} {129} (\bibinfo {year} {1974})}\BibitemShut
  {NoStop}%
\bibitem [{\citenamefont {Gross}\ and\ \citenamefont
  {Wilczek}(1973{\natexlab{a}})}]{PhysRevLett.30.1343}%
  \BibitemOpen
  \bibfield  {author} {\bibinfo {author} {\bibfnamefont {D.~J.}\ \bibnamefont
  {Gross}}\ and\ \bibinfo {author} {\bibfnamefont {F.}~\bibnamefont
  {Wilczek}},\ }\href {\doibase 10.1103/PhysRevLett.30.1343} {\bibfield
  {journal} {\bibinfo  {journal} {Phys. Rev. Lett.}\ }\textbf {\bibinfo
  {volume} {30}},\ \bibinfo {pages} {1343} (\bibinfo {year}
  {1973}{\natexlab{a}})}\BibitemShut {NoStop}%
\bibitem [{\citenamefont {Gross}\ and\ \citenamefont
  {Wilczek}(1973{\natexlab{b}})}]{PhysRevD.8.3633}%
  \BibitemOpen
  \bibfield  {author} {\bibinfo {author} {\bibfnamefont {D.~J.}\ \bibnamefont
  {Gross}}\ and\ \bibinfo {author} {\bibfnamefont {F.}~\bibnamefont
  {Wilczek}},\ }\href {\doibase 10.1103/PhysRevD.8.3633} {\bibfield  {journal}
  {\bibinfo  {journal} {Phys. Rev. D}\ }\textbf {\bibinfo {volume} {8}},\
  \bibinfo {pages} {3633} (\bibinfo {year} {1973}{\natexlab{b}})}\BibitemShut
  {NoStop}%
\bibitem [{\citenamefont {Accardi}\ \emph {et~al.}(2016)\citenamefont {Accardi}
  \emph {et~al.}}]{Accardi:2012qut}%
  \BibitemOpen
  \bibfield  {author} {\bibinfo {author} {\bibfnamefont {A.}~\bibnamefont
  {Accardi}} \emph {et~al.},\ }\href {\doibase 10.1140/epja/i2016-16268-9}
  {\bibfield  {journal} {\bibinfo  {journal} {Eur. Phys. J. A}\ }\textbf
  {\bibinfo {volume} {52}},\ \bibinfo {pages} {268} (\bibinfo {year} {2016})},\
  \Eprint {http://arxiv.org/abs/1212.1701} {arXiv:1212.1701 [nucl-ex]}
  \BibitemShut {NoStop}%
\bibitem [{\citenamefont {Abdul~Khalek}\ \emph {et~al.}(2021)\citenamefont
  {Abdul~Khalek} \emph {et~al.}}]{AbdulKhalek:2021gbh}%
  \BibitemOpen
  \bibfield  {author} {\bibinfo {author} {\bibfnamefont {R.}~\bibnamefont
  {Abdul~Khalek}} \emph {et~al.},\ }\href@noop {} {\  (\bibinfo {year}
  {2021})},\ \Eprint {http://arxiv.org/abs/2103.05419} {arXiv:2103.05419
  [physics.ins-det]} \BibitemShut {NoStop}%
\bibitem [{\citenamefont {Adam}\ \emph {et~al.}(2021)\citenamefont {Adam} \emph
  {et~al.}}]{ref:EICCDR}%
  \BibitemOpen
  \bibfield  {author} {\bibinfo {author} {\bibfnamefont {J.}~\bibnamefont
  {Adam}} \emph {et~al.},\ }\href@noop {} {\enquote {\bibinfo {title} {Electron
  ion collider conceptual design report},}\ }\bibinfo {howpublished}
  {\url{https://www.bnl.gov/ec/files/EIC_CDR_Final.pdf}} (\bibinfo {year}
  {2021})\BibitemShut {NoStop}%
\bibitem [{\citenamefont {Gelis}\ \emph {et~al.}(2010)\citenamefont {Gelis},
  \citenamefont {Iancu}, \citenamefont {Jalilian-Marian},\ and\ \citenamefont
  {Venugopalan}}]{Gelis:2010nm}%
  \BibitemOpen
  \bibfield  {author} {\bibinfo {author} {\bibfnamefont {F.}~\bibnamefont
  {Gelis}}, \bibinfo {author} {\bibfnamefont {E.}~\bibnamefont {Iancu}},
  \bibinfo {author} {\bibfnamefont {J.}~\bibnamefont {Jalilian-Marian}}, \ and\
  \bibinfo {author} {\bibfnamefont {R.}~\bibnamefont {Venugopalan}},\ }\href
  {\doibase 10.1146/annurev.nucl.010909.083629} {\bibfield  {journal} {\bibinfo
   {journal} {Ann. Rev. Nucl. Part. Sci.}\ }\textbf {\bibinfo {volume} {60}},\
  \bibinfo {pages} {463} (\bibinfo {year} {2010})},\ \Eprint
  {http://arxiv.org/abs/1002.0333} {arXiv:1002.0333 [hep-ph]} \BibitemShut
  {NoStop}%
\bibitem [{\citenamefont {Jalilian-Marian}(2014)}]{Jalilian-Marian:2014ica}%
  \BibitemOpen
  \bibfield  {author} {\bibinfo {author} {\bibfnamefont {J.}~\bibnamefont
  {Jalilian-Marian}},\ }\href {\doibase 10.1051/epjconf/20146604012} {\bibfield
   {journal} {\bibinfo  {journal} {EPJ Web Conf.}\ }\textbf {\bibinfo {volume}
  {66}},\ \bibinfo {pages} {04012} (\bibinfo {year} {2014})}\BibitemShut
  {NoStop}%
\bibitem [{\citenamefont {Aschenauer}\ \emph {et~al.}(2019)\citenamefont
  {Aschenauer}, \citenamefont {Baker}, \citenamefont {Chang}, \citenamefont
  {Lee}, \citenamefont {Tu},\ and\ \citenamefont {Zheng}}]{Beagle}%
  \BibitemOpen
  \bibfield  {author} {\bibinfo {author} {\bibfnamefont {E.}~\bibnamefont
  {Aschenauer}}, \bibinfo {author} {\bibfnamefont {M.}~\bibnamefont {Baker}},
  \bibinfo {author} {\bibfnamefont {W.}~\bibnamefont {Chang}}, \bibinfo
  {author} {\bibfnamefont {J.}~\bibnamefont {Lee}}, \bibinfo {author}
  {\bibfnamefont {Z.}~\bibnamefont {Tu}}, \ and\ \bibinfo {author}
  {\bibfnamefont {L.}~\bibnamefont {Zheng}},\ }\href@noop {} {\enquote
  {\bibinfo {title} {{BeAGLE: A Tool to Refine Detector Requirements for eA
  Collisions EIC R\&D Project eRD17: Progress Report (January-June 2019) and
  Proposal}},}\ }\bibinfo {howpublished} {\url
  {https://wiki.bnl.gov/eic/index.php/BeAGLE}} (\bibinfo {year}
  {2019})\BibitemShut {NoStop}%
\bibitem [{\citenamefont {Chang}\ \emph {et~al.}(2021)\citenamefont {Chang},
  \citenamefont {Aschenauer}, \citenamefont {Baker}, \citenamefont {Jentsch},
  \citenamefont {Lee}, \citenamefont {Tu}, \citenamefont {Yin},\ and\
  \citenamefont {Zheng}}]{Chang:2021jnu}%
  \BibitemOpen
  \bibfield  {author} {\bibinfo {author} {\bibfnamefont {W.}~\bibnamefont
  {Chang}}, \bibinfo {author} {\bibfnamefont {E.-C.}\ \bibnamefont
  {Aschenauer}}, \bibinfo {author} {\bibfnamefont {M.~D.}\ \bibnamefont
  {Baker}}, \bibinfo {author} {\bibfnamefont {A.}~\bibnamefont {Jentsch}},
  \bibinfo {author} {\bibfnamefont {J.-H.}\ \bibnamefont {Lee}}, \bibinfo
  {author} {\bibfnamefont {Z.}~\bibnamefont {Tu}}, \bibinfo {author}
  {\bibfnamefont {Z.}~\bibnamefont {Yin}}, \ and\ \bibinfo {author}
  {\bibfnamefont {L.}~\bibnamefont {Zheng}},\ }\href {\doibase
  10.1103/PhysRevD.104.114030} {\bibfield  {journal} {\bibinfo  {journal}
  {Phys. Rev. D}\ }\textbf {\bibinfo {volume} {104}},\ \bibinfo {pages}
  {114030} (\bibinfo {year} {2021})},\ \Eprint
  {http://arxiv.org/abs/2108.01694} {arXiv:2108.01694 [nucl-ex]} \BibitemShut
  {NoStop}%
\bibitem [{\citenamefont {Jentsch}\ \emph {et~al.}(2021)\citenamefont
  {Jentsch}, \citenamefont {Tu},\ and\ \citenamefont
  {Weiss}}]{Jentsch:2021qdp}%
  \BibitemOpen
  \bibfield  {author} {\bibinfo {author} {\bibfnamefont {A.}~\bibnamefont
  {Jentsch}}, \bibinfo {author} {\bibfnamefont {Z.}~\bibnamefont {Tu}}, \ and\
  \bibinfo {author} {\bibfnamefont {C.}~\bibnamefont {Weiss}},\ }\href
  {\doibase 10.1103/PhysRevC.104.065205} {\bibfield  {journal} {\bibinfo
  {journal} {Phys. Rev. C}\ }\textbf {\bibinfo {volume} {104}},\ \bibinfo
  {pages} {065205} (\bibinfo {year} {2021})},\ \Eprint
  {http://arxiv.org/abs/2108.08314} {arXiv:2108.08314 [hep-ph]} \BibitemShut
  {NoStop}%
\bibitem [{\citenamefont {Hen}\ \emph {et~al.}(2017)\citenamefont {Hen},
  \citenamefont {Miller}, \citenamefont {Piasetzky},\ and\ \citenamefont
  {Weinstein}}]{Hen:2016kwk}%
  \BibitemOpen
  \bibfield  {author} {\bibinfo {author} {\bibfnamefont {O.}~\bibnamefont
  {Hen}}, \bibinfo {author} {\bibfnamefont {G.}~\bibnamefont {Miller}},
  \bibinfo {author} {\bibfnamefont {E.}~\bibnamefont {Piasetzky}}, \ and\
  \bibinfo {author} {\bibfnamefont {L.}~\bibnamefont {Weinstein}},\ }\href
  {\doibase 10.1103/RevModPhys.89.045002} {\bibfield  {journal} {\bibinfo
  {journal} {Rev. Mod. Phys.}\ }\textbf {\bibinfo {volume} {89}},\ \bibinfo
  {pages} {045002} (\bibinfo {year} {2017})},\ \Eprint
  {http://arxiv.org/abs/1611.09748} {arXiv:1611.09748 [nucl-ex]} \BibitemShut
  {NoStop}%
\bibitem [{\citenamefont {degli Atti}(2015)}]{Atti2015InmediumSD}%
  \BibitemOpen
  \bibfield  {author} {\bibinfo {author} {\bibfnamefont {C.~C.}\ \bibnamefont
  {degli Atti}},\ }\href@noop {} {\bibfield  {journal} {\bibinfo  {journal}
  {Physics Reports}\ }\textbf {\bibinfo {volume} {590}},\ \bibinfo {pages} {1}
  (\bibinfo {year} {2015})}\BibitemShut {NoStop}%
\bibitem [{\citenamefont {Tu}\ \emph {et~al.}(2020)\citenamefont {Tu},
  \citenamefont {Jentsch}, \citenamefont {Baker}, \citenamefont {Zheng},
  \citenamefont {Lee}, \citenamefont {Venugopalan}, \citenamefont {Hen},
  \citenamefont {Higinbotham}, \citenamefont {Aschenauer},\ and\ \citenamefont
  {Ullrich}}]{Tu:2020ymk}%
  \BibitemOpen
  \bibfield  {author} {\bibinfo {author} {\bibfnamefont {Z.}~\bibnamefont
  {Tu}}, \bibinfo {author} {\bibfnamefont {A.}~\bibnamefont {Jentsch}},
  \bibinfo {author} {\bibfnamefont {M.}~\bibnamefont {Baker}}, \bibinfo
  {author} {\bibfnamefont {L.}~\bibnamefont {Zheng}}, \bibinfo {author}
  {\bibfnamefont {J.-H.}\ \bibnamefont {Lee}}, \bibinfo {author} {\bibfnamefont
  {R.}~\bibnamefont {Venugopalan}}, \bibinfo {author} {\bibfnamefont
  {O.}~\bibnamefont {Hen}}, \bibinfo {author} {\bibfnamefont {D.}~\bibnamefont
  {Higinbotham}}, \bibinfo {author} {\bibfnamefont {E.-C.}\ \bibnamefont
  {Aschenauer}}, \ and\ \bibinfo {author} {\bibfnamefont {T.}~\bibnamefont
  {Ullrich}},\ }\href {\doibase 10.1016/j.physletb.2020.135877} {\bibfield
  {journal} {\bibinfo  {journal} {Phys. Lett. B}\ }\textbf {\bibinfo {volume}
  {811}},\ \bibinfo {pages} {135877} (\bibinfo {year} {2020})},\ \Eprint
  {http://arxiv.org/abs/2005.14706} {arXiv:2005.14706 [nucl-ex]} \BibitemShut
  {NoStop}%
\bibitem [{\citenamefont {Sj\"ostrand}\ \emph {et~al.}(2006)\citenamefont
  {Sj\"ostrand}, \citenamefont {Mrenna},\ and\ \citenamefont
  {Skands}}]{Sjostrand:2006za}%
  \BibitemOpen
  \bibfield  {author} {\bibinfo {author} {\bibfnamefont {T.}~\bibnamefont
  {Sj\"ostrand}}, \bibinfo {author} {\bibfnamefont {S.}~\bibnamefont {Mrenna}},
  \ and\ \bibinfo {author} {\bibfnamefont {P.}~\bibnamefont {Skands}},\ }\href
  {\doibase 10.1088/1126-6708/2006/05/026} {\bibfield  {journal} {\bibinfo
  {journal} {JHEP}\ }\textbf {\bibinfo {volume} {05}},\ \bibinfo {pages} {026}
  (\bibinfo {year} {2006})},\ \Eprint {http://arxiv.org/abs/hep-ph/0603175}
  {arXiv:hep-ph/0603175 [hep-ph]} \BibitemShut {NoStop}%
\bibitem [{\citenamefont {Basile}\ \emph {et~al.}(1981)\citenamefont {Basile}
  \emph {et~al.}}]{basile1981leading}%
  \BibitemOpen
  \bibfield  {author} {\bibinfo {author} {\bibfnamefont {M.}~\bibnamefont
  {Basile}} \emph {et~al.},\ }\href@noop {} {\bibfield  {journal} {\bibinfo
  {journal} {Il Nuovo Cimento A (1965-1970)}\ }\textbf {\bibinfo {volume}
  {66}},\ \bibinfo {pages} {129} (\bibinfo {year} {1981})}\BibitemShut
  {NoStop}%
\bibitem [{\citenamefont {Adams}\ \emph {et~al.}(1994)\citenamefont {Adams}
  \emph {et~al.}}]{E665:1993trt}%
  \BibitemOpen
  \bibfield  {author} {\bibinfo {author} {\bibfnamefont {M.~R.}\ \bibnamefont
  {Adams}} \emph {et~al.} (\bibinfo {collaboration} {E665}),\ }\href {\doibase
  10.1007/BF01413096} {\bibfield  {journal} {\bibinfo  {journal} {Z. Phys. C}\
  }\textbf {\bibinfo {volume} {61}},\ \bibinfo {pages} {179} (\bibinfo {year}
  {1994})}\BibitemShut {NoStop}%
\bibitem [{\citenamefont {Roesler}\ \emph {et~al.}(2000)\citenamefont
  {Roesler}, \citenamefont {Engel},\ and\ \citenamefont
  {Ranft}}]{Roesler:2000he}%
  \BibitemOpen
  \bibfield  {author} {\bibinfo {author} {\bibfnamefont {S.}~\bibnamefont
  {Roesler}}, \bibinfo {author} {\bibfnamefont {R.}~\bibnamefont {Engel}}, \
  and\ \bibinfo {author} {\bibfnamefont {J.}~\bibnamefont {Ranft}},\ }in\ \href
  {\doibase 10.1007/978-3-642-18211-2\_166} {\emph {\bibinfo {booktitle}
  {{Advanced Monte Carlo for radiation physics, particle transport simulation
  and applications. Proceedings, Conference, MC2000, Lisbon, Portugal, October
  23-26, 2000}}}}\ (\bibinfo {year} {2000})\ pp.\ \bibinfo {pages}
  {1033--1038},\ \Eprint {http://arxiv.org/abs/hep-ph/0012252}
  {arXiv:hep-ph/0012252} \BibitemShut {NoStop}%
\bibitem [{\citenamefont {Dupr\'e}(2011)}]{Dupre:2011afa}%
  \BibitemOpen
  \bibfield  {author} {\bibinfo {author} {\bibfnamefont {R.}~\bibnamefont
  {Dupr\'e}},\ }\emph {\bibinfo {title} {{Quark Fragmentation and Hadron
  Formation in Nuclear Matter}}},\ \href@noop {} {Ph.D. thesis},\ \bibinfo
  {school} {Lyon, IPN} (\bibinfo {year} {2011})\BibitemShut {NoStop}%
\bibitem [{\citenamefont {Böhlen}\ \emph {et~al.}(2014)\citenamefont
  {Böhlen}, \citenamefont {Cerutti}, \citenamefont {Chin}, \citenamefont
  {Fassò}, \citenamefont {Ferrari}, \citenamefont {Ortega}, \citenamefont
  {Mairani}, \citenamefont {Sala}, \citenamefont {Smirnov},\ and\ \citenamefont
  {Vlachoudis}}]{Bohlen:2014buj}%
  \BibitemOpen
  \bibfield  {author} {\bibinfo {author} {\bibfnamefont {T.}~\bibnamefont
  {Böhlen}}, \bibinfo {author} {\bibfnamefont {F.}~\bibnamefont {Cerutti}},
  \bibinfo {author} {\bibfnamefont {M.}~\bibnamefont {Chin}}, \bibinfo {author}
  {\bibfnamefont {A.}~\bibnamefont {Fassò}}, \bibinfo {author} {\bibfnamefont
  {A.}~\bibnamefont {Ferrari}}, \bibinfo {author} {\bibfnamefont
  {P.}~\bibnamefont {Ortega}}, \bibinfo {author} {\bibfnamefont
  {A.}~\bibnamefont {Mairani}}, \bibinfo {author} {\bibfnamefont
  {P.}~\bibnamefont {Sala}}, \bibinfo {author} {\bibfnamefont {G.}~\bibnamefont
  {Smirnov}}, \ and\ \bibinfo {author} {\bibfnamefont {V.}~\bibnamefont
  {Vlachoudis}},\ }\href {\doibase 10.1016/j.nds.2014.07.049} {\bibfield
  {journal} {\bibinfo  {journal} {Nucl. Data Sheets}\ }\textbf {\bibinfo
  {volume} {120}},\ \bibinfo {pages} {211} (\bibinfo {year}
  {2014})}\BibitemShut {NoStop}%
\bibitem [{\citenamefont {Ferrari}\ \emph {et~al.}(2005)\citenamefont
  {Ferrari}, \citenamefont {Sala}, \citenamefont {Fasso},\ and\ \citenamefont
  {Ranft}}]{Ferrari:2005zk}%
  \BibitemOpen
  \bibfield  {author} {\bibinfo {author} {\bibfnamefont {A.}~\bibnamefont
  {Ferrari}}, \bibinfo {author} {\bibfnamefont {P.~R.}\ \bibnamefont {Sala}},
  \bibinfo {author} {\bibfnamefont {A.}~\bibnamefont {Fasso}}, \ and\ \bibinfo
  {author} {\bibfnamefont {J.}~\bibnamefont {Ranft}},\ }\href {\doibase
  10.2172/877507} {\  (\bibinfo {year} {2005}),\ 10.2172/877507}\BibitemShut
  {NoStop}%
\bibitem [{\citenamefont {Whalley}\ \emph {et~al.}(2005)\citenamefont
  {Whalley}, \citenamefont {Bourilkov},\ and\ \citenamefont
  {Group}}]{Whalley:2005nh}%
  \BibitemOpen
  \bibfield  {author} {\bibinfo {author} {\bibfnamefont {M.~R.}\ \bibnamefont
  {Whalley}}, \bibinfo {author} {\bibfnamefont {D.}~\bibnamefont {Bourilkov}},
  \ and\ \bibinfo {author} {\bibfnamefont {R.~C.}\ \bibnamefont {Group}},\ }in\
  \href@noop {} {\emph {\bibinfo {booktitle} {{HERA and the LHC: A Workshop on
  the Implications of HERA and LHC Physics (Startup Meeting, CERN, 26-27 March
  2004; Midterm Meeting, CERN, 11-13 October 2004)}}}}\ (\bibinfo {year}
  {2005})\ pp.\ \bibinfo {pages} {575--581},\ \Eprint
  {http://arxiv.org/abs/hep-ph/0508110} {arXiv:hep-ph/0508110} \BibitemShut
  {NoStop}%
\bibitem [{\citenamefont {Eskola}\ \emph {et~al.}(2009)\citenamefont {Eskola},
  \citenamefont {Paukkunen},\ and\ \citenamefont {Salgado}}]{Eskola:2009uj}%
  \BibitemOpen
  \bibfield  {author} {\bibinfo {author} {\bibfnamefont {K.}~\bibnamefont
  {Eskola}}, \bibinfo {author} {\bibfnamefont {H.}~\bibnamefont {Paukkunen}}, \
  and\ \bibinfo {author} {\bibfnamefont {C.}~\bibnamefont {Salgado}},\ }\href
  {\doibase 10.1088/1126-6708/2009/04/065} {\bibfield  {journal} {\bibinfo
  {journal} {JHEP}\ }\textbf {\bibinfo {volume} {04}},\ \bibinfo {pages} {065}
  (\bibinfo {year} {2009})},\ \Eprint {http://arxiv.org/abs/0902.4154}
  {arXiv:0902.4154 [hep-ph]} \BibitemShut {NoStop}%
\bibitem [{\citenamefont {Salgado}\ and\ \citenamefont
  {Wiedemann}(2003)}]{SW:2003}%
  \BibitemOpen
  \bibfield  {author} {\bibinfo {author} {\bibfnamefont {C.}~\bibnamefont
  {Salgado}}\ and\ \bibinfo {author} {\bibfnamefont {U.~A.}\ \bibnamefont
  {Wiedemann}},\ }\href {\doibase 10.1103/PhysRevD.68.014008} {\bibfield
  {journal} {\bibinfo  {journal} {Phys. Rev. D}\ }\textbf {\bibinfo {volume}
  {68}},\ \bibinfo {pages} {014008} (\bibinfo {year} {2003})},\ \Eprint
  {http://arxiv.org/abs/hep-ph/0302184} {arXiv:hep-ph/0302184} \BibitemShut
  {NoStop}%
\bibitem [{\citenamefont {Abramowicz}\ and\ \citenamefont
  {Caldwell}(1999)}]{Abramowicz:1998ii}%
  \BibitemOpen
  \bibfield  {author} {\bibinfo {author} {\bibfnamefont {H.}~\bibnamefont
  {Abramowicz}}\ and\ \bibinfo {author} {\bibfnamefont {A.}~\bibnamefont
  {Caldwell}},\ }\href {\doibase 10.1103/RevModPhys.71.1275} {\bibfield
  {journal} {\bibinfo  {journal} {Rev. Mod. Phys.}\ }\textbf {\bibinfo {volume}
  {71}},\ \bibinfo {pages} {1275} (\bibinfo {year} {1999})},\ \Eprint
  {http://arxiv.org/abs/hep-ex/9903037} {arXiv:hep-ex/9903037} \BibitemShut
  {NoStop}%
\bibitem [{\citenamefont {Toll}\ and\ \citenamefont
  {Ullrich}(2013)}]{Toll:2012mb}%
  \BibitemOpen
  \bibfield  {author} {\bibinfo {author} {\bibfnamefont {T.}~\bibnamefont
  {Toll}}\ and\ \bibinfo {author} {\bibfnamefont {T.}~\bibnamefont {Ullrich}},\
  }\href {\doibase 10.1103/PhysRevC.87.024913} {\bibfield  {journal} {\bibinfo
  {journal} {Phys. Rev. C}\ }\textbf {\bibinfo {volume} {87}},\ \bibinfo
  {pages} {024913} (\bibinfo {year} {2013})},\ \Eprint
  {http://arxiv.org/abs/1211.3048} {arXiv:1211.3048 [hep-ph]} \BibitemShut
  {NoStop}%
\bibitem [{\citenamefont {Iancu}\ \emph {et~al.}(2002)\citenamefont {Iancu},
  \citenamefont {Leonidov},\ and\ \citenamefont {McLerran}}]{Iancu:2002xk}%
  \BibitemOpen
  \bibfield  {author} {\bibinfo {author} {\bibfnamefont {E.}~\bibnamefont
  {Iancu}}, \bibinfo {author} {\bibfnamefont {A.}~\bibnamefont {Leonidov}}, \
  and\ \bibinfo {author} {\bibfnamefont {L.}~\bibnamefont {McLerran}},\ }in\
  \href@noop {} {\emph {\bibinfo {booktitle} {{Cargese Summer School on QCD
  Perspectives on Hot and Dense Matter}}}}\ (\bibinfo {year} {2002})\ pp.\
  \bibinfo {pages} {73--145},\ \Eprint {http://arxiv.org/abs/hep-ph/0202270}
  {arXiv:hep-ph/0202270} \BibitemShut {NoStop}%
\bibitem [{\citenamefont {Robles~Gajardo}\ \emph {et~al.}(2022)\citenamefont
  {Robles~Gajardo}, \citenamefont {Accardi}, \citenamefont {Baker},
  \citenamefont {Brooks}, \citenamefont {Dupr\'e}, \citenamefont {Ehrhart},
  \citenamefont {L\'opez},\ and\ \citenamefont {Tu}}]{RoblesGajardo:2022efe}%
  \BibitemOpen
  \bibfield  {author} {\bibinfo {author} {\bibfnamefont {C.~M.}\ \bibnamefont
  {Robles~Gajardo}}, \bibinfo {author} {\bibfnamefont {A.}~\bibnamefont
  {Accardi}}, \bibinfo {author} {\bibfnamefont {M.~D.}\ \bibnamefont {Baker}},
  \bibinfo {author} {\bibfnamefont {W.~K.}\ \bibnamefont {Brooks}}, \bibinfo
  {author} {\bibfnamefont {R.}~\bibnamefont {Dupr\'e}}, \bibinfo {author}
  {\bibfnamefont {M.}~\bibnamefont {Ehrhart}}, \bibinfo {author} {\bibfnamefont
  {J.~A.}\ \bibnamefont {L\'opez}}, \ and\ \bibinfo {author} {\bibfnamefont
  {Z.}~\bibnamefont {Tu}},\ }\href@noop {} {\  (\bibinfo {year} {2022})},\
  \Eprint {http://arxiv.org/abs/2203.16665} {arXiv:2203.16665 [hep-ph]}
  \BibitemShut {NoStop}%
\bibitem [{\citenamefont {Bertini}(1963)}]{PhysRev.131.1801}%
  \BibitemOpen
  \bibfield  {author} {\bibinfo {author} {\bibfnamefont {H.~W.}\ \bibnamefont
  {Bertini}},\ }\href {\doibase 10.1103/PhysRev.131.1801} {\bibfield  {journal}
  {\bibinfo  {journal} {Phys. Rev.}\ }\textbf {\bibinfo {volume} {131}},\
  \bibinfo {pages} {1801} (\bibinfo {year} {1963})}\BibitemShut {NoStop}%
\bibitem [{\citenamefont {Miller}\ \emph {et~al.}(2007)\citenamefont {Miller},
  \citenamefont {Reygers}, \citenamefont {Sanders},\ and\ \citenamefont
  {Steinberg}}]{Miller:2007ri}%
  \BibitemOpen
  \bibfield  {author} {\bibinfo {author} {\bibfnamefont {M.~L.}\ \bibnamefont
  {Miller}}, \bibinfo {author} {\bibfnamefont {K.}~\bibnamefont {Reygers}},
  \bibinfo {author} {\bibfnamefont {S.~J.}\ \bibnamefont {Sanders}}, \ and\
  \bibinfo {author} {\bibfnamefont {P.}~\bibnamefont {Steinberg}},\ }\href
  {\doibase 10.1146/annurev.nucl.57.090506.123020} {\bibfield  {journal}
  {\bibinfo  {journal} {Ann. Rev. Nucl. Part. Sci.}\ }\textbf {\bibinfo
  {volume} {57}},\ \bibinfo {pages} {205} (\bibinfo {year} {2007})},\ \Eprint
  {http://arxiv.org/abs/nucl-ex/0701025} {arXiv:nucl-ex/0701025} \BibitemShut
  {NoStop}%
\bibitem [{\citenamefont {Frankfurt}\ and\ \citenamefont
  {Strikman}(1988)}]{Frankfurt:1988nt}%
  \BibitemOpen
  \bibfield  {author} {\bibinfo {author} {\bibfnamefont {L.~L.}\ \bibnamefont
  {Frankfurt}}\ and\ \bibinfo {author} {\bibfnamefont {M.~I.}\ \bibnamefont
  {Strikman}},\ }\href {\doibase 10.1016/0370-1573(88)90179-2} {\bibfield
  {journal} {\bibinfo  {journal} {Phys. Rept.}\ }\textbf {\bibinfo {volume}
  {160}},\ \bibinfo {pages} {235} (\bibinfo {year} {1988})}\BibitemShut
  {NoStop}%
\bibitem [{\citenamefont {Frankfurt}\ \emph {et~al.}(2003)\citenamefont
  {Frankfurt}, \citenamefont {Guzey},\ and\ \citenamefont
  {Strikman}}]{Frankfurt:2003jf}%
  \BibitemOpen
  \bibfield  {author} {\bibinfo {author} {\bibfnamefont {L.}~\bibnamefont
  {Frankfurt}}, \bibinfo {author} {\bibfnamefont {V.}~\bibnamefont {Guzey}}, \
  and\ \bibinfo {author} {\bibfnamefont {M.}~\bibnamefont {Strikman}},\
  }\href@noop {} {\bibfield  {journal} {\bibinfo  {journal} {Phys. Rev. Lett.}\
  }\textbf {\bibinfo {volume} {91}},\ \bibinfo {pages} {202001} (\bibinfo
  {year} {2003})}\BibitemShut {NoStop}%
\bibitem [{\citenamefont {Frankfurt}\ \emph {et~al.}(2006)\citenamefont
  {Frankfurt}, \citenamefont {Guzey},\ and\ \citenamefont
  {Strikman}}]{Frankfurt:2006am}%
  \BibitemOpen
  \bibfield  {author} {\bibinfo {author} {\bibfnamefont {L.}~\bibnamefont
  {Frankfurt}}, \bibinfo {author} {\bibfnamefont {V.}~\bibnamefont {Guzey}}, \
  and\ \bibinfo {author} {\bibfnamefont {M.}~\bibnamefont {Strikman}},\
  }\href@noop {} {\bibfield  {journal} {\bibinfo  {journal} {Mod. Phys. Lett.
  A}\ }\textbf {\bibinfo {volume} {21}},\ \bibinfo {pages} {23} (\bibinfo
  {year} {2006})}\BibitemShut {NoStop}%
\bibitem [{\citenamefont {Bodek}\ and\ \citenamefont
  {Ritchie}(1981)}]{Bodek:1980ar}%
  \BibitemOpen
  \bibfield  {author} {\bibinfo {author} {\bibfnamefont {A.}~\bibnamefont
  {Bodek}}\ and\ \bibinfo {author} {\bibfnamefont {J.~L.}\ \bibnamefont
  {Ritchie}},\ }\href {\doibase 10.1103/PhysRevD.23.1070} {\bibfield  {journal}
  {\bibinfo  {journal} {Phys. Rev. D}\ }\textbf {\bibinfo {volume} {23}},\
  \bibinfo {pages} {1070} (\bibinfo {year} {1981})}\BibitemShut {NoStop}%
\bibitem [{\citenamefont {Ciofi~degli Atti}\ and\ \citenamefont
  {Simula}(1996)}]{CiofidegliAtti:1995qe}%
  \BibitemOpen
  \bibfield  {author} {\bibinfo {author} {\bibfnamefont {C.}~\bibnamefont
  {Ciofi~degli Atti}}\ and\ \bibinfo {author} {\bibfnamefont {S.}~\bibnamefont
  {Simula}},\ }\href {\doibase 10.1103/PhysRevC.53.1689} {\bibfield  {journal}
  {\bibinfo  {journal} {Phys. Rev. C}\ }\textbf {\bibinfo {volume} {53}},\
  \bibinfo {pages} {1689} (\bibinfo {year} {1996})},\ \Eprint
  {http://arxiv.org/abs/nucl-th/9507024} {arXiv:nucl-th/9507024} \BibitemShut
  {NoStop}%
\bibitem [{\citenamefont {Strikman}\ and\ \citenamefont
  {Weiss}(2018)}]{Strikman:2017koc}%
  \BibitemOpen
  \bibfield  {author} {\bibinfo {author} {\bibfnamefont {M.}~\bibnamefont
  {Strikman}}\ and\ \bibinfo {author} {\bibfnamefont {C.}~\bibnamefont
  {Weiss}},\ }\href {\doibase 10.1103/PhysRevC.97.035209} {\bibfield  {journal}
  {\bibinfo  {journal} {Phys. Rev. C}\ }\textbf {\bibinfo {volume} {97}},\
  \bibinfo {pages} {035209} (\bibinfo {year} {2018})},\ \Eprint
  {http://arxiv.org/abs/1706.02244} {arXiv:1706.02244 [hep-ph]} \BibitemShut
  {NoStop}%
\bibitem [{\citenamefont {Adams}\ \emph
  {et~al.}(1995{\natexlab{a}})\citenamefont {Adams} \emph
  {et~al.}}]{E665:1995utr}%
  \BibitemOpen
  \bibfield  {author} {\bibinfo {author} {\bibfnamefont {M.~R.}\ \bibnamefont
  {Adams}} \emph {et~al.} (\bibinfo {collaboration} {E665}),\ }\href {\doibase
  10.1103/PhysRevLett.74.5198} {\bibfield  {journal} {\bibinfo  {journal}
  {Phys. Rev. Lett.}\ }\textbf {\bibinfo {volume} {74}},\ \bibinfo {pages}
  {5198} (\bibinfo {year} {1995}{\natexlab{a}})},\ \bibinfo {note} {[Erratum:
  Phys.Rev.Lett. 80, 2020--2021 (1998)]}\BibitemShut {NoStop}%
\bibitem [{\citenamefont {Ferrari}\ \emph {et~al.}(1996)\citenamefont
  {Ferrari}, \citenamefont {Sala}, \citenamefont {Ranft},\ and\ \citenamefont
  {Roesler}}]{Ferrari:1995cq}%
  \BibitemOpen
  \bibfield  {author} {\bibinfo {author} {\bibfnamefont {A.}~\bibnamefont
  {Ferrari}}, \bibinfo {author} {\bibfnamefont {P.~R.}\ \bibnamefont {Sala}},
  \bibinfo {author} {\bibfnamefont {J.}~\bibnamefont {Ranft}}, \ and\ \bibinfo
  {author} {\bibfnamefont {S.}~\bibnamefont {Roesler}},\ }\href {\doibase
  10.1007/s002880050119} {\bibfield  {journal} {\bibinfo  {journal} {Z. Phys.
  C}\ }\textbf {\bibinfo {volume} {70}},\ \bibinfo {pages} {413} (\bibinfo
  {year} {1996})},\ \Eprint {http://arxiv.org/abs/nucl-th/9509039}
  {arXiv:nucl-th/9509039} \BibitemShut {NoStop}%
\bibitem [{\citenamefont {Zheng}\ \emph {et~al.}(2014)\citenamefont {Zheng},
  \citenamefont {Aschenauer},\ and\ \citenamefont {Lee}}]{Zheng:2014cha}%
  \BibitemOpen
  \bibfield  {author} {\bibinfo {author} {\bibfnamefont {L.}~\bibnamefont
  {Zheng}}, \bibinfo {author} {\bibfnamefont {E.}~\bibnamefont {Aschenauer}}, \
  and\ \bibinfo {author} {\bibfnamefont {J.}~\bibnamefont {Lee}},\ }\href
  {\doibase 10.1140/epja/i2014-14189-3} {\bibfield  {journal} {\bibinfo
  {journal} {Eur. Phys. J. A}\ }\textbf {\bibinfo {volume} {50}},\ \bibinfo
  {pages} {189} (\bibinfo {year} {2014})},\ \Eprint
  {http://arxiv.org/abs/1407.8055} {arXiv:1407.8055 [hep-ex]} \BibitemShut
  {NoStop}%
\bibitem [{\citenamefont {Wolf}(2010)}]{Wolf:2009jm}%
  \BibitemOpen
  \bibfield  {author} {\bibinfo {author} {\bibfnamefont {G.}~\bibnamefont
  {Wolf}},\ }\href {\doibase 10.1088/0034-4885/73/11/116202} {\bibfield
  {journal} {\bibinfo  {journal} {Rept. Prog. Phys.}\ }\textbf {\bibinfo
  {volume} {73}},\ \bibinfo {pages} {116202} (\bibinfo {year} {2010})},\
  \Eprint {http://arxiv.org/abs/0907.1217} {arXiv:0907.1217 [hep-ex]}
  \BibitemShut {NoStop}%
\bibitem [{\citenamefont {Armesto}\ \emph {et~al.}(2019)\citenamefont
  {Armesto}, \citenamefont {Newman}, \citenamefont {S\l{}omi\'nski},\ and\
  \citenamefont {Sta\'sto}}]{Armesto:2019gxy}%
  \BibitemOpen
  \bibfield  {author} {\bibinfo {author} {\bibfnamefont {N.}~\bibnamefont
  {Armesto}}, \bibinfo {author} {\bibfnamefont {P.~R.}\ \bibnamefont {Newman}},
  \bibinfo {author} {\bibfnamefont {W.}~\bibnamefont {S\l{}omi\'nski}}, \ and\
  \bibinfo {author} {\bibfnamefont {A.~M.}\ \bibnamefont {Sta\'sto}},\ }\href
  {\doibase 10.1103/PhysRevD.100.074022} {\bibfield  {journal} {\bibinfo
  {journal} {Phys. Rev. D}\ }\textbf {\bibinfo {volume} {100}},\ \bibinfo
  {pages} {074022} (\bibinfo {year} {2019})},\ \Eprint
  {http://arxiv.org/abs/1901.09076} {arXiv:1901.09076 [hep-ph]} \BibitemShut
  {NoStop}%
\bibitem [{\citenamefont {Toll}\ and\ \citenamefont
  {Ullrich}(2014)}]{Toll:2013gda}%
  \BibitemOpen
  \bibfield  {author} {\bibinfo {author} {\bibfnamefont {T.}~\bibnamefont
  {Toll}}\ and\ \bibinfo {author} {\bibfnamefont {T.}~\bibnamefont {Ullrich}},\
  }\href {\doibase 10.1016/j.cpc.2014.03.010} {\bibfield  {journal} {\bibinfo
  {journal} {Comput. Phys. Commun.}\ }\textbf {\bibinfo {volume} {185}},\
  \bibinfo {pages} {1835} (\bibinfo {year} {2014})},\ \Eprint
  {http://arxiv.org/abs/1307.8059} {arXiv:1307.8059 [hep-ph]} \BibitemShut
  {NoStop}%
\bibitem [{\citenamefont {Chekanov}\ \emph {et~al.}(2009)\citenamefont
  {Chekanov} \emph {et~al.}}]{ZEUS:2008ipu}%
  \BibitemOpen
  \bibfield  {author} {\bibinfo {author} {\bibfnamefont {S.}~\bibnamefont
  {Chekanov}} \emph {et~al.} (\bibinfo {collaboration} {ZEUS}),\ }\href
  {\doibase 10.1088/1126-6708/2009/06/074} {\bibfield  {journal} {\bibinfo
  {journal} {JHEP}\ }\textbf {\bibinfo {volume} {06}},\ \bibinfo {pages} {074}
  (\bibinfo {year} {2009})},\ \Eprint {http://arxiv.org/abs/0812.2416}
  {arXiv:0812.2416 [hep-ex]} \BibitemShut {NoStop}%
\bibitem [{\citenamefont {Airapetian}\ \emph {et~al.}(2010)\citenamefont
  {Airapetian} \emph {et~al.}}]{HERMES:2010nas}%
  \BibitemOpen
  \bibfield  {author} {\bibinfo {author} {\bibfnamefont {A.}~\bibnamefont
  {Airapetian}} \emph {et~al.} (\bibinfo {collaboration} {HERMES}),\ }\href
  {\doibase 10.1007/JHEP08(2010)130} {\bibfield  {journal} {\bibinfo  {journal}
  {JHEP}\ }\textbf {\bibinfo {volume} {08}},\ \bibinfo {pages} {130} (\bibinfo
  {year} {2010})},\ \Eprint {http://arxiv.org/abs/1002.3921} {arXiv:1002.3921
  [hep-ex]} \BibitemShut {NoStop}%
\bibitem [{\citenamefont {Liebing}(2004)}]{Liebing:2004us}%
  \BibitemOpen
  \bibfield  {author} {\bibinfo {author} {\bibfnamefont {P.}~\bibnamefont
  {Liebing}},\ }\emph {\bibinfo {title} {{Can the gluon polarization in the
  nucleon be extracted from HERMES data on single high-p(T) hadrons?}}},\ \href
  {\doibase 10.3204/DESY-THESIS-2004-036} {Ph.D. thesis},\ \bibinfo  {school}
  {Hamburg U.} (\bibinfo {year} {2004})\BibitemShut {NoStop}%
\bibitem [{\citenamefont {Sloan}(2019)}]{Sloan:2690197}%
  \BibitemOpen
  \bibfield  {author} {\bibinfo {author} {\bibfnamefont {T.}~\bibnamefont
  {Sloan}},\ }\href {\doibase 10.23731/CYRM-2019-005} {\emph {\bibinfo {title}
  {{History of the European Muon Collaboration (EMC)}}}},\ CERN Yellow Reports:
  Monographs\ (\bibinfo  {publisher} {CERN},\ \bibinfo {address} {Geneva},\
  \bibinfo {year} {2019})\BibitemShut {NoStop}%
\bibitem [{\citenamefont {Andreev}\ \emph {et~al.}(2021)\citenamefont {Andreev}
  \emph {et~al.}}]{H1:2020zpd}%
  \BibitemOpen
  \bibfield  {author} {\bibinfo {author} {\bibfnamefont {V.}~\bibnamefont
  {Andreev}} \emph {et~al.} (\bibinfo {collaboration} {H1}),\ }\href {\doibase
  10.1140/epjc/s10052-021-08896-1} {\bibfield  {journal} {\bibinfo  {journal}
  {Eur. Phys. J. C}\ }\textbf {\bibinfo {volume} {81}},\ \bibinfo {pages} {212}
  (\bibinfo {year} {2021})},\ \Eprint {http://arxiv.org/abs/2011.01812}
  {arXiv:2011.01812 [hep-ex]} \BibitemShut {NoStop}%
\bibitem [{\citenamefont {Sjöstrand}\ \emph {et~al.}(2008)\citenamefont
  {Sjöstrand}, \citenamefont {Mrenna},\ and\ \citenamefont
  {Skands}}]{Sjostrand:2007gs}%
  \BibitemOpen
  \bibfield  {author} {\bibinfo {author} {\bibfnamefont {T.}~\bibnamefont
  {Sjöstrand}}, \bibinfo {author} {\bibfnamefont {S.}~\bibnamefont {Mrenna}},
  \ and\ \bibinfo {author} {\bibfnamefont {P.~Z.}\ \bibnamefont {Skands}},\
  }\href {\doibase 10.1016/j.cpc.2008.01.036} {\bibfield  {journal} {\bibinfo
  {journal} {Comput. Phys. Commun.}\ }\textbf {\bibinfo {volume} {178}},\
  \bibinfo {pages} {852} (\bibinfo {year} {2008})},\ \Eprint
  {http://arxiv.org/abs/0710.3820} {arXiv:0710.3820 [hep-ph]} \BibitemShut
  {NoStop}%
\bibitem [{\citenamefont {Sjöstrand}\ \emph {et~al.}(2015)\citenamefont
  {Sjöstrand}, \citenamefont {Ask}, \citenamefont {Christiansen},
  \citenamefont {Corke}, \citenamefont {Desai}, \citenamefont {Ilten},
  \citenamefont {Mrenna}, \citenamefont {Prestel}, \citenamefont {Rasmussen},\
  and\ \citenamefont {Skands}}]{Sjostrand:2014zea}%
  \BibitemOpen
  \bibfield  {author} {\bibinfo {author} {\bibfnamefont {T.}~\bibnamefont
  {Sjöstrand}}, \bibinfo {author} {\bibfnamefont {S.}~\bibnamefont {Ask}},
  \bibinfo {author} {\bibfnamefont {J.~R.}\ \bibnamefont {Christiansen}},
  \bibinfo {author} {\bibfnamefont {R.}~\bibnamefont {Corke}}, \bibinfo
  {author} {\bibfnamefont {N.}~\bibnamefont {Desai}}, \bibinfo {author}
  {\bibfnamefont {P.}~\bibnamefont {Ilten}}, \bibinfo {author} {\bibfnamefont
  {S.}~\bibnamefont {Mrenna}}, \bibinfo {author} {\bibfnamefont
  {S.}~\bibnamefont {Prestel}}, \bibinfo {author} {\bibfnamefont {C.~O.}\
  \bibnamefont {Rasmussen}}, \ and\ \bibinfo {author} {\bibfnamefont {P.~Z.}\
  \bibnamefont {Skands}},\ }\href {\doibase 10.1016/j.cpc.2015.01.024}
  {\bibfield  {journal} {\bibinfo  {journal} {Comput. Phys. Commun.}\ }\textbf
  {\bibinfo {volume} {191}},\ \bibinfo {pages} {159} (\bibinfo {year}
  {2015})},\ \Eprint {http://arxiv.org/abs/1410.3012} {arXiv:1410.3012
  [hep-ph]} \BibitemShut {NoStop}%
\bibitem [{\citenamefont {Bierlich}\ \emph {et~al.}(2020)\citenamefont
  {Bierlich} \emph {et~al.}}]{Bierlich:2019rhm}%
  \BibitemOpen
  \bibfield  {author} {\bibinfo {author} {\bibfnamefont {C.}~\bibnamefont
  {Bierlich}} \emph {et~al.},\ }\href {\doibase 10.21468/SciPostPhys.8.2.026}
  {\bibfield  {journal} {\bibinfo  {journal} {SciPost Phys.}\ }\textbf
  {\bibinfo {volume} {8}},\ \bibinfo {pages} {026} (\bibinfo {year} {2020})},\
  \Eprint {http://arxiv.org/abs/1912.05451} {arXiv:1912.05451 [hep-ph]}
  \BibitemShut {NoStop}%
\bibitem [{\citenamefont {Adams}\ \emph
  {et~al.}(1995{\natexlab{b}})\citenamefont {Adams} \emph
  {et~al.}}]{adams1995nuclear}%
  \BibitemOpen
  \bibfield  {author} {\bibinfo {author} {\bibfnamefont {M.~R.}\ \bibnamefont
  {Adams}} \emph {et~al.} (\bibinfo {collaboration} {E665}),\ }\href {\doibase
  10.1007/BF01571879} {\bibfield  {journal} {\bibinfo  {journal} {Zeitschrift
  f{\"u}r Physik C Particles and Fields}\ }\textbf {\bibinfo {volume} {65}},\
  \bibinfo {pages} {225} (\bibinfo {year} {1995}{\natexlab{b}})}\BibitemShut
  {NoStop}%
\bibitem [{\citenamefont {Alver}\ \emph {et~al.}(2008)\citenamefont {Alver},
  \citenamefont {Baker}, \citenamefont {Loizides},\ and\ \citenamefont
  {Steinberg}}]{Alver:2008aq}%
  \BibitemOpen
  \bibfield  {author} {\bibinfo {author} {\bibfnamefont {B.}~\bibnamefont
  {Alver}}, \bibinfo {author} {\bibfnamefont {M.}~\bibnamefont {Baker}},
  \bibinfo {author} {\bibfnamefont {C.}~\bibnamefont {Loizides}}, \ and\
  \bibinfo {author} {\bibfnamefont {P.}~\bibnamefont {Steinberg}},\ }\href@noop
  {} {\  (\bibinfo {year} {2008})},\ \Eprint {http://arxiv.org/abs/0805.4411}
  {arXiv:0805.4411 [nucl-ex]} \BibitemShut {NoStop}%
\bibitem [{\citenamefont {Abdallah}\ \emph {et~al.}(2022)\citenamefont
  {Abdallah} \emph {et~al.}}]{STAR:2021wwq}%
  \BibitemOpen
  \bibfield  {author} {\bibinfo {author} {\bibfnamefont {M.}~\bibnamefont
  {Abdallah}} \emph {et~al.} (\bibinfo {collaboration} {STAR}),\ }\href
  {\doibase 10.1103/PhysRevLett.128.122303} {\bibfield  {journal} {\bibinfo
  {journal} {Phys. Rev. Lett.}\ }\textbf {\bibinfo {volume} {128}},\ \bibinfo
  {pages} {122303} (\bibinfo {year} {2022})},\ \Eprint
  {http://arxiv.org/abs/2109.07625} {arXiv:2109.07625 [nucl-ex]} \BibitemShut
  {NoStop}%
\bibitem [{\citenamefont {Hen}\ \emph {et~al.}(2013)\citenamefont {Hen},
  \citenamefont {Higinbotham}, \citenamefont {Miller}, \citenamefont
  {Piasetzky},\ and\ \citenamefont {Weinstein}}]{Hen:2013oha}%
  \BibitemOpen
  \bibfield  {author} {\bibinfo {author} {\bibfnamefont {O.}~\bibnamefont
  {Hen}}, \bibinfo {author} {\bibfnamefont {D.~W.}\ \bibnamefont
  {Higinbotham}}, \bibinfo {author} {\bibfnamefont {G.~A.}\ \bibnamefont
  {Miller}}, \bibinfo {author} {\bibfnamefont {E.}~\bibnamefont {Piasetzky}}, \
  and\ \bibinfo {author} {\bibfnamefont {L.~B.}\ \bibnamefont {Weinstein}},\
  }\href {\doibase 10.1142/S0218301313300178} {\bibfield  {journal} {\bibinfo
  {journal} {Int. J. Mod. Phys.}\ }\textbf {\bibinfo {volume} {E22}},\ \bibinfo
  {pages} {1330017} (\bibinfo {year} {2013})},\ \Eprint
  {http://arxiv.org/abs/1304.2813} {arXiv:1304.2813 [nucl-th]} \BibitemShut
  {NoStop}%
\bibitem [{\citenamefont {Weinstein}\ \emph {et~al.}(2011)\citenamefont
  {Weinstein}, \citenamefont {Piasetzky}, \citenamefont {Higinbotham},
  \citenamefont {Gomez}, \citenamefont {Hen},\ and\ \citenamefont
  {Shneor}}]{weinstein11}%
  \BibitemOpen
  \bibfield  {author} {\bibinfo {author} {\bibfnamefont {L.~B.}\ \bibnamefont
  {Weinstein}}, \bibinfo {author} {\bibfnamefont {E.}~\bibnamefont
  {Piasetzky}}, \bibinfo {author} {\bibfnamefont {D.~W.}\ \bibnamefont
  {Higinbotham}}, \bibinfo {author} {\bibfnamefont {J.}~\bibnamefont {Gomez}},
  \bibinfo {author} {\bibfnamefont {O.}~\bibnamefont {Hen}}, \ and\ \bibinfo
  {author} {\bibfnamefont {R.}~\bibnamefont {Shneor}},\ }\href {\doibase
  10.1103/PhysRevLett.106.052301} {\bibfield  {journal} {\bibinfo  {journal}
  {Phys. Rev. Lett.}\ }\textbf {\bibinfo {volume} {106}},\ \bibinfo {pages}
  {052301} (\bibinfo {year} {2011})}\BibitemShut {NoStop}%
\bibitem [{\citenamefont {Seely}\ \emph {et~al.}(2009)\citenamefont {Seely}
  \emph {et~al.}}]{Seely:2009gt}%
  \BibitemOpen
  \bibfield  {author} {\bibinfo {author} {\bibfnamefont {J.}~\bibnamefont
  {Seely}} \emph {et~al.},\ }\href {\doibase 10.1103/PhysRevLett.103.202301}
  {\bibfield  {journal} {\bibinfo  {journal} {Phys. Rev. Lett.}\ }\textbf
  {\bibinfo {volume} {103}},\ \bibinfo {pages} {202301} (\bibinfo {year}
  {2009})},\ \Eprint {http://arxiv.org/abs/0904.4448} {arXiv:0904.4448
  [nucl-ex]} \BibitemShut {NoStop}%
\bibitem [{\citenamefont {Norton}(2003)}]{Norton:2003cb}%
  \BibitemOpen
  \bibfield  {author} {\bibinfo {author} {\bibfnamefont {P.~R.}\ \bibnamefont
  {Norton}},\ }\href {\doibase 10.1088/0034-4885/66/8/201} {\bibfield
  {journal} {\bibinfo  {journal} {Rept. Prog. Phys.}\ }\textbf {\bibinfo
  {volume} {66}},\ \bibinfo {pages} {1253} (\bibinfo {year}
  {2003})}\BibitemShut {NoStop}%
\bibitem [{\citenamefont {Malace}\ \emph {et~al.}(2014)\citenamefont {Malace},
  \citenamefont {Gaskell}, \citenamefont {Higinbotham},\ and\ \citenamefont
  {Cloet}}]{Malace:2014uea}%
  \BibitemOpen
  \bibfield  {author} {\bibinfo {author} {\bibfnamefont {S.}~\bibnamefont
  {Malace}}, \bibinfo {author} {\bibfnamefont {D.}~\bibnamefont {Gaskell}},
  \bibinfo {author} {\bibfnamefont {D.~W.}\ \bibnamefont {Higinbotham}}, \ and\
  \bibinfo {author} {\bibfnamefont {I.}~\bibnamefont {Cloet}},\ }\href
  {\doibase 10.1142/S0218301314300136} {\bibfield  {journal} {\bibinfo
  {journal} {Int. J. Mod. Phys. E}\ }\textbf {\bibinfo {volume} {23}},\
  \bibinfo {pages} {1430013} (\bibinfo {year} {2014})},\ \Eprint
  {http://arxiv.org/abs/1405.1270} {arXiv:1405.1270 [nucl-ex]} \BibitemShut
  {NoStop}%
\bibitem [{\citenamefont {Weiss}\ \emph {et~al.}(2015)\citenamefont {Weiss},
  \citenamefont {Bazak},\ and\ \citenamefont {Barnea}}]{PhysRevC.92.054311}%
  \BibitemOpen
  \bibfield  {author} {\bibinfo {author} {\bibfnamefont {R.}~\bibnamefont
  {Weiss}}, \bibinfo {author} {\bibfnamefont {B.}~\bibnamefont {Bazak}}, \ and\
  \bibinfo {author} {\bibfnamefont {N.}~\bibnamefont {Barnea}},\ }\href
  {\doibase 10.1103/PhysRevC.92.054311} {\bibfield  {journal} {\bibinfo
  {journal} {Phys. Rev. C}\ }\textbf {\bibinfo {volume} {92}},\ \bibinfo
  {pages} {054311} (\bibinfo {year} {2015})}\BibitemShut {NoStop}%
\bibitem [{\citenamefont {Cruz-Torres}\ \emph {et~al.}(2021)\citenamefont
  {Cruz-Torres} \emph {et~al.}}]{Cruz-Torres:2019fum}%
  \BibitemOpen
  \bibfield  {author} {\bibinfo {author} {\bibfnamefont {R.}~\bibnamefont
  {Cruz-Torres}} \emph {et~al.},\ }\href {\doibase 10.1038/s41567-020-01053-7}
  {\bibfield  {journal} {\bibinfo  {journal} {Nature Phys.}\ }\textbf {\bibinfo
  {volume} {17}},\ \bibinfo {pages} {306} (\bibinfo {year} {2021})},\ \Eprint
  {http://arxiv.org/abs/1907.03658} {arXiv:1907.03658 [nucl-th]} \BibitemShut
  {NoStop}%
\bibitem [{\citenamefont {Hauenstein}\ \emph {et~al.}(2021)\citenamefont
  {Hauenstein} \emph {et~al.}}]{Hauenstein:2021zql}%
  \BibitemOpen
  \bibfield  {author} {\bibinfo {author} {\bibfnamefont {F.}~\bibnamefont
  {Hauenstein}} \emph {et~al.},\ }\href@noop {} {\  (\bibinfo {year} {2021})},\
  \Eprint {http://arxiv.org/abs/2109.09509} {arXiv:2109.09509
  [physics.ins-det]} \BibitemShut {NoStop}%
\end{thebibliography}%

\end{document}